%% file: IOX_with_supplement.tex
\documentclass[12pt]{article}
\usepackage{amsmath, amsthm}
\usepackage{graphicx}
\usepackage{enumerate}
\usepackage{natbib}
\usepackage{url} % not crucial - just used below for the URL 
\usepackage{proof-at-the-end}
\usepackage{makecell}
\usepackage{subfig}

\newtheorem*{prop*}{Proposition}

\newtheorem*{coro*}{Corollary}

\input{preamble}
\usepackage{amsthm}
\theoremstyle{definition}

\newtheorem{definition}{Definition}[section]

\newtheorem{proposition}{Proposition}[section]
\newtheorem{corollary}{Corollary}[section]

\newcommand{\modelname}{IOX}%{\textsc{iox}}
\newcommand{\modelnames}{IOX}%{\textsc{iox}}

\newcommand{\lmc}{LMC}%{\textsc{lmc}}
\newcommand{\lmcs}{LMCs}%{\textsc{lmc}s}

\newtheorem{innercustomthm}{Proposition}
\newenvironment{customprop}[1]
  {\renewcommand\theinnercustomthm{#1}\innercustomthm}
  {\endinnercustomthm}

\newtheorem{innercustomcoro}{Corollary}
\newenvironment{customcoro}[1]
  {\renewcommand\theinnercustomcoro{#1}\innercustomcoro}
  {\endinnercustomcoro}

\newcommand{\repourl}{\if0\blind
	{\url{github.com/mkln/spiox}}\fi
	\if1\blind
	{\url{github.com/}\texttt{[url redacted in blinded version]}}\fi}

\newcommand{\blind}{0}

\begin{document}

\def\spacingset#1{\renewcommand{\baselinestretch}%
{#1}\small\normalsize} \spacingset{1}

\date{}

\newcommand{\footremember}[2]{%
    \footnote{#2}
    \newcounter{#1}
    \setcounter{#1}{\value{footnote}}%
}
\newcommand{\footrecall}[1]{%
    \footnotemark[\value{#1}]%
} 
%%%%%%%%%%%%%%%%%%%%%%%%%%%%%%%%%%%%%%%%%%%%%%%%%%%%%%%%%%%%%%%%%%%%%%%%%%%%%%
\newcommand{\mytitle}{Inside-out cross-covariance \\ \vspace{.2cm} for spatial multivariate data}  

\if0\blind
{
  \title{\bf \mytitle}
  \author{Michele Peruzzi\footremember{alley}{Department of Biostatistics, University of Michigan--Ann Arbor.}
  }
  \maketitle
} \fi

\if1\blind
{
  \bigskip
  \bigskip
  \bigskip
  \begin{center}
    {\LARGE\bf \mytitle}
\end{center}
  \medskip
} \fi

\bigskip
\begin{abstract}
As the spatial features of multivariate data are increasingly central in researchers' applied problems, there is a growing demand for novel spatially aware methods that are flexible, easily interpretable, and scalable to large data. We develop inside-out cross-covariance (IOX) models for multivariate spatial likelihood-based inference. IOX leads to valid cross-covariance matrix functions which we interpret as inducing spatial dependence on independent replicates of a correlated random vector. The resulting sample cross-covariance matrices are ``inside-out'' relative to the ubiquitous linear model of coregionalization (LMC). However, unlike \lmcs, our methods offer direct marginal inference, easy prior elicitation of covariance parameters, the ability to model outcomes with unequal smoothness, and flexible dimension reduction. As a covariance model for a $q$-variate Gaussian process, IOX leads to scalable models for noisy vector data as well as flexible latent models. For large $n$ cases, IOX complements Vecchia approximations and related process-based methods relying on sparse graphical models. We demonstrate competitive performance of IOX on synthetic datasets as well as on colorectal cancer proteomics data. An R package implementing the proposed methods is available at \repourl.
\end{abstract}

\noindent%
 
%\newpage
\spacingset{1.9} % DON'T change the spacing. JASA template: 1.9!

\input{IOX_contents}

\subsection*{Acknowledgements}
The author thanks D. Dunson and B. Jin for helpful comments and suggestions, and the Associate Editor and referees for constructive feedback that improved both the manuscript and the reproducibility code.

\subsection*{Disclosure statement}
The author reports there are no competing interests to declare.

\newpage

\bigskip
\begin{center}
{\large\bf SUPPLEMENTARY MATERIAL}
\end{center}

\appendix
\input{IOX_supplement_contents}

\newpage

\spacingset{1}
\bibliographystyle{agsm}

\bibliography{biblio}
\end{document}

%% file: IOX_contents.tex
% !TEX root = article_organization.tex
\section{Introduction}
%The typical aim of multivariate spatial analysis in these contexts is to estimate the covariance function of each variable as well as the cross-covariance of each pair of variables. Because of the large number of inferential objects involved ($q(q+1)/2$ functions), fully flexible models for multivariate spatial data involve complicated validity conditions to verify and do not scale to large $q$. On the other hand, routine dimension reduction methods often fail to deliver greater insights than univariate spatial models, despite their increased computational cost. In this article, we introduce new methods for analyzing noisy multivariate spatial data that enable direct marginal inference of variables with possibly unequal smoothness, seamless integration within Bayesian hierarchies, easy prior elicitation, and without requiring the user to check complicated validity conditions. Our model can be paired with state-of-the-art methods for scalable process-based inference to analyze data at large scales.

Multivariate data with spatial dependence are increasingly common in fields such as ecology, environmental science, remote sensing, epidemiology, with growing interest in applying spatially aware methods to `omics' data, such as proteomics and genomics. Researchers aim to understand how the joint dependence across multiple variables changes as a function of space.
For example, in community ecology, joint species distribution models assume that species interactions are influenced by both their spatial context and the co-location with other species. 
Similarly, cancer development impacts the spatial arrangement of cells of different types within the tumor microenvironment. 
In these contexts, we expect multivariate spatial models to uncover deeper insights than univariate or non-spatial multivariate models, but their increased computational cost often limits their practical utility. In fact, any method for multivariate spatial data must balance flexibility and interpretability with scalability and parsimony. Achieving this balance is particularly difficult when we have a large number of spatially resolved variables (e.g., $q>10$) observed at many spatial sites ($n$ in the thousands). In such cases, fully flexible methods struggle to scale, whereas dimension reduction techniques can lead to interpretability or performance issues relative to univariate or non-spatial models. 

These challenges can be addressed in principle by modeling spatial dependence across multiple variables via a $q$-variate Gaussian process (GP). 
The GP is a flexible probabilistic framework for modeling spatial dependence of vector-valued outcomes. It can also act as a prior process in a Bayesian hierarchical regression for inferring spatial and cross-variable dependence, estimation of non-linear exposure effects, and predictions at unobserved locations. A GP is fully specified by its mean and cross-covariance matrix functions. The mean can typically be assumed to be zero or modeled as a separate component of the Bayesian hierarchy, if necessary. Cross-covariance matrix functions vary in terms of parsimony, interpretability, modeling flexibility, parameter identifiability, and computational feasibility. %, especially in ``large $n$, large $q$'' settings on which we focus. 
The choice of cross-covariance matrix function thus reflects the challenges we mentioned above. 
Let $\boldsymbol{M}_{q\times q}$ be the set of $q \times q$ real-valued matrices and $\calD \subset \Re^d$ the spatial domain of dimension $d$, then a cross-covariance matrix function is $\Cov: \calD\times\calD \rightarrow \boldsymbol{M}_{q\times q}$ such that $\Cov(\bl,\bl') = \Cov(\bl',\bl)^{\top}$ and $\sum_{i=1}^n\sum_{j=1}^n \bx_i^{\top}\Cov(\bl_i,\bl_j)\bx_j > 0$ for any integer $n$, any finite set $\{\bl_1,\bl_2,\ldots,\bl_n\}$, and for all $\bx_i,\bx_j \in \Re^q\setminus \{\bzero\}$. 
Then, if $\{\by(\bl), \bl \in \calD\}$ is a $q$-variate GP with cross-covariance matrix function $\Cov(\cdot,\cdot)$, the $(i,j)$th entry of $\Cov(\bl,\bl')$ evaluates to $C_{ij}(\bl, \bl')=\covf(y_i(\bl),y_j(\bl'))$. Following \cite{genton_ccov}, we call $C_{ii}(\cdot, \cdot)$ the marginal covariance and $C_{ij}(\cdot, \cdot), i\neq j$ as the cross-covariance functions. Building a cross-covariance matrix function boils down to specifying $q^2$ distinct covariance and cross-covariance functions and establishing the validity of the overall model; both tasks are increasingly difficult as $q$ grows. 

The linear model of coregionalization (\lmc, \citealt{matheron82, wackernagel03, schmidtgelfand}, see also \citealt{alie24}) writes $C_{ij}(\cdot, \cdot)$ as a linear combination of $k\leq q \ll q^2$ univariate correlation functions, i.e., $C_{ij}(\bl, \bl') = \sum_{r=1}^k a_{ir}a_{rj} \rho_r(\bl, \bl')$. If $k<q$, we refer to the \lmc\ as a spatial factor model. The \lmc\ is the most popular cross-covariance model for multivariate data due to its computational tractability: it can be extended to model nonstationarity \citep{gelnonstat}, spatially-varying regression coefficients  (\citealt{gelfand2003spatial} and \citealt{reich_bayesian_2010}, both assuming separability), space-time data \citep{berrocal2010bivariate, de_iaco_choosing_2019}; it can be included as a latent process in Bayesian hierarchies for non-Gaussian data \citep{melange}, used as a dimension reduction tool in the large $n$, large $q$ setting by letting $k \ll q$ \citep{taylor2019spatial, zhangbanerjee20}. \lmcs\ and methods based on linear combinations of independent GPs are very popular in many fields (see, e.g., \citealt{Teh2005Semiparametric, finley2008, fricker2013multivariate, alvarez2011jmlr,  moreno18neurips, lmc_neural, townes2023engelhardt}). Statistical software packages for multivariate spatial data typically implement \lmcs\ \citep{gstat, spbayes15, hmsc_package, finazzifasso, krainski_advanced_2019}. 

The \lmc\ has several major limitations: it is inadequate in modeling outcomes with different smoothness \citep{genton_ccov}; some of its parameters lead to difficult interpretations and prior elicitation, e.g., the spatial range of outcome $j$ is a non-linear function of all the $k$ spatial range parameters (\citealt{schmidtgelfand}, Section 3.4); \lmcs\ do not facilitate the inclusion of independent measurement error in response models. Finally, the infill asymptotics of \lmcs\ are poorly understood (\citealt{zhang2007environmetrics}, see also \citealt{velandia17}). 
These issues limit the flexibility of \lmcs\ and reduce their inferential usefulness. 
Alternative methods that solve some of the limitations of the \lmc\ include the multivariate Mat\'ern model %, which specifies all $C_{ij}(\cdot,\cdot)$ as Mat\'ern functions 
(\citealt{gneiting2010}, see also \citealt{apanasovich2012, Emery2022, yarger24matern}), approaches based on latent dimensions \citep{apanasovich_genton2010}, and convolution methods \citep{Gaspari1999Construction, Majumdar2007convol}. However, these methods are only suitable to very small $q$ as they do not lead to exploitable structure in the sample covariance or facilitate dimension reduction, or require numerical integration for computing $C_{ij}(\cdot,\cdot)$. The Supplement provides more details on the limitations of \lmcs\ and includes a more detailed discussion on the multivariate Mat\'ern model. 

\begin{figure}%[h!]
    \centering
    \includegraphics[width=0.8\textwidth]{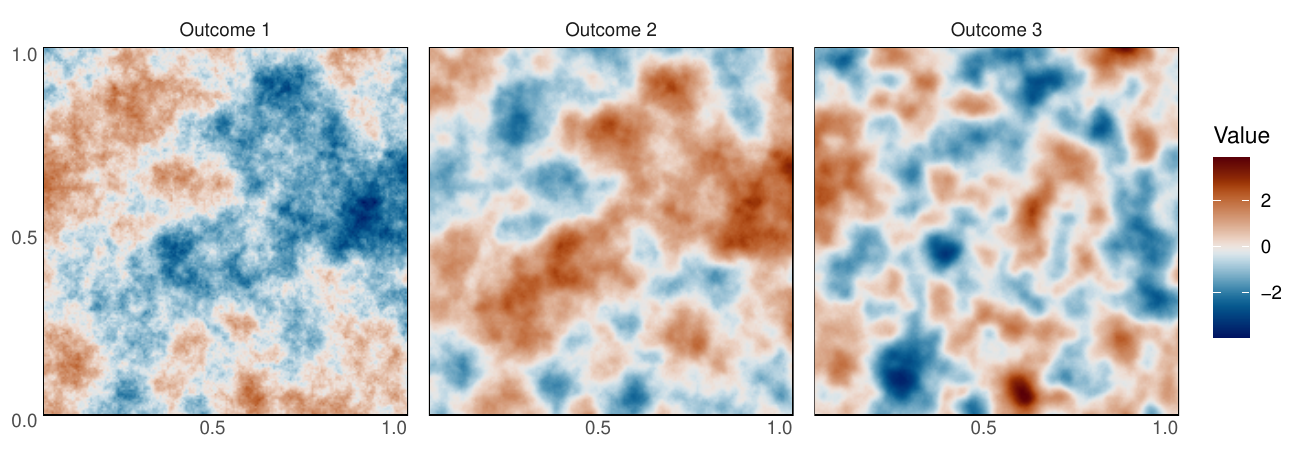}
    \caption{\footnotesize Three spatially correlated outcomes generated via GP-\modelnames\ at $n=$ 40,000 gridded locations. We let $\sigma_{ii}=1$ for $i\in\{1,2,3\}$ and $\sigma_{12}=-0.9, \sigma_{13}=0.7, \sigma_{23}=-0.5$ and choose $\rho_i(\cdot, \cdot)$ as Mat\'ern (Vecchia-approximated with $m=50$ neighbors) with range $1/5, 1/15, 1/30$ and smoothness $0.5, 1.2, 1.9$, respectively. \normalsize }
    \label{fig:prior_sample}
\end{figure}

In this article, we introduce the Inside-Out Cross-covariance (\modelname) model for multivariate spatial data. %large $n$, large $q$ data. 
\modelname\ leads to parsimonious cross-covariance matrix functions defined on the basis of $q$ univariate correlation functions and a covariance matrix $\bSigma$. 
Constructively, \modelname\ induces multivariate spatial dependence by applying outcome-specific spatial Cholesky transforms to $n$ i.i.d. $q$-vectors, thereby introducing within-outcome spatial correlation while preserving cross-outcome dependence.
\modelname\ offers novel avenues for dimension reduction via clustering or low-rank assumptions. 
%Dimension reduction using \modelname\ may proceed by clustering the outcomes or via a low-rank assumption on $\bSigma$.  
Our model resolves several major flaws of \lmcs: in \modelname, there is a one-to-one link between the marginal covariance of the $j$th variable and the $j$th univariate correlation function used to define it, facilitating prior elicitation and interpretable process-based inference. Furthermore, \modelname\ is flexible in allowing different smoothness, nugget effects, or nonstationarity of one or more outcomes. Figure \ref{fig:prior_sample} shows a realization of a noise-free GP-\modelname\ where three Mat\'ern outcomes have different smoothness. 
\begin{figure}%[h!]
    \centering
    \includegraphics[width=0.99\textwidth]{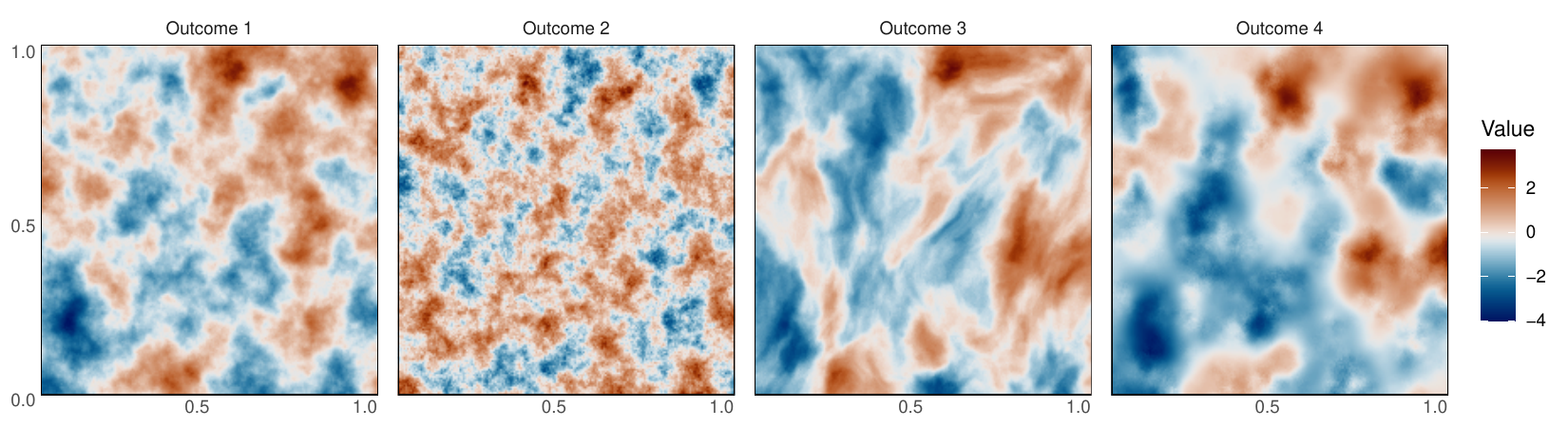}
    \caption{\footnotesize Four spatially correlated outcomes generated via GP-\modelnames\ at $n=$ 40,000 gridded locations. Outcomes 1 and 2 have stationary marginal covariances, whereas 3 and 4 are nonstationary. \normalsize }
    \label{fig:prior_sample2}
\end{figure}
In Figure \ref{fig:prior_sample2}, we generate four spatially correlated outcomes via \modelname; one of them is nonstationary via deformation of the spatial domain as in \cite{sampsonguttorp1992}, another has a spatially-varying sill. More details about Figure \ref{fig:prior_sample2} are in the Supplement. 
Relative to multivariate Mat\'ern models, \modelname\ is more flexible in allowing the marginal covariances to be defined freely (e.g., not Mat\'ern), and more scalable because it results in sample covariance matrices whose structure we may exploit for likelihood-based inference in large scale settings. When building \modelname\ using $q$ stationary Mat\'ern covariances, \modelname\ is more flexible than a parsimonious multivariate Mat\'ern \citep{gneiting2010} because it allows the marginal ranges to be defined freely, and more parsimonious than a general multivariate Mat\'ern because \modelname\ does not directly parametrize the cross-covariances.

Because IOX leads to structured covariance matrices, we envision it as a novel tool for developing flexible and interpretable models for high-dimensional multivariate spatial data. In  large $q$ settings, scalable methods have been developed for multivariate areal or count-valued data \citep{bradley_multivariate_2015, bradley_computationally_2018}. Recently, \cite{krock_modeling_2023} modeled multivariate spatial data using linear combinations of fixed basis functions. Their low-rank approach achieves scalability by assuming sparsity in an underlying graphical model of the basis function weights but remains limited in its ability to flexibly characterize cross-covariance structures. \cite{deyetal20} introduce Graphical Gaussian Processes (GGPs), which model multivariate spatial data under sparse network dependency assumptions. Constructing a GGP requires users to pre-specify a cross-covariance matrix function, whose properties and validity conditions are inherited by the resulting GGP. For instance, a LMC-based GGP cannot accommodate outcomes with unequal smoothness, and a multivariate Mat\'ern GGP must satisfy validity conditions as in \cite{Emery2022}. In introducing IOX as a novel cross-covariance model, we offer new ways for future developments of flexible GGPs.

Our definition of \modelnames\ requires the specification of a fixed set of \textit{reference} locations $\calS$. As a consequence, our model is nonstationary even when built using $q$ stationary correlation functions. However, this has minimal practical impact, as \modelnames\ is no more complex than other cross-covariance models when $n$ is large. In these settings, GPs scale poorly and they must be replaced with scalable alternatives. Vecchia approximations \citep{vecchia88} and related methods based on sparse directed acyclic graphs (DAGs; see, e.g., \citealt{nngp, conjnngp, prates, meshedgp}) have desirable theoretical properties and excellent practical performance \citep{radgp, schafer2024sparse, Heaton2019}. Several other scalable GPs can be cast as ``sparse DAG'' methods \citep{gp_predictive_process, gp_pp_biasadj, katzfuss_jasa17, katzfussgong2019, spamtrees}. The main advantage of sparse DAG methods is that they enable process-based inference and can thus be seamlessly embedded into Bayesian hierarchies for added flexibility. 
Process-based inference using sparse DAGs \textit{always} leads to nonstationary GPs, as one must specify a set $\calS$ when implementing any sparse DAG GP. \modelname\ can simply use the same $\calS$. As is standard practice in implementing Vecchia-related methods, we choose $\calS$ as the set of observed locations in all our applications.

We outline \modelname\ in Section \ref{sec:intromethod} and prove its validity. We develop the GP-\modelname\ in Section \ref{sec:ioxgp} and the related posterior sampling algorithms in Section \ref{sec:scalable_algorithms}. %We show how \modelname\ can be exploited for scalable operations in the ``large $n$, large $q$'' settings of interest in Sections \ref{sec:largen} and \ref{sec:qclustering}. 
Section \ref{sec:applications} applies IOX on simulated and colorectal cancer data. 
Proofs, as well as additional details, comparisons, and algorithms, are available in the Supplement. Software for fitting Bayesian spatial regression models using \modelname\ is available at \repourl.

\section{Inside-out cross-covariance}
\label{sec:intromethod}
Let the spatial domain $\calD \subset \Re^d$ be of dimension $d$, typically $d=2$ for longitude and latitude or x,y coordinates on an image. Introduce $q$ univariate correlation functions $\rho_j(\cdot, \cdot):\calD \times \calD \to \Re$. For example, we may let $\rho_j(\bl, \bl') = \exp \{ -\phi_j\|\bl-\bl'\|\}$ or a more complex Mat\'ern correlation parametrized by vector $\btheta_j$. Letting $\btheta_j = (\phi_j, \nu_j)$, the Mat\'ern correlation function is $\calM_{\nu_j}(\bl, \bl'; \phi_j) = \frac{2^{1-\nu_j}}{\Gamma(\nu_j)} \phi_j^{\nu_j} \| \bl - \bl' \|^{\nu_j} K_{\nu_j}\left( \phi_j \| \bl - \bl' \|\right)$ where $K_{\nu_j}$ is the modified Bessel function of the second kind of order $\nu_j$. The inclusion of nugget effects is possible, as we discuss later. 
Suppose $\calA=\{\ba_1, \dots, \ba_{n_A}\} \subset \calD$ and $\calB = \{\bb_1, \dots, \bb_{n_B}\} \subset \calD$, then we denote as $\rho_h(\calA, \calB)$ the $n_A \times n_B$ matrix whose $(i,j)$ element is $\rho_h(\ba_i, \bb_j)$, and let $\rho_h(\calA):=\rho_h(\calA, \calA)$ and $\rho_h(\bl, \calA) := \rho_h(\{\bl\},\calA)$. 
Choose a set of \textit{reference} locations $\calS=\{\bl_1, \dots, \bl_n\} \subset \calD$. In applications, we choose $\calS$ as the sample locations. The set of non-reference locations $\calS^c = \calD\setminus \calS$ may be thought of as a superset of the prediction locations. We do not make any distributional assumption in this section. After defining our cross-covariance model, we study its properties and prove the validity of the resulting cross-covariance matrix function. All proofs and derivations are in the Supplement.  %Finally, denote by $y_j(\bl)$ the value of the $j$th spatially-indexed outcome at location $\bl \in \calD$, for $j=1,\dots,q$. 

%% definition
%\begin{defin}[Inside-out cross-covariance]\label{def:ioc_xcov} 
\begin{definition}[Inside-out cross-covariance]\label{def:ioc_xcov} 
    Let $\bSigma = (\sigma_{ij})_{i,j=1}^q$ be a positive semidefinite matrix, $\rho_i(\cdot, \cdot)$ for $i=1,\dots,q$ and $\calS$ as above. We define the \modelname\ cross-covariance as:
\begin{equation}\label{eq:ioc_xcov}
C_{ij}(\bl, \bl') = \sigma_{ij} \left[ \bh_i(\bl) \bL_i \bL_j^\top \bh_j(\bl')^\top + \xi_{ij}(\bl, \bl') \right],
\end{equation}
where $\bh_i(\bl) = \rho_i(\bl, \calS) \rho_i(\calS)^{-1}$, $\bL_i$ is the lower Cholesky factor of $\rho_i(\calS)$, i.e. it is lower triangular and such that $\bL_i \bL_i^\top = \rho_i(\calS)$, $\xi_{ij}(\bl, \bl') = \mathbb{1}_{\{ \bl=\bl'\} }\sqrt{r_i(\bl) r_j(\bl')}$, and $r_i(\bl)=\rho_i(\bl,\bl)-\bh_i(\bl)\rho_i(\calS,\bl)$. We let $\mathbb{1}_{\{c\}}= 1$ if $c$ is true, $0$ otherwise.  %The properties of \modelnames\ can be clarified by analyzing special cases of \eqref{eq:ioc_xcov}. %$\bR_{i,\bl} = 1 - \rho_i(\bl, \calS) \rho_i(\calS)^{-1} \rho_i(\calS,\bl)$, and $\mathbb{1}_{\{c\}}$ is the indicator function equal to zero if $c$ is false. 
%Equivalently, letting $\bh_i(\bl) = \rho_i(\bl, \calS) \rho_i(\calS)^{-1}$, and similarly for $\bh_j(\bl')$, we can write, for $\bl\neq\bl$ or $i\neq j$, $C_{ij} = \sigma_{ij} \bh_i(\bl) \bL_i \bL_j^\top \bh_j(\bl)^\top$. 
\end{definition}
%% definition

\noindent  A constructive sampling scheme highlights why IOX is ``inside-out'' relative to the LMC. Suppose we sample at $\calS$. We begin by computing $\bL_j = \texttt{chol}(\rho_j(\calS))$ for $j=1, \dots, q$. Sample $U_{r}, r=1, \dots, nq$ as uncorrelated white noise with unit variance and build the $n\times q$ matrix $\bU$; denote its $j$th column as $\bu_j$. Set $\bV = \bU \bLambda^T$ where $\bLambda \bLambda^T = \bSigma$. Let $\bY$ be the $n\times q$ matrix whose $j$th column is $\by_j = \bL_j \bv_j$. Then, $\covf\{\bY\rowcol{r}{i}, \bY\rowcol{c}{j}\} = C_{ij}(\bl_r, \bl_c)$ as in \eqref{eq:ioc_xcov}. We have first introduced cross-variable dependence on $\bU$, then injected spatial dependence on the columns of $\bV$. The LMC performs operations in reverse: it starts by injecting spatial dependence via $\bv_j = \bL_j \bu_j$, then couples the outcomes by setting $\bY = \bV\bLambda^T$.
Refer to Section C in the Supplement for additional details. 
Next, we show that \modelname\ leads to easily interpretable inference and prior elicitation of marginal covariance parameters. 
%\begin{theoremEnd}[end, restate, text link=]{prop}[Marginal covariance]\label{prop:marginal} 
\begin{proposition}[Marginal covariance]\label{prop:marginal} 
\begin{equation*}
C_{ii}(\bl, \bl') = \begin{cases} 
\sigma_{ii} \rho_i(\bl, \bl') \qquad & \text{if } \bl \in \calS \text{ or } \bl'\in \calS \text{ or } \bl = \bl', \\
\sigma_{ii} \rho_i(\bl, \calS) \rho_i(\calS)^{-1} \rho_i(\calS, \bl') \qquad& \text{if } \bl,\bl' \in \calS^c \text{ and } \bl\neq \bl'.\end{cases}
\end{equation*}
\end{proposition}
\begin{comment}\begin{proof} 
Letting $i=j$ we write \eqref{def:ioc_xcov} as
\begin{align*} 
C_{ii}(\bl, \bl') &= \sigma_{ii} \left[ \bh_i(\bl) \rho_i(\calS) \bh_i(\bl')^\top + \mathbb{1}_{\{ \bl=\bl'\} }r_i(\bl)  \right] \\
&= \sigma_{ii} \left[ \rho_i(\bl, \calS) \rho_i(\calS)^{-1} \rho_i(\calS,\bl') + \mathbb{1}_{\{ \bl=\bl'\} } \left( \rho_i(\bl,\bl) - \rho_i(\bl, \calS) \rho_i(\calS)^{-1} \rho_i(\calS,\bl)\right)  \right].
\end{align*}
We immediately notice that if $\bl=\bl'$ then $C_{ii}(\bl,\bl) = \sigma_{ii} \rho_i(\bl, \bl) = \sigma_{ii}$. If $\bl\neq\bl'$ and $\bl,\bl' \in \calS^c$ then we obtain the stated result by just dropping the indicator term. Finally, suppose $\bl \in \calS$ (the case $\bl'\in\calS$ is analogous). This means there is $r \in \{1, \dots, n\}$ such that $\bl =\bl_r$, which implies the vector  $\bh_i(\bl) = \be_{\bl} = (e_1, \dots, e_n)^\top$ has $e_r=1$ and $e_h=0$ for $h\neq r$. We see this e.g. by letting $\bx = (x_1, \dots, x_n)^\top$, $E[\bx]=0$ , and $\covf(\bx,\bx) = \rho_i(\calS)$. Then, $\bh_i(\bl)\bx = E[x_r \mid \bx] = x_r$. In other words, if $\bl = \bl_r \in \calS$ then $\bh_i(\bl)$ selects the $r$th row from the matrix to which it is premultiplied. Therefore, we select the $r$th row of $\rho_i(\calS, \bl')$, which is equal to $\rho_i(\bl_r, \bl') = \rho_i(\bl, \bl')$. 
\end{proof}\end{comment} 

\noindent The first case ($\bl \in \calS$ or $\bl'\in \calS$ or $\bl = \bl'$) is the most relevant in practice and motivates our choice of $\calS$ as the observed locations in Section \ref{sec:ioxgp}. In fact, by letting $\calS$ correspond to the observed locations, the tasks of parameter estimation and prediction at new locations will both solely depend on the cross-covariance model when at least one of $\bl$ or $\bl'$ are in $\calS$. In this key scenario, Proposition \ref{prop:marginal} shows that \modelname\ retains $\rho_i(\bl, \bl')$ as the marginal covariance for outcome $i$. 
%In the residual case $\bl, \bl'\in \calS^c$, the marginal covariance takes the form of the predictive process (PP) covariance on $\rho_i(\cdot, \cdot)$ with knots $\calS$ \citep{gp_predictive_process}, which is known to approximate $\rho_i(\cdot, \cdot)$ increasingly well as the size of $\calS$ increases. 
More in general, for all $\bl, \bl'\in \calD$, $C_{ii}(\cdot, \cdot)$ is still a function of $\rho_i(\cdot, \cdot)$ only, leading to direct interpretation of its parameters. $C_{ii}(\bl, \bl')$ is equivalent to the modified predictive process covariance \citep{modifiedpp} on knots $\calS$, which is known to approximate $\rho_i(\cdot, \cdot)$ increasingly well as the size of $\calS$ increases, and reduces to the predictive process covariance \citep{gp_predictive_process} only in the residual case $\bl, \bl'\in \calS^c$. 
From a Bayesian perspective, Proposition \ref{prop:marginal} simplifies prior elicitation, as one can take advantage of exploratory tools like variograms to build informative priors. %The one-to-one link between $C_{ii}(\cdot, \cdot)$ and $\rho_i(\cdot, \cdot)$ also suggests that \modelname\ directly inherits the infill asymptotic behavior of the $\rho_i(\cdot, \cdot)$. 

% begin norm<1 proof
Next, we study the role of $\sigma_{ij}$ for $i\neq j$ and show that two variables have a cross-covariance of $\sigma_{ij}$ at zero distance if their marginal correlations functions are the same. Otherwise, $\sigma_{ij}$ bounds the cross-covariance $C_{ij}(\bl,\bl')$ from above---this behavior also occurs in the multivariate Mat\'ern model due to the necessary validity conditions (refer to the Supplement for more details). %Figure \ref{fig:cij_average} visualizes the resulting scaling factor.
%\begin{theoremEnd}[end, restate, text link=]{prop}[Cross-covariance and $\sigma_{ij}$]\label{prop:cij_norm}
\begin{proposition}[Cross-covariance and $\sigma_{ij}$]\label{prop:cij_norm}
For all $\bl, \bl' \in \calD$ and all $i, j = 1,\dots, q$, the cross-covariance is $C_{ij}(\bl, \bl') \leq \sigma_{ij}$. If $\rho_i(\cdot, \cdot) = \rho_j(\cdot, \cdot)$, then $C_{ij}(\bl, \bl) = \sigma_{ij}$.
\end{proposition}
\begin{comment}\begin{proof}
We start by proving that $\| \bh_i(\bl) \bL_i \| \leq 1$. Because $\rho_i(\cdot, \cdot)$ is a correlation function, then $r_i(\bl) = \rho_i(\bl, \bl) - \rho(\bl, \calS) \rho_i(\calS)^{-1} \rho_i(\calS, \bl) = 1 - \bh_i(\bl) \rho_i(\calS) \bh_i(\bl)^\top  \ge 0$ which implies $\bh_i(\bl) \rho_i(\calS) \bh_i(\bl)^\top = \bh_i(\bl) \bL_i \bL_i^\top \bh_i(\bl)^\top = \| \bh_i(\bl) \bL_i \|^2 \leq 1$.

Let $\ba^\top = \bh_i(\bl) \bL_i$ and $\bb^\top = \bh_j(\bl') \bL_j$. We showed above that $\| \ba \|\leq 1$ and $\|\bb \| \leq 1$. Then
\begin{align*} C_{ij}(\bl, \bl') &= \sigma_{ij} \left[ \bh_i(\bl) \bL_i \bL_j^\top \bh_j(\bl')^\top + \sqrt{ r_i(\bl) r_j(\bl') } \right] \\
&= \sigma_{ij} \left[ \ba^\top \bb + \sqrt{ (1 - \ba^\top \ba) (1 - \bb^\top \bb) } \right] \\
&\leq \sigma_{ij} \left[ \|\ba\| \|\bb\| + \sqrt{ (1-\|\ba\|^2)(1-\|\bb\|^2) }  \right] \leq \sigma_{ij}.
\end{align*}
In particular, consider the case $\bl=\bl'$. If $\rho_i(\cdot, \cdot) = \rho_j(\cdot, \cdot)$ then we obtain $C_{ij}(\bl, \bl') = \sigma_{ij}$, which implies that $\sigma_{ij}$ is the zero-distance correlation between variables with the same marginal correlation. 
\end{proof}\end{comment}

% end norm<1 proof
\begin{figure}%[h!]
    \centering
    \includegraphics[width=0.99\textwidth]{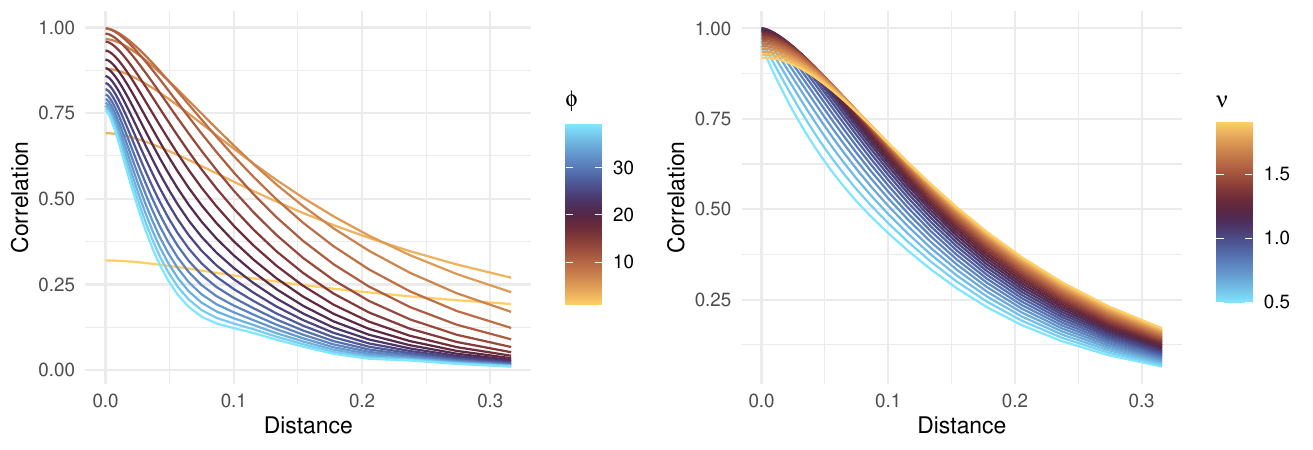}
    \caption{\footnotesize We plot $C_{ij}(\bl, \bl + \bh)/\sigma_{ij} = \bh_i(\bl) \bL_i \bL_j^\top \bh_j(\bl + \bh)^\top$ at varying distances $\|\bh\|$ and taking $\rho_i(\cdot, \cdot)$ as a Mat\'ern correlation with $\nu_i=1$, $\phi_i=10$ and $\rho_j(\cdot, \cdot)$ as a Mat\'ern with $\nu_j=1$ and varying the values of $\phi_j$ (left), or with $\phi_j=10$ and varying the values of $\nu_j$ (right), averaged over locations of a gridded set $\calS$ of size 400.\normalsize}
    \label{fig:cij_average}
\end{figure}

\begin{figure}%[h!]
    \centering
    \includegraphics[width=0.99\textwidth]{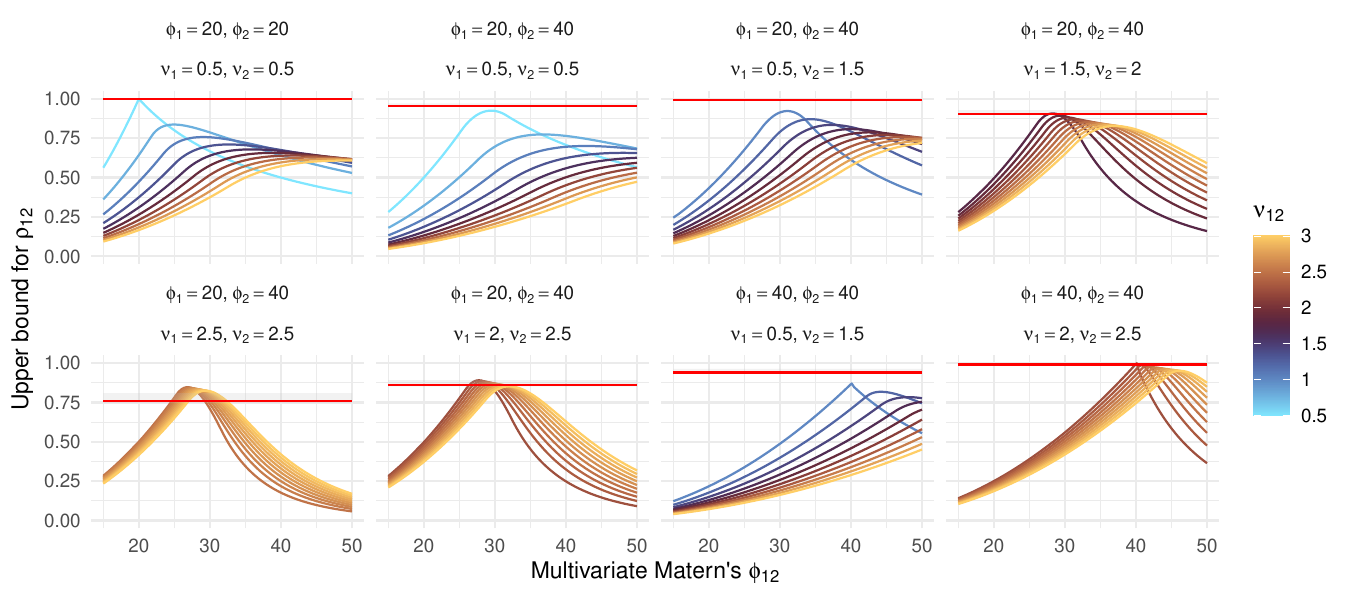}
    \caption{\footnotesize Maximum allowable cross-correlation at zero distance of a bivariate \modelname\ on two Mat\'ern margins (red horizontal lines), compared to a bivariate Mat\'ern model, at different values of $\phi_1, \nu_1, \phi_2, \nu_2$ (subplots), $\phi_{12}$ (x-axes), and $\nu_{12}$ (by line color). \modelname\ cross-correlations are averaged over locations of a gridded set $\calS$ of size 400. \normalsize}
    \label{fig:cij_mmatern_compare}
\end{figure}

\noindent $C_{ij}(\cdot, \cdot)$ is nonstationary even when each of the $\rho_i(\cdot,\cdot)$ are stationary and isotropic due to its dependence on $\calS$. We showed in Proposition \ref{prop:marginal} that we can rely on $\rho_i(\cdot, \cdot)$ and its parameters to interpret $C_{ii}(\cdot, \cdot)$. For $i\neq j$, we can easily draw interpretable covariance curves by averaging the values of $C_{ij}(\cdot, \cdot)$ across locations. In Figure \ref{fig:cij_average}, we compute and visualize the scaling factor, i.e., $C_{ij}(\cdot, \cdot)/\sigma_{ij}$, when $\rho_i(\cdot, \cdot)$ and $\rho_j(\cdot, \cdot)$ are Mat\'ern. Each curve is monotonically decreasing and corresponds to a different value of $\phi_j$ (left) or $\nu_j$ (right). For example, in the left subfigure, different lines correspond to different values of $\phi_j$, keeping $\phi_i=10$ and $\nu_i=1$; the highest correlation is achieved at $\| \bh\|=0$ by setting $\phi_j=\phi_i=10$, in which case $C_{ij}(\bl, \bl)/\sigma_{ij}=1$. In other words, cross-correlation between the two margins is highest at $\|\bh \|=0$ when the two marginal correlation functions coincide. It then decreases when we pick $\phi_j \neq \phi_i$.
Figure \ref{fig:cij_mmatern_compare} compares IOX with the bivariate Mat\'ern model on the maximum allowable cross-correlation at zero distance. For IOX, the colocated correlation coefficient can be computed by averaging values of $C_{ij}(\bl, \bl)/\sigma_{ij}$ across the spatial domain. For the bivariate Mat\'ern, we compute the upper bound given by equation (13) of \cite{gneiting2010} at several combinations of $\phi_1, \phi_2, \nu_1, \nu_2, \phi_{12}, \nu_{12}$. While some combinations of the bivariate Mat\'ern parameters achieve higher correlations, IOX is competitive in all cases.  We give more details and analyze the role of $\calS$ in shaping $C_{ij}(\bl, \bl')$ in the Supplement.

We now outline two equivalent definitions of the cross-covariance matrix function of \modelname. The first does not involve $nq \times nq$ matrices, whereas the second facilitates likelihood inference. We prove the validity of this cross-covariance matrix function in Proposition \ref{prop:valid}. 

%\begin{theoremEnd}[end, restate, text link=]{prop}[\modelname\ cross-covariance matrix function]\label{prop:xcovmatf}
\begin{proposition}[\modelname\ cross-covariance matrix function]\label{prop:xcovmatf}
Let $\bl, \bl'\in \calS$.
\begin{equation}\label{eq:xcovmatf}
    \begin{aligned}
\Cov(\bl, \bl') &= \bSigma \odot \left[ \bK(\bl, \bl') + \bd(\bl, \bl')\bd(\bl, \bl' )^\top \right] \\
&= \{ \oplus \bh_i(\bl) \bL_i \} (\bSigma \otimes \bI_n) \{ \oplus \bL_i^\top \bh_i(\bl')^\top \} + \bD(\bl, \bl') \bSigma \bD(\bl, \bl'),
\end{aligned} \end{equation}
where $\bK(\bl, \bl')$ is the $q\times q$ matrix whose $i,j$ element is $\bh_i(\bl) \bL_i \bL_j^\top \bh_j(\bl')^\top$, $\bd(\bl, \bl')$ is a vector of dimension $q$ whose $i$th element is $\mathbb{1}_{\{ \bl=\bl'\} }\sqrt{r_i(\bl)}$, $\bD(\bl, \bl') = \text{diag}\{\bd(\bl, \bl')\}$, ``$\oplus$'' is the direct sum operator and we denote with $\{ \oplus \bh_i(\bl) \bL_i \}$ the $q \times nq$ block-diagonal matrix with $\bh_i(\bl) \bL_i$ as its $i$th block, while ``$\odot$'' is the Hadamard element-by-element product. 
\end{proposition}

\noindent The following corollary specifies the structure of \modelnames\ at locations in $\calS$.
%\begin{theoremEnd}[end, restate, text link=]{coro} 
\begin{corollary} 
If $\bl \in \calS$ then $\bh_i(\bl) \bL_i = \bL_i\row{\bl}$, the row of $\bL_i$ corresponding to $\bl$. Then,
\begin{equation} \label{eq:xc_S} \begin{aligned}
\Cov(\calS) = (\bSigma \otimes \mathbf{1}_{n,n}) \odot \bK = \{ \oplus \bL_i \} (\bSigma \otimes \bI_n) \{ \oplus \bL_i^\top \},
\end{aligned} \end{equation}
where $\bK$ is the $nq \times nq$ matrix whose $(i,j)$th block is $\bL_i \bL_j^\top$; $\mathbf{1}_{n,n}$ is the $n\times n$ matrix of $1$s.
\end{corollary}
\begin{comment}\begin{proof}
If $\bl = \bl_r$ for some $r\in \{1, \dots, n\}$, then $\bh_i(\bl) = \be_{\bl} = (e_1, \dots, e_n)^\top$ has $e_r=1$ and $e_h=0$ for $h\neq r$. We see this e.g. by letting $\bx = (x_1, \dots, x_n)^\top$, $E[\bx]=0$ , and $\covf(\bx,\bx) = \rho_i(\calS)$. Then, $\bh_i(\bl)\bx = E[x_r \mid \bx] = x_r$. In other words, if $\bl = \bl_r \in \calS$ then $\bh_i(\bl)$ selects the $r$th row from the matrix to which it is premultiplied. Similarly, if $\bl\in \calS$ then $r_i(\bl)=0$ for all $i=1,\dots,q$. The rest follows.
\end{proof}\end{comment}

%BEFORE
%\noindent A \lmc\ corresponds to $\Cov(\calS) = (\bLambda \otimes \bI_n) \{\oplus \rho_i(\calS)\} (\bLambda^\top \otimes \bI_n)$, where $\bLambda \bLambda^\top = \bSigma$. 
%We call our cross-covariance model ``inside-out'' because it rearranges the way in which spatial and cross-outcome dependence appear in $\Cov(\calS)$ relative to \lmcs.
%AFTER
\noindent In \eqref{eq:xc_S}, cross-variable dependence appears \textit{inside}, spatial dependence is \textit{outside}. In the \lmc, where $\Cov(\calS) = (\bLambda \otimes \bI_n) \{\oplus \rho_i(\calS)\} (\bLambda^\top \otimes \bI_n)$, where $\bLambda \bLambda^\top = \bSigma$, cross-variable dependence appears \textit{outside}, spatial dependence is \textit{inside}.

%We call our cross-covariance model ``inside-out'' because it rearranges the way in which spatial and cross-outcome dependence appear in $\Cov(\calS)$ relative to \lmcs.

%\begin{theoremEnd}[end, restate, text link=]{prop}\label{prop:valid}
\begin{proposition}\label{prop:valid}
$\Cov(\bl, \bl')$ is a valid cross-covariance matrix function.
\end{proposition}

\noindent The main take-away of Proposition \ref{prop:valid} is that \modelname\ can be built upon any off-the-shelf univariate correlation functions without additional conditions or constraints. This yields flexibility in modeling outcome-specific features such as nonstationarities (as in Figure \ref{fig:prior_sample2}) or covariate dependence. The lack of additional restrictions distinguishes \modelname\ from multivariate Mat\'erns. We provide a more detailed comparison in the Supplement. 

We establish a connection between \modelname\ and \lmcs\ in the separable or intrinsic specification, i.e. $\rho_i(\cdot, \cdot) = \rho(\cdot, \cdot)$ for all $i$. Separability is restrictive in forcing all outcomes to be modeled via the same spatial covariance (with the same range and smoothness).
%\begin{theoremEnd}[end, restate, text link=]{prop}\label{prop:separable}
\begin{proposition}\label{prop:separable}
In the separable specification, i.e., $\rho_i(\cdot, \cdot) = \rho(\cdot, \cdot)$ for all $i$, \modelname\ and \lmc\ coincide at $\calS$.
\end{proposition}
\begin{comment}\begin{proof}
For \modelnames,
\begin{align*}
\Cov(\calS) &= \{\oplus \bL_i \} (\bSigma \otimes \bI_n) \{ \oplus \bL_i^\top \} \\
&= (\bI_q \otimes \bL) (\bSigma \otimes \bI_n) (\bI_q \otimes \bL^\top ) \\
&= \bSigma \otimes \bL\bL^\top = \bSigma \otimes \rho(\calS).
\end{align*}
For \lmc,
\begin{align*}
\Cov(\calS) &= (\bA \otimes \bI_n) \{ \oplus \rho_i(\calS) \} (\bA^\top \otimes \bI_n)\\
&= (\bA \otimes \bI_n) ( \bI_q \otimes \rho(\calS) ) (\bA^\top \otimes \bI_n)\\
&= \bA\bA^\top \otimes \rho(\calS) = \bSigma \otimes \rho(\calS).
\end{align*}
\end{proof}\end{comment}

% begin conditional covariance
\noindent Finally, define $\calT = \calS \cup \{ \bl, \bl' \}$ with $\bl\neq \bl'$. The following proposition shows that the Schur complement of $\Cov(\calS)$ in $\Cov(\calT)$ is the zero matrix. If we assume a joint Gaussian distribution for all variables in $\calT$, then this proposition implies that the observation vectors at $\bl$ and $\bl'$ are conditionally independent given their values at $\calS$.
%\begin{theoremEnd}[end, restate, text link=]{prop}\label{prop:schur}
\begin{proposition}\label{prop:schur}
\noindent Suppose $\bl,\bl' \in \calS^c$, $\bl\neq \bl'$. Then, $\Cov(\bl, \bl') - \Cov(\bl, \calS) \Cov(\calS)^{-1} \Cov(\calS, \bl')  = \bzero$. %Let $\bX$ be a random $(n+2)\times q$ matrix whose $j$th column is $\bx_j = (x_j(\bl_1), \dots, x_j(\bl_n), x_j(\bl), x_j(\bl'))^\top$, and let $\bx = \text{vec}(\bX)$ with $E(\bx) = 0$ and $\covf(\bx) = \Cov(\calT)$ as in \eqref{prop:xcovmatf}. Then
\end{proposition}
\begin{comment}\begin{proof}
We have $\Cov(\bl, \calS) =\{ \oplus \bh_i(\bl) \bL_i \} (\bSigma \otimes \bI_n) \{ \oplus \bL_i^\top \} $ and $\Cov(\calS)^{-1} = \{ \oplus \bL_i^{-\top} \} (\bSigma^{-1} \otimes \bI_n) \{ \oplus \bL_i^{-1} \}$. Together, these imply:
\begin{align*}
\Cov(\bl, \calS)\Cov(\calS)^{-1}\Cov(\calS, \bl') &= \{ \oplus \bh_i(\bl) \bL_i \} (\bSigma \otimes \bI_n) \{ \oplus \bL_i^\top  \bh_i(\bl')^\top \} = \Cov(\bl, \bl'). \qedhere
\end{align*}
\end{proof}\end{comment}
% end conditional covariance

\noindent Finally, \modelname\ allows outcome-specific nugget effects without impacting the validity of the resulting cross-covariance matrix function. %We refer to the nugget effect as the variance of spatially uncorrelated measurement error.
Because the measurement error process may be unrelated to the underlying multivariate spatial process, we seek flexibility in modeling nugget effects independently from other model components. 
% OLD
%Suppose for each $j=1,\dots, q$ we let $\rho_j(\bl, \bl') = \rho_j^*(\bl, \bl') + \tau^2_j \mathbb{1}_{\{\bl=\bl'\}}$. Then, if $\bl=\bl'$ we have $C_{jj}(\bl, \bl) = \sigma_{jj} + \sigma_{jj} \tau^2_j$, therefore we directly interpret $\sigma_{jj}\tau^2_j$ as the nugget effect term for the $j$th variable.
%The fact that we are defining $\rho_j(\cdot,\cdot)$ as a covariance function rather than a correlation function has no impact on the validity of the resulting \modelname\ model; in the Supplement, we show that \modelname\ can be defined on the basis of $q$ univariate covariance functions without compromising the validity of the approach. 
% NEW
Suppose for each $j=1,\dots, q$ we let $\rho_j(\bl, \bl') = (1-\alpha) \rho_j^*(\bl, \bl') + \alpha\mathbb{1}_{\{\bl=\bl'\}}$, where $\alpha\in(0,1)$. This parametrization leads to a decomposition of $C_{jj}(\bl, \bl)=\sigma_{ii}$ into purely spatial variance $\sigma_{ii}(1-\alpha)$ and nugget effect $\sigma_{ii}\alpha$. It is possible to parametrize each $\rho_j(\bl, \bl')$ more freely; in the Supplement, we show that \modelname\ can be defined on the basis of $q$ univariate covariance functions without compromising the validity of the approach. 

Because \modelname\ allows outcome-specific nugget effects, it is more flexible than a \lmc. Refer to the Supplement for a more in-depth discussion of nugget effects and \lmcs.
%In multivariate models of non-Gaussian data, the inflexibility of \lmcs\ in incorporating flexible nugget effects can be circumvented with an additional set of non-spatial random effects only at a steep cost in terms of computational efficiency.

\noindent We have shown that \modelname\ is a valid cross-covariance matrix function. We can define a GP based on it, which we call the GP-\modelname. The next section details properties and develops models based on GP-\modelname.

\section{\modelname\ Gaussian process models}\label{sec:ioxgp}
Let $\Cov(\cdot, \cdot)$ be the \modelname\ cross-covariance matrix function \eqref{eq:xcovmatf}, denote $\bQ = \bSigma^{-1}$, and and let $\calS = \{ \bl_1, \dots, \bl_n \}$ be the set of observed locations. To simplify exposition, we assume we observe all outcomes at all locations. Let $\bY$ as the $n \times q$ matrix whose $(i,j)$ entry is $y_j(\bl_i)$. Finally, $\by = \vecop(\bY) = (\by_1^\top, \dots, \by_q^\top)^\top$ and $\by(\bl_i) = (y_1(\bl_i), \dots, y_q(\bl_i))^\top$. Define $\bW, \bw_j$, and $\bw(\bl_i)$ similarly. We begin this section by studying the joint density induced by GP-IOX on $\calS$ as well the density of one process component conditional on all others. We then move to defining response and latent models using GP-IOX. 
%We outline general algorithms for prior and posterior sampling of the models we present, as well as analyze posterior predictive distributions. 

% OLD
%Throughout this section, we choose $\calS = \{ \bl_1, \dots, \bl_n \}$ as the set of observed locations, and let $\by_j$ be the $n\times 1$ vector whose $i$th element is $y_j(\bl_i)$. We assume the outcomes are aligned and $\bY$ is a $n \times q$ matrix with $(i,j)$ entry $y_j(\bl_i)$. We let $\by_j = (y_j(\bl_1), \dots, y_j(\bl_n))^\top$, $\by(\bl_i) = (y_1(\bl_i), \dots, y_q(\bl_i))^\top$, $\by = \text{vec}(\bY) = (\by_1^\top, \dots, \by_q^\top)^\top$. 
%Unless otherwise specified, we assume $\by$ has mean zero for ease of exposition. More in general, we can simply condition on $\bbeta$ and work with $\tilde{\by} = \by-\bX_q\bbeta$, where after letting $\bX$ be the $n\times p$ matrix of predictors at the $n$ observed locations, we denote $\bX_q = \bI_q \otimes \bX$. Throughout the article, we denote $\bQ = \bSigma^{-1}$, $\bv_j = \bL_j^{-1} \by_j$, and let $\bV$ be the $n\times q$ matrix whose $j$th column is $\bv_j$. 
% NEW

%\begin{theoremEnd}[proof here, restate, text link=]{prop}[Log-likelihood of GP-\modelname]\label{prop:density}
\begin{proposition}\label{prop:density} 
Let $\bw(\cdot) \sim GP(\bzero, \Cov(\cdot, \cdot)).$ The GP-IOX log-density is
    \begin{equation}\label{eq:ioxdens} \log p(\bw \mid \btheta, \bQ) = \text{const} + \frac{n}{2}\log\det(\bQ) + \sum_{ij}\log \bL_j^{-1}\rowcol{i}{i} -\frac{1}{2} \text{Tr}\left( \bV \bQ \bV^\top \right), \end{equation}
    where $\bL_j^{-1}\rowcol{i}{i}$ is the $i$th diagonal element of $\bL_j^{-1}$, and $\bV$ is the $n\times q$ matrix whose $j$th row is $\bv_j = \bL_j^{-1}\bw_j$. 
\end{proposition}

\noindent A consequence of Proposition \ref{prop:density} is that, for a given $\bV$ and without additional assumptions on $\bQ$, evaluating \eqref{eq:ioxdens} scales as $O(nq^2)$. Therefore, the scalability of \modelname\ to large $n$ depends on the complexity of computing $\bV$, which is the matrix we obtain by spatial ``whitening'' each column of $\bW$. Section \ref{sec:largen} details how to scalably compute $\bV$. 
\modelname\ also leads to easy-to-evaluate conditional densities $p(\bw_j \mid \bw_{j^c}, \btheta, \bQ)$, where $j^c = \{1, \dots, q \} \setminus \{j\}$. We use the following result to develop posterior sampling algorithms in Sections \ref{sec:response_sampling} and \ref{sec:latent_sampling}.

%\begin{theoremEnd}[end, restate, text link=]{prop}[Conditional density]\label{prop:conditionals}
\begin{proposition}[Conditional density]\label{prop:conditionals}
$p(\bw_j \mid \bw_{j^c}, \btheta, \bQ) = N(\bw_j; \bm_j, \bM_j)$, where $
\bm_j = - \bL_j \sum_{r\in j^c} \frac{Q_{jr}}{Q_{jj}} \bv_r$ and $\bM_j^{-1} = Q_{jj} \rho_j(\calS)^{-1}$. Then, we evaluate $\log N(\bw_j; \bm_j, \bM_j)$ as:
\begin{equation}\label{eq:ioxconditional} 
    \begin{aligned}
\log N(\bw_j; \bm_j, \bM_j)
    &\propto  -\frac{1}{2} \log |\bM_j| -\frac{1}{2}(\bw_j - \bm_j)^\top \bM_j^{-1} (\bw_j - \bm_j) \\ 
    & = \frac{n}{2}\log Q_{jj} + \sum_{i=1}^n \log \bL_j^{-1}[i,i] -\frac{1}{2Q_{jj}} \bQ_{j\cdot} \bV^\top \bV \bQ_{j\cdot}^\top ,
\end{aligned}    
\end{equation}
where $Q_{jr}$ is the $(j,r)$th element and $\bQ_{j\cdot}$ is the $j$th row of $\bQ$.
\end{proposition}

\subsection{Multivariate response model}\label{sec:response_full}
We model the observed process directly as a GP-IOX. We assume all outcomes share a common set of $p$ predictors; generalizations to outcome-specific sets of predictors are straightforward.  
Specifically, we let $\by(\cdot) \sim GP(\bB^T\bx(\cdot) , \Cov(\cdot, \cdot))$, where $\bx(\bl)$ is the $p\times 1$ vector of predictors at any $\bl\in \calD$, and $\bB$ is a $p \times q$ matrix whose $j$th column, $\bbeta_j$, represents the linear covariate effects on the $j$th outcome. Denote $\bbeta = \vecop(\bB)$ and $\bX$ the $n\times p$ matrix storing the covariates at $\calS$.   
Propositions \ref{prop:density} and \ref{prop:conditionals} apply to the centered process $\tilde{\by}(\cdot) = \by(\cdot) - \bB^T\bx(\cdot)$.   
For simplicity, we let $\rho_i(\cdot, \cdot) = \rho(\cdot, \cdot; \btheta_i)$, $i=1, \dots, q$, and $\btheta = (\btheta_1^\top, \dots, \btheta_q^\top)^\top$.  We obtain
\begin{align}\label{eq:response_model}
\by &\sim N(\bX_q \bbeta, \Cov),  \qquad \Cov = \{ \oplus \bL_i \}(\bSigma \otimes \bI_n) \{ \oplus \bL_i^\top \},
\end{align}
where $\bX_q = \bI_q \otimes \bX$. We assign the priors $\bSigma \sim \calW^{-1}(\nu_{\Sigma}, \bPsi^{-1})$ (the inverse Wishart distribution with scale matrix $\bPsi^{-1}$ and $\nu_{\Sigma}$ degrees of freedom), $\bbeta \sim N(\bm_{\beta}, \bM_{\beta})$, and $\btheta \sim p(\btheta)$, yielding conditionally conjugate updates for $\bbeta$ and $\bSigma$ (see Section \ref{sec:response_sampling}).

\subsection{Latent model}\label{sec:latent_full}
We specify the latent model as $\by(\cdot) = \bB^T\bx(\cdot) + \bw(\cdot) + \beps(\cdot)$, where $\bw(\cdot) \sim GP(\bzero, \Cov(\cdot, \cdot))$, $\beps(\bl) \iidsim N(\bzero, \bDelta)$, and $\bDelta$ is a $q\times q$ covariance matrix. 
This yields:
\begin{equation}\label{eq:latent_model}
\begin{aligned}
\by &= \bX_q \bbeta + \bw + \beps, \qquad \beps \sim N(\bzero, \bDelta \otimes \bI_n),\\
\bw &\sim N(\bzero, \Cov),  \qquad \Cov = \{ \oplus \bL_i \}(\bSigma \otimes \bI_n) \{ \oplus \bL_i^\top \},
\end{aligned}
\end{equation}
where all parameters that appear in both \eqref{eq:response_model} and \eqref{eq:latent_model} can be assigned the same prior distributions. Although $\bDelta$ can be dense, we assume it is diagonal for simplicity, and assign the priors $\delta_{ii}\iidsim Inv.G(a_d, b_d)$ for $i=1,\dots,q$; therefore, $\bw$ captures all cross-outcome dependence as a $q$-variate noise-free GP-\modelname. The latent model can be extended further: in the Supplement, we propose a latent factor model which assumes $\bSigma$ is low-rank. 

Although \eqref{eq:latent_model} has a Gaussian first stage, we can straightforwardly generalize to binomial and negative-binomial data via data augmentation schemes \citep{albertchib93, polyagamma}. Our hierarchical setup remains useful in more general settings: efficient Hamiltonian Monte Carlo methods that use second order information about the target density have connections with Gibbs sampling of Gaussian models (see \citealt{girolamicalderhead11}, or \citealt{melange} for the multivariate spatial setting).

\subsection{Predictions at new locations}
Having chosen $\calS$ as the set of observed locations, denote $\calT = \{\bt_1, \dots, \bt_N \}$ as the test set. For any zero-mean multivariate GP $\bw(\cdot)$ with cross-covariance matrix function $\bK(\cdot, \cdot)$, the distribution of the $Nq$-dimensional vector $\bw(\calT)$ conditional on $\bw=\bw(\calS)$ is Gaussian, with mean $\bK(\calT, \calS) \bK(\calS)^{-1} \bw$ and covariance $\bK(\calT) - \bK(\calT, \calS)\bK(\calS)^{-1}\bK(\calS, \calT)$. In GP-\modelname, Proposition \ref{prop:schur} implies that $\covf(\bw(\bt_r), \bw(\bt_s) \mid \bw) = 0$ if $\bt_r \neq \bt_s$, i.e., $\bw(\bt_r)$ and $\bw(\bt_s)$ are conditionally independent given the data; therefore, we can focus on marginal predictions of the outcome vector at individual locations. 

For a new location $\bt \notin \calS$, let $\bH(\bt) =  \Cov(\bt, \calS) \Cov(\calS)^{-1}$
and $\bR(\bt) = \Cov(\bt) - \bH(\bt) \Cov(\calS, \bt)$. Then, in the response model, for every posterior sample of $\bSigma$, $\bbeta$, $\btheta$ we draw from $p(\by(\bt) \mid \by, \bbeta, \bSigma, \btheta) = N(\bm_y(\bt), \bR(\bt) )$ for each $\bt \in \calT$, where $\bm_y(\bt) = \bB^T\bx(\bt) + \bH(\bt) (\by - \bX_q \bbeta)$. In the latent model, for every posterior sample of $\bbeta,\bw,\bSigma,\btheta, \bDelta$ we sample $\bw(\bt)$ from $p(\bw(\bt) \mid \bw, \bSigma) = N(\bm_w(\bt), \bR(\bt))$, where $\bm_w(\bt) = \bH(\bt) \bw$, then let $\by(\bt) = \bB^T\bx(\bt) + \bw(\bt) + \bDelta^{\frac{1}{2}} \bu$ where $\bu \sim N(\bzero, \bI_q)$.
The following proposition clarifies that \modelnames\ maintains for predictions the same marginal structure we outlined in Proposition \ref{prop:marginal}; spatial cross-variable dependence in predictions is induced by the off-diagonal elements of $\bR(\bt)$. 
%\begin{theoremEnd}[end, restate, text link=]{prop}\label{prop:predictions}
\begin{proposition}\label{prop:predictions}
\noindent $\bH(\bt) = \{ \oplus \bh_i(\bt)\}$ and $\bR(\bt)= \bD(\bt, \bt) \bSigma \bD(\bt, \bt)$. For $\bt \notin \calS$, $p(w_i(\bt) \mid \bw, \btheta, \bSigma)$ only depends on $\rho_i(\cdot, \cdot)$ at $\calS$, $\sigma_{ii}$ and $\bw_i$. 
\end{proposition}
\begin{comment}\begin{proof}
\begin{align*}
\bH(\bt) &= \Cov(\bt, \calS) \Cov(\calS)^{-1} \\
&= \{\oplus \bh_i(\bt) \bL_i \}(\bSigma\otimes \bI_n) \{ \oplus \bL_i^\top \} \{ \oplus \bL_i^{-\top} \} (\bSigma^{-1} \otimes \bI_n) \{ \oplus \bL_i^{-1} \} \\
&= \{\oplus \bh_i(\bt) \bL_i \} \{ \oplus \bL_i^{-1} \}  = \{ \oplus \bh_i(\bt) \}
\end{align*}
\begin{align*}
\bR(\bt) &= \Cov(\bt) - \bH(\bt) \Cov(\calS, \bt) \\
&= \{\oplus \bh_i(\bt) \bL_i \}(\bSigma\otimes \bI_n) \{\oplus \bL_i^\top \bh_i(\bt)^\top \} + \bD(\bt, \bt) \bSigma \bD(\bt, \bt) -\\ & \qquad\qquad - \{ \oplus \bh_i(\bt) \} \{ \oplus \bL_i \} (\bSigma\otimes \bI_n) \{\oplus \bL_i^\top \bh_i(\bt)^\top \} \\
&= \{\oplus \bh_i(\bt) \bL_i \}(\bSigma\otimes \bI_n) \{\oplus \bL_i^\top \bh_i(\bt)^\top \} + \bD(\bt, \bt) \bSigma \bD(\bt, \bt) -\\ & \qquad\qquad - \{ \oplus \bh_i(\bt)\bL_i \} (\bSigma\otimes \bI_n) \{\oplus \bL_i^\top \bh_i(\bt)^\top \} \\
&= \bD(\bt, \bt) \bSigma \bD(\bt, \bt) = \bD(\bt, \bt) \bA \bA^\top \bD(\bt, \bt)
\end{align*}
\end{proof}\end{comment}

%% NEW 
\noindent For predictions on $w_i(\bt)$ to depend on the entire $\bW$, we can extend $\calS$ to include $\bt$, then condition on data at $\calS \setminus \{ \bt \}$. We demonstrate this via Proposition \ref{prop:predictions_S} for the case $q=2$ and $\calS = \{\bl_1, \bl_2\}$ without loss of generality.

\begin{proposition}\label{prop:predictions_S}
Suppose $q=2$ and $\calS = \{\bl_1, \bl_2\}$. Then $p(w_i(\bl_r) \mid \bw(\bl_s))$ depends on both $w_i(\bl_s)$ and $w_j(\bl_s)$, for each choice of $i,j,r,s \in \{1,2\}$, $i\neq j$ and $r\neq s$. 
\end{proposition}

\noindent Proposition~\ref{prop:predictions_S} clarifies how in the latent model (i.e., noisy observation of a smooth process), IOX borrows information across all observed data $\by$ to recover the underlying signal $\bw$. Although the set $\mathcal{S}$ can be augmented with unobserved locations to allow marginal predictions to depend on all data, computational cost scales with the size of $\mathcal{S}$ (see Supplement), exposing a trade-off between flexibility in predictions and computational efficiency.

\noindent Finally, we study $p(\bw_m(\bt) \mid \bw_{o}(\bt), \bw)$, where $o\subset \{1,\dots,q\}$ (for \textit{observed}) and $m = \{1,\dots,q\}\setminus o$ (for \textit{missing}) its complement. Intuitively, we want to predict $\bw_m(\bt)$ based not only on $\bw$ but also on $\bw_{o}(\bt)$. 
 
%\begin{theoremEnd}[end, restate, text link=]{prop}
\begin{proposition}
$p(\bw_m(\bt) \mid \bw_{o}(\bt), \bw) = N(\bw_m(\bt); \bh_{m\mid o}, \bR_{m \mid o}(\bt))$, where $\bh_{m\mid o} = \bH_m(\bt)\bw + \bH_{m\mid o}(\bt) (\bw_{o}(\bt) - \bH_{o}(\bt) \bw)$, $\bH_{o}(\bt)$ is the submatrix of $\bH(\bt)$ where we take the rows corresponding to $o$ (similarly $\bH_{m}(\bt)$ with rows $m$), and 
\begin{align*}
\bH_{m \mid o}(\bt) &= \bD_m(\bt, \bt) \bSigma_{m, o}  \bSigma_{o}^{-1} \bD^{-1}_{o}(\bt, \bt) = - \bD_m(\bt, \bt) \bQ^{-1}_m \bQ_{m,o} \bD^{-1}_{o}(\bt, \bt)\\
\bR_{m \mid o}(\bt) &= \bD_m(\bt, \bt) (\bSigma_{m} - \bSigma_{m, o} \bSigma_{o}^{-1} \bSigma_{o,m}) \bD_m(\bt, \bt) = \bD_m(\bt, \bt) \bQ_{m}^{-1} \bD_m(\bt, \bt),
\end{align*}
where $\bSigma_{m, o}$ subsets $\bSigma$ to its $m$ rows and $o$ columns, and similarly for $\bQ$. 
\end{proposition}
\begin{comment}\begin{proof}
All the results are a consequence of the properties of multivariate Gaussians and Proposition \ref{prop:predictions}.
\end{proof}
\end{comment}
\noindent In particular, if $m = \{j \}$, $\bR_{m\mid o}(\bt) = \frac{r_j(\bt)}{Q_{jj}}$. In other terms, the predictive variance at an unobserved location when conditioning on all outcomes except $j$ decreases if $\bt$ is close to $\calS$ based on $\rho_j(\cdot, \cdot)$, or when the other variables ``explain'' the $j$th variable (large $Q_{jj}$).

%\noindent Finally, we note that in the above discussion the \modelname\ cross-covariance matrix function is always invoked with at least one argument being the set $\calS$, which implies that estimation and prediction using GP-\modelname\ leads to exact inference about marginal covariances, as shown in Proposition \ref{prop:marginal}.

\section{Computations with GP-\modelname\ models} \label{sec:scalable_algorithms}

We outline the key details of posterior sampling for GP-\modelname. Additional details, as well as methods for prior sampling of GP-\modelname, are in the Supplement.

\subsection{Posterior sampling of the multivariate response model} \label{sec:response_sampling}
We target the posterior distribution
%\begin{align}\label{eq:response_posterior}
$p(\bSigma, \bbeta, \btheta \mid \by) \propto p(\by \mid \bSigma, \bbeta, \btheta) p(\bSigma) p(\btheta) p(\bbeta)$.
%\end{align}
The sampling algorithm detailed in the Supplement is straightforward: $p(\bbeta \mid \by, \bSigma, \btheta)$ is Gaussian; $p(\bSigma \mid \by, \bbeta, \btheta)$ is inverse Wishart; $\btheta$ is updated via Metropolis-Hastings. 
There is a computational bottleneck in updating $\btheta$. A joint update targets $p(\btheta \mid \by, \bbeta, \bSigma) \propto p(\by \mid \btheta, \bbeta, \bSigma) p(\btheta)$, which we can evaluate using Proposition \ref{prop:density}. A block update of $\btheta_j$---the marginal correlation parameters for outcome $j$---targets $p(\btheta_j \mid \btheta_{j^c}, \by)$. We assume independent prior $p(\btheta) = \prod_j p(\btheta_j)$, then (omitting other model parameters for simplicity) we find
\[ p(\btheta_j \mid \btheta_{j^c}, \by) \propto p(\by \mid \btheta_j, \btheta_{j^c}) p(\btheta_j \mid \btheta_{j^c}) = p(\by_j \mid \by_{j^c}, \btheta)p(\by_{j^c} \mid \btheta)  p(\btheta_j) \propto p(\by_j \mid \by_{j^c}, \btheta) p(\btheta_j), \]
where we evaluate $p(\by_j \mid \by_{j^c}, \btheta)$ via Proposition \ref{prop:conditionals}. We use joint updates of $\btheta$ when %using dimension reduction via clustering or when 
$q$ is small (e.g., Section \ref{sec:app:toy}), and block updates in all other settings.

\subsection{Posterior sampling of the latent model} \label{sec:latent_sampling}
We sample from the joint posterior $p(\bw, \bbeta, \bSigma, \bDelta, \btheta \given \by)$ by iterating through all full conditional distributions. In particular, the full conditional posterior distribution of $\bw$ is $p(\bw \mid \by, \bbeta, \bSigma, \bDelta, \btheta) = N(\bm_{w\mid y}, \bM_{w\mid y})$, where
\begin{align*}%\label{eq:latent_wcondpost}
\bM_{w\mid y}^{-1} &= \bC^{-1} + (\bDelta^{-1}\otimes \bI_n) \quad\text{and}\quad \bM_{w\mid y}^{-1} \bm_{w\mid y} = (\bDelta^{-1} \otimes \bI_n) (\by - \bX_q \bbeta),
\end{align*}
which requires to solve a linear system on $\bM_{w\mid y}^{-1}$ to compute $\bm_{w\mid y}$; this operation is costly since $\bM_{w\mid y}^{-1}$ has dimension $nq \times nq$. Alternatively, we can update $\bw_j$ given $\bw_{j^c}$ and the data for a sequential single-outcome sampler: we write the hierarchical model as
\begin{align*}
\bw_j \mid \bw_{j^c} &\sim N(\bm_j, \bM_j), \qquad \by_j = \bX \bbeta_j + \bw_j + \beps_j, \quad \beps_j \sim N(\bzero, \delta_{jj} \bI_n),
\end{align*}
where $\bM_j$ and $\bm_j$ are the same as in Proposition \ref{prop:conditionals}, except we replace $\by$ with $\bw$. 
%Because $\bM_j^{-1}$ is the $(j,j)$th block of the precision matrix $\Cov^{-1}$, we find $\bM_j^{-1} = Q_{jj} \rho_j(\calS)^{-1}$ and $\bm_j = \Cov_{j, j^c} \Cov_{j^c}^{-1} \bw_{j^c}$ can be computed directly as
%\begin{align*}
%\bm_j &= \bL_j (\bSigma_{j, j^c} \bSigma_{j^c}^{-1} \otimes \bI_n) \{ \oplus_{j^c} \bL_r^{-1} \} \bw_{j^c} = -\bL_j \left(\frac{\bQ_{j, j^c}}{Q_{jj}} \otimes \bI_n \right) \{ \oplus_{j^c} \bL_r^{-1} \} \bw_{j^c},
%\end{align*}
%where $Q_{jj}$ is the $j$th diagonal element of $\bQ = \bSigma^{-1}$ and we denote with $``\oplus_{j^c}''$ the direct sum operator over $j^c$. 
Then, we update $\bw_j$ via its conditional posterior $N(\bw_j; \bm_{w_j \mid y_j}, \bM_{w_j \mid y_j})$, where $\bM_{w_j \mid y_j} = (\bM_j^{-1} + \frac{1}{\delta_{jj}}\bI_n)^{-1}$ and $\bM_{w_j \mid y_j}^{-1}\bm_{w_j \mid y_j} = \bM_j^{-1} \bm_j + \frac{1}{\delta_{jj}}(\by_j - \bX\bbeta_j)$. We compute $\bM_j^{-1}\bm_j$ directly as
\begin{align*}
    \bM_j^{-1}\bm_j &= Q_{jj} \bL_j^{-\top} \tilde{\bW}_{j^c}\bSigma_{j^c}^{-1} \bSigma_{j^c, j} = - \bL_j^{-\top} \tilde{\bW}_{j^c}\bQ_{j^c, j},
\end{align*}
where $\tilde{\bW}_{j^c}$ is the $n \times q-1$ matrix such that $\vecop(\tilde{\bW}_{j^c} ) =\{ \oplus_{j^c} \bL_r^{-1} \} \bw_{j^c}$. The bottleneck here is computing the conditional covariances $\bM_{w_j \mid y_j}$ (i.e., solving the linear system $\bM_{w_j\mid y_j}^{-1} \bx = \bb$), but each of them can be computed efficiently in parallel -- for large $n$, we use the methods of Section \ref{sec:largen}. The sequential portion of the algorithm is the computation of $\bM_j^{-1}\bm_j$ and subsequent sampling of $\bw_j$. 
The remainder of our posterior sampling algorithm requires to sample $\bbeta$ from its Gaussian full conditional posterior distribution, whereas $\bSigma$ is conditionally sampled from an inverse Wishart distribution and the diagonal elements of $\bDelta$ from an inverse Gamma distribution. Updating $\btheta$ proceeds like in the response model after replacing $\by$ with $\bw$. 
 Additional details are in the Supplement, which also includes the single-site sampler for updating $\bw(\bl_i)$ given $\bw( \calS \setminus \{ \bl_i\})$.

%\modelname\ resolves a limitation of common spatial latent factor models based on \lmcs\ \citep{Ren2013, taylor2019spatial, zhangbanerjee20} in choosing the number of factors. With \lmcs, one has two options. First, we set $k_2$ large and infer about the ``effective'' number of factors a posteriori. Second, we adjust $k_2$ on-the-fly. Both these options are computationally intensive as $k_2$ equals the number of latent GPs. In \modelname, we can use any off-the-shelf prior for $\bSigma$ with no change to the spatial dependence structure. We see this in \eqref{eq:factorq} because we operate changes of $k_2$ on spatially pre-``whitened'' data.

%Finally, the posterior predictive covariance is $\bD(\bt, \bt) \bA \bA^\top \bD(\bt, \bt)$ from Proposition \ref{prop:predictions}, sampling from the posterior predictive distribution also depends on $k_2$.

\subsection{Scaling \modelnames\ to high dimensional data} \label{sec:largen}\label{sec:qclustering}

If $n$ is large, computing $\bL_i$ or $\bL_i^{-1}$ for each $i=1,\dots,q$ from $\rho_i(\calS)$ is prohibitively costly. We can easily construct GP-\modelname\ based on Vecchia approximations and sparse DAGs, resulting in fast likelihood-based inference. 
Suppose $p(\bz)$ is the joint density of a $n\times 1$ vector $\bz$ at locations $\calS= \{\bl_1, \dots, \bl_n\}$. \cite{vecchia88} proposed the approximation $p(\bz) \approx \prod_i p(z_i \mid \bz_{N_m(i)})$, where $\bz_{N_m(i)}$ is the subset of $\bz$ at the $m$ locations closest to $\bl_i$ within the set $\{\bl_1, \dots, \bl_{i-1}\}$. This approximation yields a valid standalone stochastic process %that satisfies Kolmogorov's consistency conditions by assuming conditional independence at all other locations 
\citep{nngp}, and can be generalized to any underlying DAG across $\calS$ locations, see, e.g., \cite{katzfuss_vecchia, meshedgp, radgp}. We can understand a DAG-based GP such as a nearest-neighbor GP \citep{nngp} or any similarly constructed GP in two manners. First, they are approximations of a parent GP with correlation function $\rho(\cdot, \cdot)$. Equivalently, they are exact GPs with (valid) correlation function $\tilde{\rho}(\cdot, \cdot)$, constructed from $\rho(\cdot, \cdot)$. We can thus build \modelname\ and the corresponding GP-\modelname\ directly using a set of $\tilde{\rho}_i(\cdot, \cdot)$, $i=1\dots, q$. By Proposition \ref{prop:valid}, the resulting \modelname\ cross-covariance matrix function inherits its validity from the validity of each $\tilde{\rho}_i(\cdot, \cdot)$. 
Working with $\tilde{\rho}(\cdot, \cdot)$, we have $\tilde{\rho}(\calS) = \tilde{\bL} \tilde{\bL}^{\top} = (\bI_n - \bH)^{-1} \bR^{\frac{1}{2}} \bR^{\frac{1}{2}} (\bI_n - \bH)^{-\top}$, where $\bH$ is a lower triangular $n\times n$ matrix with zero diagonal whose $(i,j)$ entry is nonzero if and only if $j\to i$ in the DAG. Letting $[i]$ be the set of indices of the non-zero columns of $\bH$ at row $i$, the nonzeros at row $i$ are $\bh_i = \rho(\bl_i, \bl_{[i]}) \rho(\bl_{[i]})^{-1}$, i.e., the value of $\bH$ at $(i, [i]_r)$ is the $r$th element of $\bh_i$. Building $\bH$ and $\tilde{\bL}^{-1} = \bR^{-\frac{1}{2}} (\bI_n - \bH)$ requires $O(n m^3)$ flops. The posterior sampling algorithms we outlined in Sections \ref{sec:response_full} and \ref{sec:latent_full} only involve $\bL^{-1}_i$ for $i=1,\dots,q$, therefore they scale to large $n$ when using $\tilde{\bL}_i^{-1}$. Both response and latent models benefit from these methods in evaluating the likelihood for updating $\btheta$ (either the joint of \eqref{eq:ioxdens} or the conditional of \eqref{eq:ioxconditional}).
The sparse DAG yields efficient, single-site sampling of $\bw$, as we detail in the Supplement.

Finally, if some outcomes require conditioning sets that include locations farther in space \citep{steinetal2004}, one just needs to choose the appropriate outcome-specific DAG and build $\tilde{\bL}_i^{-1}$ accordingly for each $i=1,\dots,q$. Doing so in \lmcs\ is not as straightforward because a factor-specific DAG does not translate to the same DAG for any of the outcomes.

If $q$ is large, computing and storing $\bL_i$ for each $i=1,\dots,q$ and each MCMC iteration becomes computationally intensive, even with methods based on sparse DAG as outlined above. %Additionally, the dimension of $\btheta$ increases with $q$ and joint updates of $\btheta$ (say, for an adaptive Metropolis step) become increasingly complex, with likely loss of efficiency. 
We suggest two possible ways to reduce the computational overhead in large $q$ settings. First, we can assume that not all outcome have distinct marginal correlation parameters. If outcomes can be grouped into $G$ groups, then the number of distinct $\bL_i$ that need to be computed at each MCMC iteration is $G$ or less. Second, we may assign $\btheta$ a categorical distribution and precompute all corresponding $\bL_i$'s once. Even if the number of options is large, these matrices are never updated during MCMC, leading to massive computational savings. 
Choosing the correlation hyperparameters from a fixed grid is common in cross-validatory approaches, see, e.g., \cite{conjnngp, zbf21, zhangbanerjee20}. %Translating \eqref{eq:qclustering} for the latent model of Section \ref{sec:latent_full} is straightforward. 
We refer to this method as \textit{\modelname\ Grid} in Section \ref{sec:applications}.

\section{Applications} \label{sec:applications}
We test GP-\modelnames\ in three settings: first, we consider simulated datasets with $q=3$ outcomes to compare \modelname\ to state-of-the-art methods for fitting multivariate Mat\'ern models. Then, we increase the number of outcomes to $q=24$ in order to showcase the performance of \modelname\ and related dimension reduction methods in higher dimensional settings. Finally, we analyze a spatial proteomics dataset collected as part of a colorectal cancer study.

\subsection{Toy examples on trivariate data} \label{sec:app:toy}
We analyze two scenarios. First, we sample from GP-\modelname\ with Mat\'ern margins, whereas in the second, we sample from a GP with multivariate Mat\'ern cross-covariance. When we build \modelname\ with univariate Mat\'ern margins, the marginal covariance parameters of both \modelname\ and multivariate Mat\'ern models have the same interpretations. Therefore, we can cross-test the performance of \modelname\ in estimating marginal parameters of multivariate Mat\'erns, and, vice versa, the performance of a multivariate Mat\'ern model in estimating marginal parameters of an \modelname\ with Mat\'ern margins. We include a \lmc\ for completeness.

We repeat the same analysis on 60 datasets for each scenario, for a total of 120 datasets. Each dataset measures three outcomes at a different set of 2,500 spatial locations which are sampled uniformly on $[0,1]^2$. %In all 120 datasets, we let $\bSigma$ be a correlation matrix with off-diagonal entries $\sigma_{12} = -0.9, \sigma_{13} = 0.7, \sigma_{23} = -0.5$. The decay, smoothness, and nugget effect parameters of the three outcomes are set as $\phi_1=\phi_2=\phi_3=30$, $\nu_1=0.5, \nu_2=0.8, \nu_3=1.2$, and $\tau^2_1=\tau^2_2=\tau^2_3=10^{-3}$. No other parameters are necessary for \modelname\ datasets. For the multivariate Mat\'ern, we let the cross-variable decay be fixed at $\phi_{ij}=30$ and the smoothness as $\nu_{ij}=\frac{\nu_i + \nu_j}{2}$ for all $i,j$ and we fix the cross-variable nugget effects at $\tau^2_{ij}=10^{-3}$ for all $i,j$, resulting in a parsimonious Mat\'ern specification as in \cite{gneiting2010}. 
Refer to the Supplement for additional setup details.  
We evaluate the performance in estimating the marginal parameters as well as the cross-correlations at zero distance $\rho_{ij}$.%, which are computed in the multivariate Mat\'ern cases as $\rho_{ij} = \sigma_{ij} \frac{\sqrt{ \Gamma(\nu_i+1) }}{\sqrt{ \Gamma(\nu_i)}} \frac{\sqrt{ \Gamma(\nu_j+1) }}{\sqrt{\Gamma(\nu_j)}}  

\begin{table}
\centering
\resizebox{0.95\textwidth}{!}{%
\begin{tabular}{|c|r|r|r|r|r|r|r|r|r|r|r|r|r|}
\hline
\cellcolor{gray!10} \textbf{IOX data} & \cellcolor{gray!10} $\rho_{21}$ & \cellcolor{gray!10} $\rho_{31}$ & \cellcolor{gray!10} $\rho_{32}$ & \cellcolor{gray!10} $\nu_1$ & \cellcolor{gray!10} $\nu_2$ & \cellcolor{gray!10} $\nu_3$ & \cellcolor{gray!10}  $\phi_1$ & \cellcolor{gray!10} $\phi_2$ & \cellcolor{gray!10} $\phi_3$ & \cellcolor{gray!10} $\tau^2_1$ & \cellcolor{gray!10} $\tau^2_2$ & \cellcolor{gray!10} $\tau^2_3$ & \cellcolor{gray!10} Time \\ \hline
\modelname\ Response & \textbf{0.0045} & 0.0208 & 0.0188 & 0.1100 & 0.0335 & \textbf{0.0474} & \textbf{2.89} & \textbf{2.01} & \textbf{2.28} & 0.0089 & 0.0007 & \textbf{0.0005} & 12 \\ \hline
\makecell{\modelname\ Latent \\[-3ex] {\scriptsize Sequential single-site}} & 0.0065 & \textbf{0.0198} & 0.0187 & 0.0803 & 0.0293 & 0.0836 & 3.25 & 2.11 & 3.90 & 0.0013 & \textbf{0.0004} & \textbf{0.0005} & 22 \\ \hline
\makecell{\modelname\ Latent \\[-3ex] {\scriptsize Sequential single-outcome}} & 0.0058 & 0.0197 & \textbf{0.0184} & \textbf{0.0763} & \textbf{0.0285} & 0.0842 & 3.36 & 2.13 & 3.87 & \textbf{0.0006} & 0.0005 & \textbf{0.0005} & 41 \\ \hline
Mult. Mat\'ern & 0.0098 & 0.0246 & 0.0226 & 0.1170 & 0.0620 & 0.0616 & 7.53 & 3.31 & 2.32 & 0.0209 & 0.0026 & 0.0006 & 3 \\ \hline
%\textbf{meshed} & 0.0936 & 0.3510 & 0.4020 & 0.1080 & 0.3780 & 0.3710 & 4.90 & 6.43 & 2.71 & 0.0252 & 0.0025 & 0.0020 \\ \hline
LMC & 0.0936 & 0.3510 & 0.4020 & \multicolumn{6}{c|}{ }  & 0.0252 & 0.0025 & 0.0020 & 13 \\ \hline
\multicolumn{14}{c}{ }\\
%\end{tabular}
%    \\
%   \begin{tabular}{|c|r|r|r|r|r|r|r|r|r|r|r|r|r|}
\hline
\cellcolor{gray!10} \textbf{Mult. Mat\'ern data} & \cellcolor{gray!10} $\rho_{21}$ & \cellcolor{gray!10} $\rho_{31}$ & \cellcolor{gray!10} $\rho_{32}$ & \cellcolor{gray!10} $\nu_1$ & \cellcolor{gray!10} $\nu_2$ & \cellcolor{gray!10} $\nu_3$ & \cellcolor{gray!10} $\phi_1$ & \cellcolor{gray!10} $\phi_2$ & \cellcolor{gray!10} $\phi_3$ & \cellcolor{gray!10} $\tau^2_1$ & \cellcolor{gray!10} $\tau^2_2$ & \cellcolor{gray!10} $\tau^2_3$ & \cellcolor{gray!10} Time \\ \hline
\modelname\ Response & 0.0228 & 0.0533 & 0.0506 & 0.1030 & 0.0465 & 0.0539 & \textbf{2.84} & \textbf{2.23} & 2.21 & 0.0219 & 0.0016 & 0.0009 & 11 \\ \hline
\makecell{\modelname\ Latent \\[-3ex] {\scriptsize Sequential single-site}} & 0.0100 & 0.0431 & 0.0440 & 0.0351 & 0.0462 & 0.0998 & 3.92 & 2.34 & 3.45 & 0.0056 & 0.0002 & \textbf{0.0004} & 21 \\ \hline
\makecell{\modelname\ Latent \\[-3ex] {\scriptsize Sequential single-outcome}} & 0.0129 & 0.0452 & 0.0454 & \textbf{0.0258} & 0.0551 & 0.1050 & 4.29 & 2.53 & 3.43 & \textbf{0.0037} & \textbf{0.0001} & \textbf{0.0004} & 40 \\ \hline
Mult. Mat\'ern & \textbf{0.0074} & \textbf{0.0180} & \textbf{0.0234} & 0.0527 & \textbf{0.0436} & \textbf{0.0473} & 4.03 & 2.92 & \textbf{2.10} & 0.0109 & 0.0013 & \textbf{0.0004} & 7 \\ \hline
%\textbf{meshed} & 0.0643 & 0.3450 & 0.3920 & 0.1720 & 0.2920 & 0.3580 & 6.67 & 5.49 & 2.13 & 0.0269 & 0.0032 & 0.0024 \\ \hline
LMC & 0.0643 & 0.3450 & 0.3920 & \multicolumn{6}{c|}{ }  & 0.0269 & 0.0032 & 0.0024 & 12 \\ \hline
\end{tabular}
}
\caption{\footnotesize Average RMSE and time in minutes for estimating model parameters across 60 \modelname-generated datasets (top) and 60 multivariate Mat\'ern-generated datasets (bottom). Lowest RMSEs in bold. Full box plots are available in the Supplement.\normalsize}
\label{tab:trivariate}
\end{table}

We fit an \modelname\ response model with joint updates of $\btheta$ (Section \ref{sec:response_sampling}) and test both the single-outcome sampler (Section \ref{sec:latent_sampling}) and single-site sampler (Supplement) for the latent model in Section \ref{sec:latent_full}. All \modelname\ models use a Vecchia approximation with $m=15$ neighbors for scalability.
A Vecchia-approximated ($m=15$) multivariate Matérn model, as in \cite{Emery2022}, is fit via maximum likelihood using \texttt{GpGpm} v0.4.0 \citep{fahmy2022vecchia}, available on GitHub. Posterior sampling for the \lmc\ is performed with the \texttt{meshed} R package, version 0.3, available at \url{https://github.com/mkln/meshed}.
Table \ref{tab:trivariate} reports the root mean squared error (RMSE) for parameter estimation (averaged across datasets) and average wall clock time for model fitting. The Supplement includes summary box plots providing additional details. Timing refers to maximum likelihood for the Mat\'ern model and a 10,000-iteration MCMC chain for all others. The \lmc\ model does not estimate outcome-specific spatial decay or smoothness, leaving those table cells blank.

Models based on GP-\modelname\ outperform others in the estimation of all parameters in the \modelname\ data scenario and are competitive in the multivariate Mat\'ern scenario, achieving very similar when not better performance than the correctly specified multivariate Mat\'ern model. The \lmc\ lacks the flexibility in modeling these data, resulting in inferior performance in estimating the cross-correlation at zero. 

\subsection{Comparisons on spatial data with 24 outcomes} \label{sec:compare24}

\begin{figure}%[h!]
    \centering
    \includegraphics[width=0.9\textwidth]{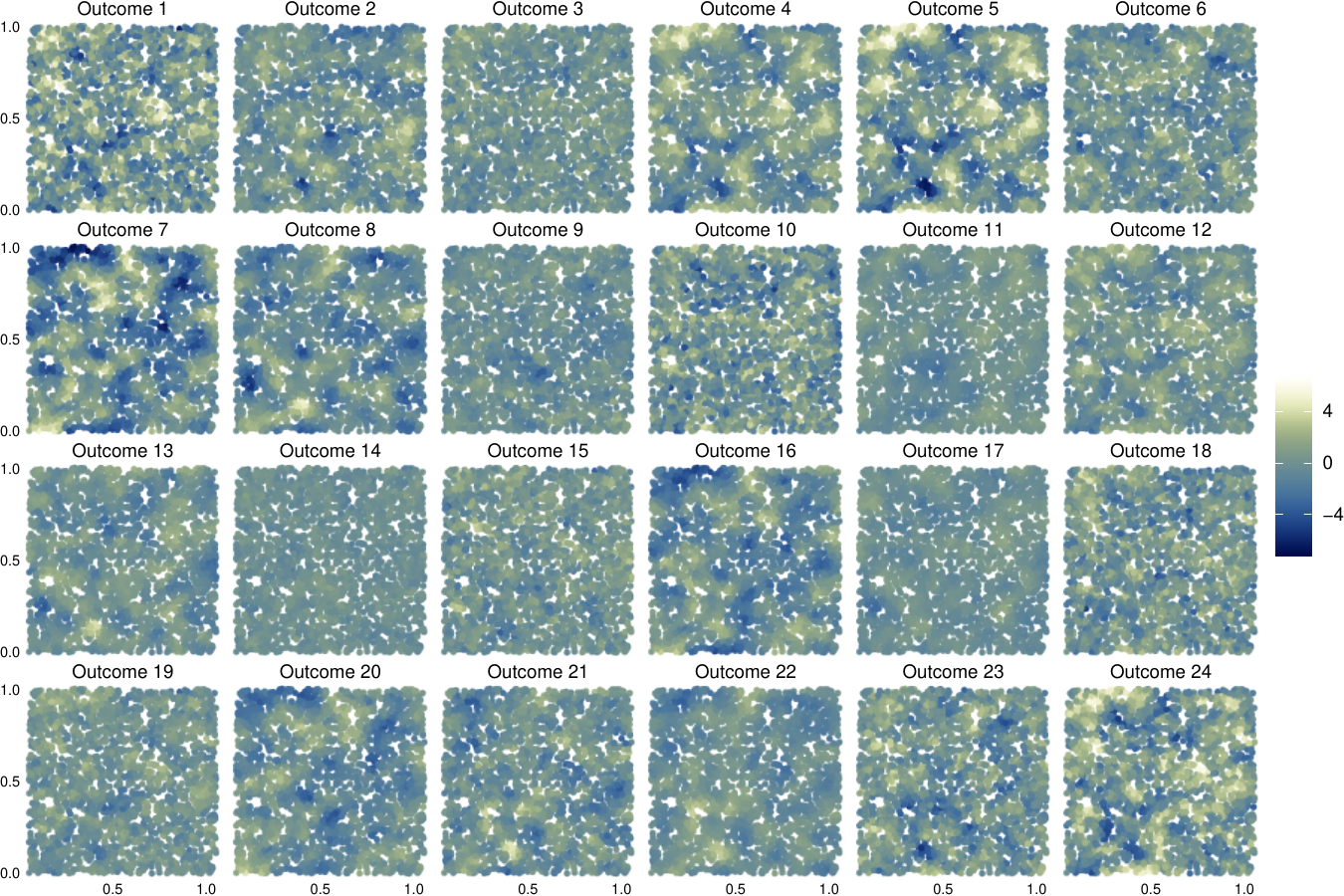}
    \caption{\footnotesize The first of 20 datasets sampled from GP-\modelname\ for the comparison of Section \ref{sec:compare24}. \normalsize}
    \label{fig:q24_example}
\end{figure}
We move to the much more challenging problem of fitting multivariate models on data with $q=24$ spatially correlated outcomes. We simulate 40 datasets, each of which includes synthetic data at a different set of $n = $ 2,500 spatial locations distributed uniformly on $[0,1]^2$. The effective dimension of each dataset is $nq = $ 60,000. Of the 40 datasets, we generate 20 by sampling from a GP-\modelname, % such that the $j$th variable has Mat\'ern marginal correlation with range $\phi_j=30$, nugget effect parameter $\tau^2_j=10^{-3}$; we sample $\nu_j$ independently from a discrete uniform prior on $\{0.5, 0.8, 1.1, 1.4, 1.7, 2.0\}$ and $\bSigma \sim \calW^{-1}_{q+1}\left(\frac{1}{2} \bI_q \right)$. Figure \ref{fig:q24_example} shows one of the \modelname\ synthetic datasets. The other 20 datasets are generated as GPs with \lmc\ cross-covariance with rank $k=8$
20 from a \lmc\ built on $k=8$ latent processes. Refer to the Supplement for additional setup details. 
%such that each of the $8$ constituent processes is Mat\'ern with smoothness $\nu_j=1$, no nugget effect, and spatial range $\phi_j$ sampled uniformly for each $j=1,\dots,8$ from a discrete distribution on a sequence of length 10 starting from $5$ and ending at $20$ with equal spacing. We generate the $24\times 8$ loading matrix $\bA$ by independently sampling each of its elements as $a_{ij} \sim N(0,1)$. Finally, we add independent Gaussian measurement error with unit variance to each of the outcomes. 

\begin{table}%[ht]
    \centering
    \resizebox{\textwidth}{!}{%
    \begin{tabular}{|c|r|r|r|r|r||r|r|r|r|}
    \multicolumn{1}{c}{ }& \multicolumn{5}{c}{\textbf{\modelname\ data}} & \multicolumn{4}{c}{\textbf{\lmc\ data}} \\ \hline
    \cellcolor{gray!10} \textbf{Method} & \cellcolor{gray!10}  $\rho_{ij}$ & \cellcolor{gray!10} $\nu_j$ & \cellcolor{gray!10} \makecell{Predictions \\[-3ex] {\scriptsize (full)}} & \cellcolor{gray!10} \makecell{Predictions \\[-3ex] {\scriptsize (partial)}} & \cellcolor{gray!10} Time &\cellcolor{gray!10}  $\rho_{ij}$ & \cellcolor{gray!10} \makecell{Predictions \\[-3ex] {\scriptsize (full)}} & \cellcolor{gray!10} \makecell{Predictions \\[-3ex] {\scriptsize (partial)}} & \cellcolor{gray!10} Time \\ \hline
    \modelname\ Full & \textbf{0.0167} & \textbf{0.0692} & \textbf{0.482} & \textbf{0.123} & 40 & 0.162 & 1.22 & 1.15 & 66 \\ \hline
    \modelname\ Grid & 0.0250 & 0.169 & 0.490 & 0.140 & 4.1 & 0.234 & 1.45 & 1.46 & 11 \\ \hline
    %\modelname\ Cluster & 0.0191 & 0.250 & 0.493 & 0.138 & 12 & 0.163 & 1.23 & 1.16 & 20 \\ \hline
    \lmc\ & 0.270 &  & 0.685 & 0.631 & 15 & 0.312 & \textbf{1.15} & \textbf{1.08}  & 34 \\ \hline
    \makecell{NNGP \\[-3ex] {\scriptsize Indep. univariate}}  & 0.106 & 0.124 & 0.483 &  & 76 & 0.123 & 1.21 &  & 55 \\ \hline
    Non-spatial model & 0.0610 & & & 0.386 & 3 & \textbf{0.0921} & & 1.27 & 3 \\ \hline
    \end{tabular}
    }
    \caption{\footnotesize RMSE in estimating $\rho_{ij}, i<j$, RMSPE in predicting $\by(\cdot)$ at 400 new locations (full vector or a random set of 4 variables, given the others), and wall clock time (in minutes) in two data scenarios, averaged across outcomes and 20 datasets for each of the two scenarios. Lowest RMSE or RMSPE in bold.  Full box plots are available in the Supplement.\normalsize}
    \label{tab:bigq_performance}
\end{table}
In both scenarios, we target the estimation of the cross-correlation between the 24 outcomes at zero spatial distance, as well as predictions at a test set of 400 locations located on a $20\times 20$ grid (the same across all 40 datasets). We compare models based on RMSE in estimating the zero-distance correlations $\rho_{ij}, i<j$, averaged across datasets, as well as the marginal smoothness in the \modelname-generated data. 
We split the prediction task into two subtasks. In the first (labeled ``full''), we perform predictions of the entire vector of $24$ outcomes. In the second (``partial''), we make predictions for $4$ outcomes (randomly selected at each location in the test set) using data about the other $20$. In both subtasks, none of the data in the test set is used to train any of the models; we compare models via root mean square prediction error (RMSPE), averaged across datasets and variables. 

We test estimation and prediction performance of several variants of the GP-\modelname\ response model: \textit{\modelname\ Full} estimates $\phi_j, \nu_j, \tau^2_j$ in the \modelname-generated data, whereas fixes $\nu_j=1$ in the \lmc-generated data, using Metropolis-within-Gibbs updates of $\btheta$ as described in Section \ref{sec:response_sampling}; \textit{\modelname\ Grid} fixes $\phi_j=30, \tau^2=10^{-3}$ for all $j$ in the \modelname-generated data scenario, and assumes a uniform prior on $\nu_j$ on a sequence of 30 equally-spaced numbers starting at 0.5 and ending at 2. %; \textit{\modelname\ Cluster} fixes $\phi_j=30, \tau^2=10^{-3}$ but implements the clustering method of Section \ref{sec:qclustering}, fixing the number of clusters at $6$. 
We fit a scalable \lmc\ with R package \texttt{meshed} (v0.3), fixing $k=6$ in the \modelname-generated data and $k=8$ for \lmc-generated data. Finally, we compare with spatial univariate methods and non-spatial multivariate methods. For the former, we fit 24 independent univariate NNGP models to each of the 40 datasets using R package \texttt{spNNGP} \citep{spnngp_rpack}: in  \modelname-generated datasets, we fix $\phi_j=30$ for all $j$ and estimate smoothness, nugget effect, and variance. In \lmc-generated datasets, we fix $\nu_j=1$ for all $j$ and estimate decay, nugget effect, and variance. We compute $\rho_{ij}$ for the NNGP models as the correlation between samples from the posterior predictive distributions. The nonspatial multivariate method simply assumes $\by(\bl) \iidsim N(\bzero, \bSigma)$. 
Timing for all methods is calculated as model fitting plus prediction time. Model fitting for all methods involves MCMC for 10,000 iterations, discarding the first half as burn-in.

Table \ref{tab:bigq_performance} summarises the result of this comparison; box plots are available in the Supplement. In the \modelname\ data scenario, \textit{\modelname\ Full} outperforms all other methods and the \lmc\ offers inferior performance to both spatial univariate and nonspatial multivariate methods. This result confirms the theoretical finding that \lmcs\ are a poor fit for variables with different smoothness. In the \lmc\ scenario, \modelname-based models outperform the \lmc\ in estimating zero-distance correlations; however, the non-spatial model is associated with the smallest average RMSE. In the prediction tasks, \textit{\modelname\ Full} exhibits similar performance to the \lmc, demonstrating the flexibility of our proposed approach in diverse scenarios. 
In terms of compute time, the \textit{\modelname\ Full} model is comparable to a set of independent univariate models, demonstrating that \modelname\ scales to large data settings in practice. %The \textit{\modelname\ Cluster} method was associated with only a slight deterioration of performance but a sizeable reduction in compute time. 
The \textit{\modelname\ Grid} is computationally inexpensive but exhibits inferior overall performance relative to \textit{Full}; therefore, it could be used as an exploratory tool in practice.

\begin{figure}%[h!]
    \centering
    \includegraphics[width=0.9\textwidth]{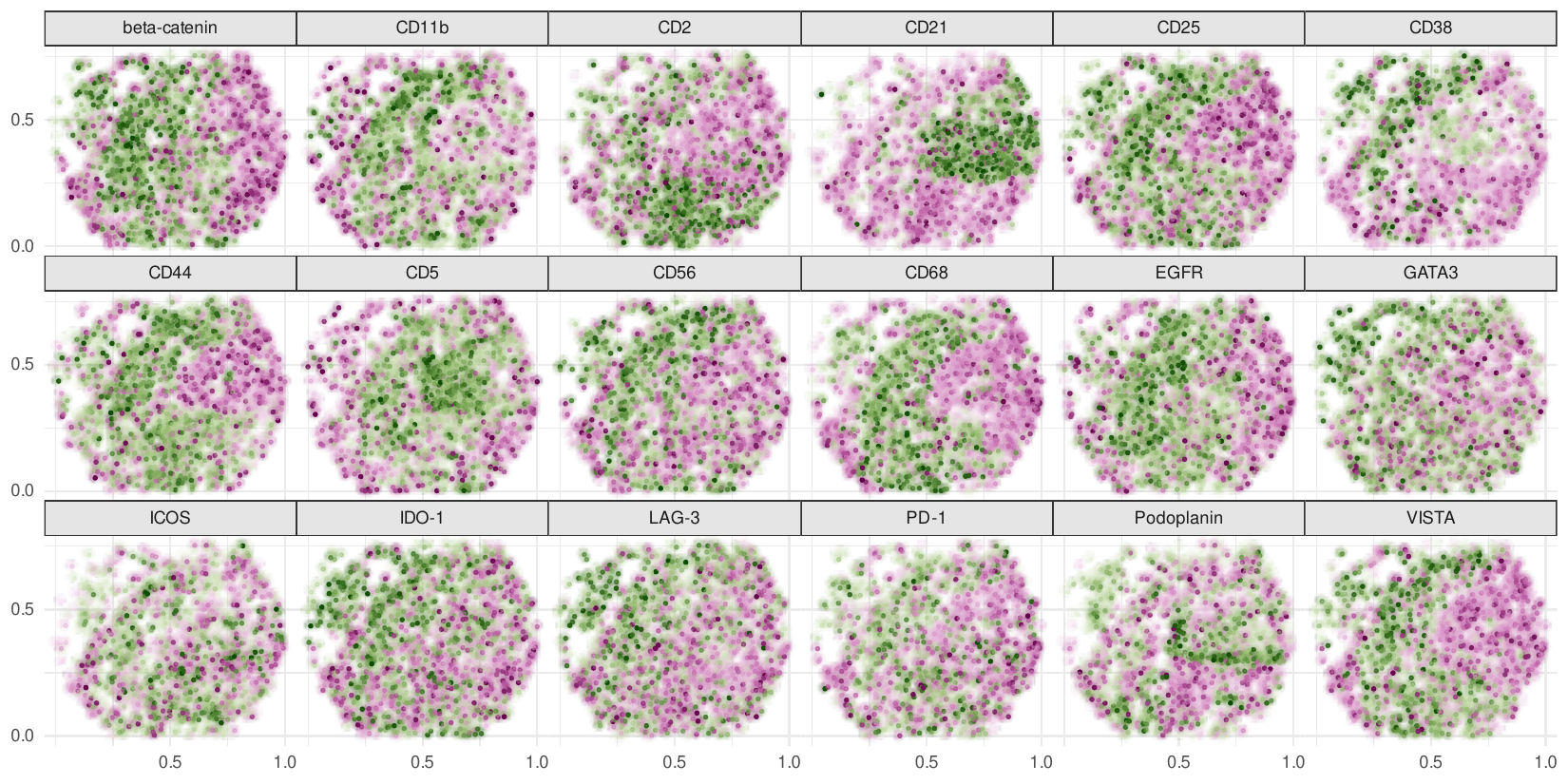}
    \caption{\footnotesize CRC CODEX data on 18 protein markers of a CRC patient. Green color corresponds to higher log-intensity of the marker. \normalsize}
    \label{fig:codex}
\end{figure}

\subsection{Cancer proteomics CODEX data}
The tumor microenvironment (TME) is a heterogeneous ecosystem of cancer, stromal, and immune cells within extracellular matrix. The TME profoundly shapes oncogenesis, metastasis, and therapy response \citep{hanahan_hallmarks_2011, de_visser_evolving_2023}. Patient-derived tissues analyzed by advanced multiplexed imaging yield spatially resolved single-cell data linking protein markers, cellular phenotypes, and clinical variables. Within the TME, disease-specific co-localization patterns inform immune dynamics and prognosis \citep{yuan_spatial_2016, giraldo_clinical_2019, tsujikawa_prognostic_2020}.

We consider a colorectal cancer dataset previously analyzed in \cite{Schurch2020}. This dataset was generated using co-detection by indexing (CODEX) technology optimized for formalin-fixed, paraffin-embedded tissue and tissue microarrays, leading to immunofluorescence imaging of the immune TME on 140 tissue regions of 35 advanced-stage colorectal cancer (CRC) patients. We restrict our analysis to a single tissue image (spot \textit{55A}) from patient 28, and consider the 18 most common protein markers in this tissue. At each spatial location, the data include the intensity of expression of each of the 18 protein markers. After log-transforming the data, we obtain an aligned dataset of 18 continuous variables at 2,873 spatial locations, for a total dimension of 51,714. Figure \ref{fig:codex} visualizes the dataset. 

%https://pubmed.ncbi.nlm.nih.gov/32763154/
%https://dx.doi.org/10.17632/mpjzbtfgfr.1

\begin{figure}%[h!]
    \centering
    \includegraphics[width=0.85\textwidth]{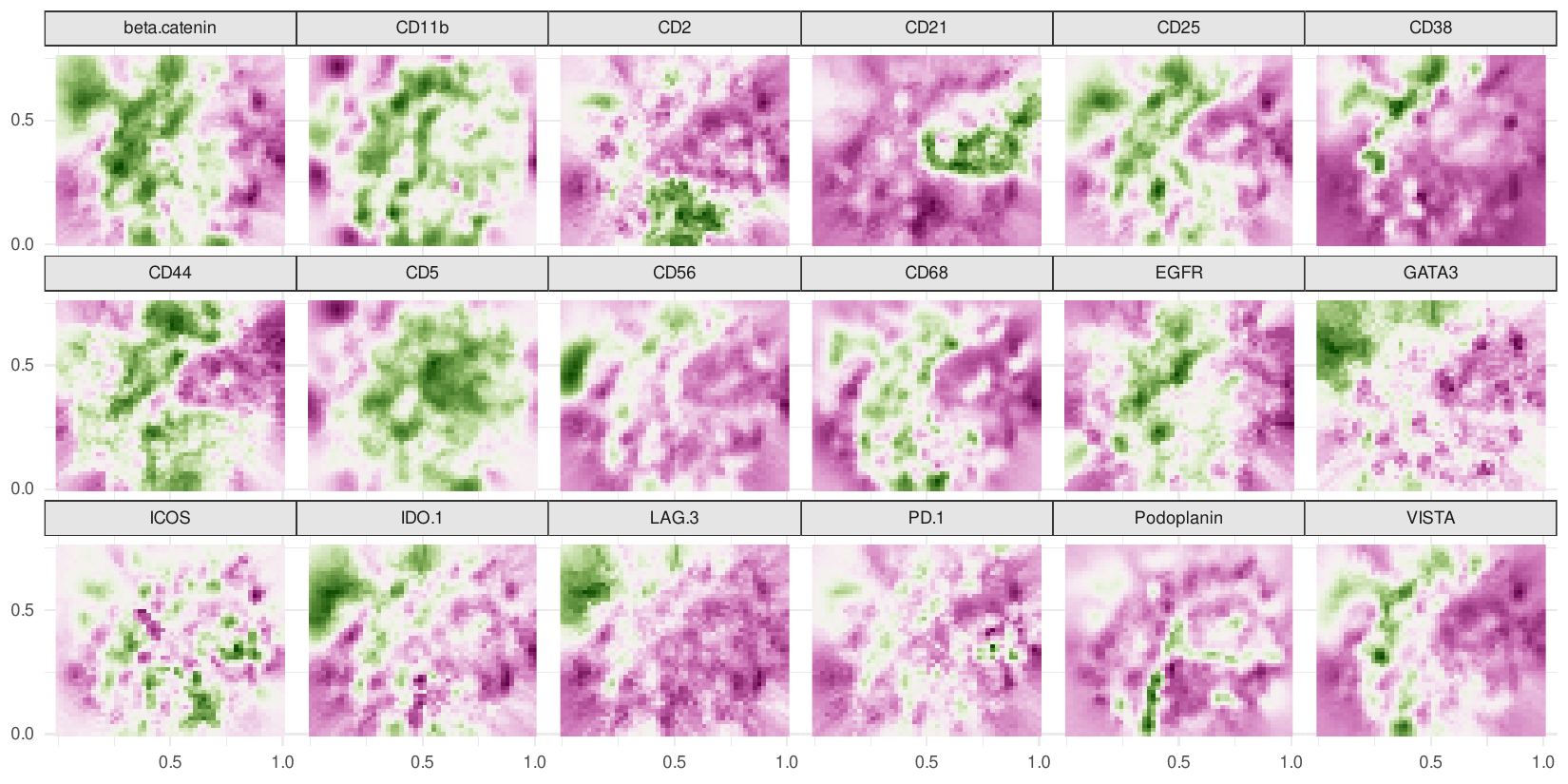}
    \caption{\footnotesize Intensity maps of the 18 protein markers from CRC CODEX data obtained via \modelname\ (\textit{Full} variant) by predicting the entire vector of variables at 2500 equally-spaced locations. %Green color corresponds to higher log-intensity of the marker.
    \normalsize}
    \label{fig:codex_results}
\end{figure}
\begin{figure}%
    \centering
    \subfloat[\centering Posterior means of $\rho_{ij}$]{{\includegraphics[width=0.55\linewidth]{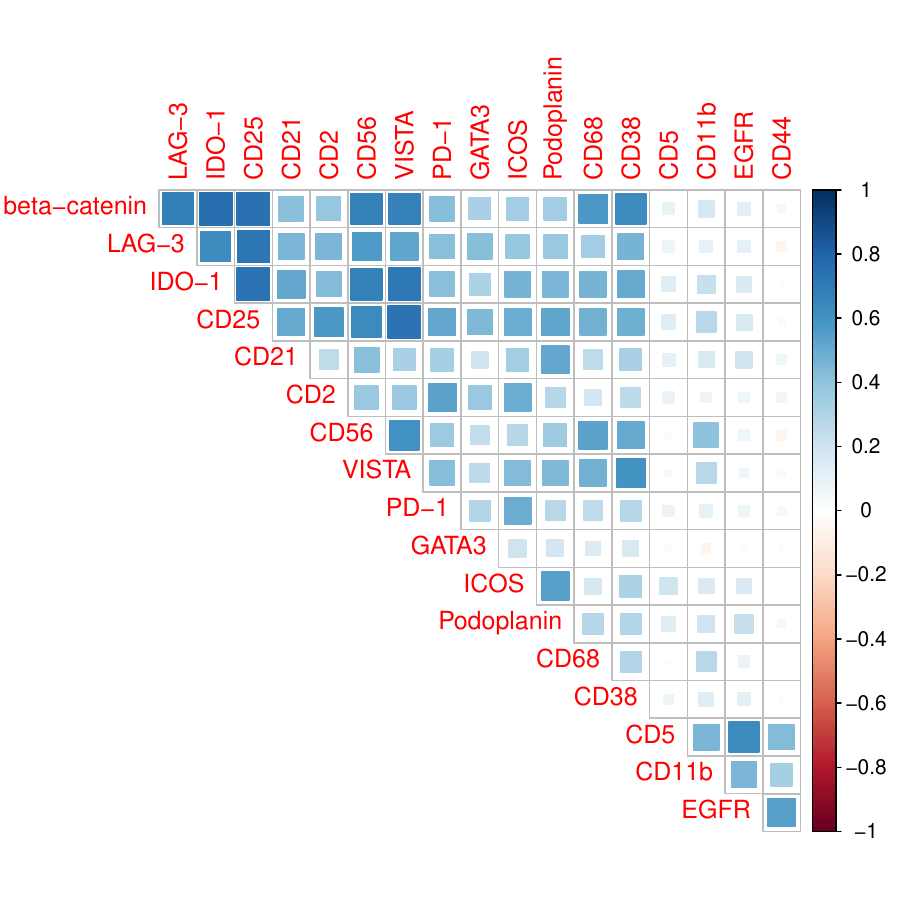} }}%
    \qquad
    \subfloat[\centering Posterior summary of $\phi_j, \nu_{j}, \tau^2_j$]{{\includegraphics[width=.35\linewidth]{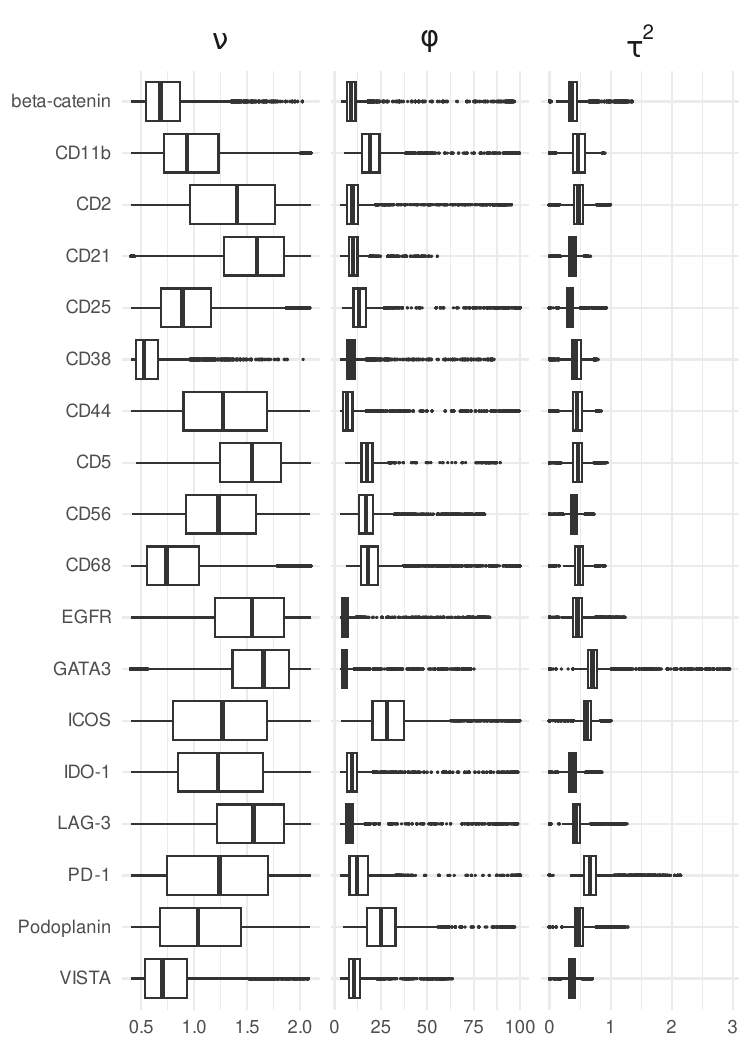} }}%
    \caption{\footnotesize Posterior summaries for \modelname\ parameters in the CODEX CRC analysis. \normalsize}%
    \label{fig:example}%
\end{figure}
\begin{table}
    \centering
    \resizebox{0.7\textwidth}{!}{%
    \begin{tabular}{|c|r|r|r|}
        \hline
        \cellcolor{gray!10}\textbf{Method} & \cellcolor{gray!10}\textbf{APE (s.e.)} & \cellcolor{gray!10}\textbf{CRPS (s.e.)} & \cellcolor{gray!10}\textbf{Time} {\footnotesize (min) } \\ \hline
        \modelname\ & \textbf{0.063} {\footnotesize (0.0065)} & \textbf{0.1823} {\footnotesize (0.0085)} & 59 \\\hline
        %\modelname\ Cluster & \textbf{0.0635} {\footnotesize (0.0064)} & 0.1826 {\footnotesize (0.0085)} & \textbf{22}     \\ \hline
        \lmc\ $k=6$           & 0.0699 {\footnotesize (0.0076)} & 0.1962 {\footnotesize (0.0089)} & 37     \\ \hline
        \lmc\ $k=8$           & 0.0675 {\footnotesize (0.0074)} & 0.1922 {\footnotesize (0.0090)} & 52           \\ \hline
        Non-spatial model     &   0.0686 {\footnotesize (0.0066)} & 0.1978 {\footnotesize (0.0089)} & 1  \\ \hline
        \end{tabular}
    }
\caption{\footnotesize Average percentage error, continuous rank probability score, and runtime in minutes of two \modelname models, two \lmcs, and a non-spatial model, on the CODEX CRC data analysis. \normalsize}\label{tab:proteomics_performance}
\end{table}
We estimate $\phi_j, \nu_j, \tau^2_j$ for all $j$ by fitting \modelname\ using Metropolis-within-Gibbs updates of $\btheta$ as described in Section \ref{sec:response_sampling}. % as well as the clustering method of Section \ref{sec:qclustering}, fixing the number of clusters at 6 (\textit{\modelname\ Cluster}). 
In fitting a multivariate spatial model to these data, we seek to understand how the TME organizes spatially, rather than making predictions within the imaged tissue. However, a model should demonstrate predictive value exceeding a non-spatial baseline to justify any conclusions about spatial dependence; otherwise, the inferred spatial structure may be attributed to overfitting.  
Therefore, we validate our approaches by comparing their predictive performance to \lmcs\ (with 6 or 8 factors) and a nonspatial multivariate model. We use a test set $\calL_{\text{test}}$ of 400 locations which we built by removing 2 randomly selected outcomes from the data at each of the locations in $\calL_{\text{test}}$. Therefore, we predict the unobserved part of $\by(\bl)$ at each $\bl \in \calL_{\text{test}}$. 
Table \ref{tab:proteomics_performance} summarizes our findings: \modelname\ outperformed other methods in out-of-sample predictions. % while demonstrating scalability to large data settings. %In particular, dimension reduction using \modelname\ (the \textit{Cluster} variant) resulted in a three-fold reduction of compute time at no cost of predictive performance. 
Dimension reduction using \lmcs\ resulted in a deterioration of predictive performance that led to worse predictive performance relative to a non-spatial model. 
Our analysis evidences tight co-localization of several protein markers, suggesting this patient's immune system is present and stimulated but locally restrained. This TME appears to exhibit micro-regions with both activation and restraint, pointing to spatially heterogeneous immune pressure within the tumor.

\section{Discussion}
We have introduced \modelname\ as a novel flexible class of cross-covariance functions for modeling multivariate spatial data. % in the ``large $n$, large $q$'' setting. 
\modelname\ leads to direct marginal inference and provides multiple avenues for flexible likelihood-based modeling and dimension reduction. 
In prioritizing inference of marginal covariances as well as overall parsimony, our method restricts $C_{ij}(\cdot, \cdot)$ through a product of the marginal Cholesky factors. Similar restrictions are common; see, e.g., \cite{Emery2022} for the multivariate Mat\'ern. Extending \modelname\ to more flexible parametrizations of $C_{ij}(\cdot, \cdot)$ is possible, as we show in the Supplement. 
There are multiple avenues for dimension reduction within the \modelname\ framework. First, if some outcomes exhibit similar spatial dependence, we can group them and fit a single correlation model per cluster. This reduces computational cost by limiting the number of distinct $\bL_i$ factors. A second avenue for dimension reduction follows from a low-rank assumption on $\bSigma$. Refer to the Supplement for implications of this assumption.
%We have outlined clustering methods for dimension reduction, but more flexible hierarchical priors may offer better performance in some settings---this is an interesting area for future research. 
%%%%%%%Another possible future direction is the exploration of spatial extensions of recent factor models which have shown promising performance \citep{chandra2021bayesian, fable}. 
%Alternatively, we can introduce more informative clustering priors for the $q$ outcomes for varying degrees of borrowing strength. 

Nonstationarities and covariate-dependence can be introduced into \modelname\ by choosing appropriate univariate correlations functions. Alternatively, one can directly take advantage of the structure of \modelname. For example, if $\bw(\cdot)$ is a $q$-variate GP, $\bw_i$ the vector of values of the $i$th margin of $\bw(\cdot)$, and $\bD_{w_i} = \text{diag}( \bw_i )$, then we can model $q$ outcomes via $C_{ij}(\bl, \bl') = \sigma_{ij} w_i(\bl) w_j(\bl') ( \bh_i(\bl) \bL_i \bL_j^\top \bh_j(\bl') + \xi(\bl, \bl') )$, leading to $\Cov(\calS) = \{ \oplus \bL_i \bD_{w_i} \} (\bSigma \otimes \bI_n) \{ \oplus  \bD_{w_i} \bL_i^\top \}$, in which case $\bw_i^2$ models the spatially-varying sill of outcome $i$. \modelname\ may also facilitate fitting multivariate extensions of the deformation method of \cite{sampsonguttorp1992} by assuming multiple outcomes share the same latent input space, which is realistic with ``omics'' data.

%Cross-outcomes spatial dependence of high-dimensional non-Gaussian data can be addressed via Bayesian hierarchical models with latent \modelname\ GPs and take advantage of the conditionally conjugate update of $\bSigma$ or use gradient-based methods such as those introduced in \cite{melange}.

Future methodological development may lead to novel models for spatial time series data based on IOX. We consider a single spatial outcome measured at $T$ discrete time points and $n$ sites. At each $t=1,\dots,T$, we can let $\rho_t(\cdot, \cdot)$ evolve with $t$ according to some dynamics, as well as impose structure on $\bQ = \bSigma^{-1}$ (its dimension is now $T \times T$), e.g., via Markovian assumptions leading to a sparse and banded $\bQ$.  %For multivariate spatial time series, we may model the temporal evolution of cross-dependence via a time-varying precision matrix $\bQ_t$. 

%% file: IOX_supplement_contents.tex
% !TEX root = appendix_standalone..tex

\section{Proofs}

\begin{customprop}{2.1}[Marginal covariance]%\label{prop:marginal} 
\begin{equation*}
C_{ii}(\bl, \bl') = \begin{cases} 
\sigma_{ii} \rho_i(\bl, \bl') \qquad & \text{if } \bl \in \calS \text{ or } \bl'\in \calS \text{ or } \bl = \bl', \\
\sigma_{ii} \rho_i(\bl, \calS) \rho_i(\calS)^{-1} \rho_i(\calS, \bl') \qquad& \text{if } \bl,\bl' \in \calS^c \text{ and } \bl\neq \bl'.\end{cases}
\end{equation*}
\end{customprop}
\textit{Proof. }
Letting $i=j$ we have
\begin{align*} 
C_{ii}(\bl, \bl') &= \sigma_{ii} \left[ \bh_i(\bl) \rho_i(\calS) \bh_i(\bl')^\top + \mathbb{1}_{\{ \bl=\bl'\} }r_i(\bl)  \right] \\
&= \sigma_{ii} \left[ \rho_i(\bl, \calS) \rho_i(\calS)^{-1} \rho_i(\calS,\bl') + \mathbb{1}_{\{ \bl=\bl'\} } \left( \rho_i(\bl,\bl) - \rho_i(\bl, \calS) \rho_i(\calS)^{-1} \rho_i(\calS,\bl)\right)  \right].
\end{align*}
We immediately notice that if $\bl=\bl'$ then $C_{ii}(\bl,\bl) = \sigma_{ii} \rho_i(\bl, \bl) = \sigma_{ii}$. If $\bl\neq\bl'$ and $\bl,\bl' \in \calS^c$ then we obtain the stated result by just dropping the indicator term. Finally, suppose $\bl \in \calS$ (the case $\bl'\in\calS$ is analogous). This means there is $r \in \{1, \dots, n\}$ such that $\bl =\bl_r$, which implies the vector  $\bh_i(\bl) = \be_{\bl} = (e_1, \dots, e_n)^\top$ has $e_r=1$ and $e_h=0$ for $h\neq r$. We see this e.g. by letting $\bx = (x_1, \dots, x_n)^\top$, $E[\bx]=0$ , and $\covf(\bx,\bx) = \rho_i(\calS)$. Then, $\bh_i(\bl)\bx = E[x_r \mid \bx] = x_r$. In other words, if $\bl = \bl_r \in \calS$ then $\bh_i(\bl)$ selects the $r$th row from the matrix to which it is premultiplied. Therefore, we select the $r$th row of $\rho_i(\calS, \bl')$, which is equal to $\rho_i(\bl_r, \bl') = \rho_i(\bl, \bl')$.

\begin{customprop}{2.2}[Cross-covariance and $\sigma_{ij}$]%\label{prop:cij_norm}
For all $\bl, \bl' \in \calD$ and all $i, j = 1,\dots, q$, the cross-covariance is $C_{ij}(\bl, \bl') \leq \sigma_{ij}$. If $\rho_i(\cdot, \cdot) = \rho_j(\cdot, \cdot)$, then $C_{ij}(\bl, \bl) = \sigma_{ij}$.
\end{customprop}
\textit{Proof. }
We start by proving that $\| \bh_i(\bl) \bL_i \| \leq 1$. Because $\rho_i(\cdot, \cdot)$ is a correlation function, then $r_i(\bl) = \rho_i(\bl, \bl) - \rho(\bl, \calS) \rho_i(\calS)^{-1} \rho_i(\calS, \bl) = 1 - \bh_i(\bl) \rho_i(\calS) \bh_i(\bl)^\top  \ge 0$ which implies $\bh_i(\bl) \rho_i(\calS) \bh_i(\bl)^\top = \bh_i(\bl) \bL_i \bL_i^\top \bh_i(\bl)^\top = \| \bh_i(\bl) \bL_i \|^2 \leq 1$.

Let $\ba^\top = \bh_i(\bl) \bL_i$ and $\bb^\top = \bh_j(\bl') \bL_j$. We showed above that $\| \ba \|\leq 1$ and $\|\bb \| \leq 1$. Then
\begin{align*} C_{ij}(\bl, \bl') &= \sigma_{ij} \left[ \bh_i(\bl) \bL_i \bL_j^\top \bh_j(\bl')^\top + \sqrt{ r_i(\bl) r_j(\bl') } \right] \\
&= \sigma_{ij} \left[ \ba^\top \bb + \sqrt{ (1 - \ba^\top \ba) (1 - \bb^\top \bb) } \right] \\
&\leq \sigma_{ij} \left[ \|\ba\| \|\bb\| + \sqrt{ (1-\|\ba\|^2)(1-\|\bb\|^2) }  \right] \leq \sigma_{ij}.
\end{align*}
In particular, consider the case $\bl=\bl'$. If $\rho_i(\cdot, \cdot) = \rho_j(\cdot, \cdot)$ then we obtain $C_{ij}(\bl, \bl') = \sigma_{ij}$, which implies that $\sigma_{ij}$ is the zero-distance correlation between variables with the same marginal correlation. 
%\end{proof}

\begin{customprop}{2.3}[\modelname\ cross-covariance matrix function]%\label{prop:xcovmatf}
Let $\bl, \bl'\in \calS$.
\begin{equation}\label{eq:xcovmatf1}
    \begin{aligned}
\Cov(\bl, \bl') &= \bSigma \odot \left[ \bK(\bl, \bl') + \bd(\bl, \bl')\bd(\bl, \bl' )^\top \right] \\
&= \{ \oplus \bh_i(\bl) \bL_i \} (\bSigma \otimes \bI_n) \{ \oplus \bL_i^\top \bh_i(\bl')^\top \} + \bD(\bl, \bl') \bSigma \bD(\bl, \bl'),
\end{aligned} \end{equation}
where $\bK(\bl, \bl')$ is the $q\times q$ matrix whose $i,j$ element is $\bh_i(\bl) \bL_i \bL_j^\top \bh_j(\bl')^\top$, $\bd(\bl, \bl')$ is a vector of dimension $q$ whose $i$th element is $\mathbb{1}_{\{ \bl=\bl'\} }\sqrt{r_i(\bl)}$, $\bD(\bl, \bl') = \text{diag}\{\bd(\bl, \bl')\}$, ``$\oplus$'' is the direct sum operator and we denote with $\{ \oplus \bh_i(\bl) \bL_i \}$ the $q \times nq$ block-diagonal matrix with $\bh_i(\bl) \bL_i$ as its $i$th block, while ``$\odot$'' is the Hadamard element-by-element product. 
\end{customprop}
\textit{Proof. }
Let $\bl \neq \bl'$. Then
\begin{align*}
\Cov(\bl, \bl') &= \begin{bmatrix} C_{ij}(\bl, \bl') \end{bmatrix}_{i,j=1, \dots, q}  \\
&= \begin{bmatrix} \sigma_{ij} \bh_i(\bl) \bL_i \bL_j^\top \bh_j(\bl')^\top \end{bmatrix}_{i,j=1, \dots, q} = \bSigma \odot \begin{bmatrix} \bh_i(\bl) \bL_i \bL_j^\top \bh_j(\bl')^\top \end{bmatrix}_{i,j=1, \dots, q}  \\
&=\begin{bmatrix}
   \sigma_{11} \bh_1(\bl) \bL_1 \bL_1^\top \bh_1(\bl')^\top & \cdots & \sigma_{1q} \bh_1(\bl)\bL_1 \bL_q^\top \bh_q(\bl')^\top \\
    & \ddots & \\
   \sigma_{q1} \bh_q(\bl)\bL_q \bL_1^\top \bh_1(\bl')^\top & & \sigma_{qq} \bh_q(\bl) \bL_q \bL_q^\top \bh_q(\bl')^\top \end{bmatrix}\\
&= \begin{bmatrix}
   \sigma_{11} \bh_1(\bl) \bL_1 & \cdots & \sigma_{1q} \bh_1(\bl)\bL_1 \\
    & \ddots & \\
   \sigma_{q1} \bh_q(\bl)\bL_q & & \sigma_{qq} \bh_q(\bl) \bL_q
\end{bmatrix} \begin{bmatrix}
     \bL_1^\top \bh_1(\bl')^\top & & \\
    & \ddots & \\
    & & \bL_q^\top \bh_q(\bl')^\top
\end{bmatrix} \\
&= \begin{bmatrix}
    \bh_1(\bl) \bL_1 & & \\
    & \ddots & \\
    & & \bh_q(\bl) \bL_q
\end{bmatrix} \begin{bmatrix}
    \sigma_{11} \bI_n & \cdots & \sigma_{1q} \bI_n \\
  \vdots  & \ddots & \vdots \\
  \sigma_{q1} \bI_n & \cdots & \sigma_{qq} \bI_n \end{bmatrix} 
  \begin{bmatrix}
     \bL_1^\top \bh_1(\bl')^\top & & \\
    & \ddots & \\
    & & \bL_q^\top \bh_q(\bl')^\top
\end{bmatrix} \\
&= \{ \oplus \bh_i(\bl) \bL_i \} (\bSigma \otimes \bI_n) \{ \oplus \bL_i^\top \bh_i(\bl')^\top \}.
\end{align*}
If $\bl = \bl'$ then we need to add $\sqrt{r_i(\bl)r_j(\bl)}$ to the $(i,j)$th element of $\Cov(\bl, \bl')$, which is the purpose of $\bd(\bl, \bl') \bd(\bl, \bl')^\top$ by construction. We conclude by noting that $\bSigma \odot ( \bd(\bl, \bl') \bd(\bl, \bl')^\top ) = \bD(\bl, \bl') \bSigma \bD(\bl, \bl')$.
%\end{proof}

\begin{customcoro}{2.1} 
If $\bl \in \calS$ then $\bh_i(\bl) \bL_i = \bL_i\row{\bl}$, the row of $\bL_i$ corresponding to $\bl$. Then,
\begin{equation} %\label{eq:xc_S} 
    \begin{aligned}
\Cov(\calS) = (\bSigma \otimes \mathbf{1}_{n,n}) \odot \bK = \{ \oplus \bL_i \} (\bSigma \otimes \bI_n) \{ \oplus \bL_i^\top \},
\end{aligned} \end{equation}
where $\bK$ is the $nq \times nq$ matrix whose $(i,j)$th block is $\bL_i \bL_j^\top$; $\mathbf{1}_{n,n}$ is the $n\times n$ matrix of $1$s.
\end{customcoro}
\textit{Proof. }%\begin{proof}
If $\bl = \bl_r$ for some $r\in \{1, \dots, n\}$, then $\bh_i(\bl) = \be_{\bl} = (e_1, \dots, e_n)^\top$ has $e_r=1$ and $e_h=0$ for $h\neq r$. We see this e.g. by letting $\bx = (x_1, \dots, x_n)^\top$, $E[\bx]=0$ , and $\covf(\bx,\bx) = \rho_i(\calS)$. Then, $\bh_i(\bl)\bx = E[x_r \mid \bx] = x_r$. In other words, if $\bl = \bl_r \in \calS$ then $\bh_i(\bl)$ selects the $r$th row from the matrix to which it is premultiplied. Similarly, if $\bl\in \calS$ then $r_i(\bl)=0$ for all $i=1,\dots,q$. The rest follows.
%\end{proof}

\begin{customprop}{2.4} %\label{prop:valid}
$\Cov(\bl, \bl')$ is a valid cross-covariance matrix function.
\end{customprop}
\textit{Proof. }%\begin{proof}
The conditions stated in the introduction can be expressed as follows. Suppose $\bZ$ is a $m \times q$ random matrix whose $i$th row is $\bz(\bl_i)^\top$, $\bl_i\in \calT = \{\bt_1, \dots, \bt_m \}$ and $j$th column $\bz_j$. Then, a cross-covariance matrix function must be such that the assumed covariance matrix of $\bz = \text{vec}(\bZ)$ (or, equivalently, of $\text{vec}(\bZ^\top)$) is symmetric positive semidefinite for any $\calT$ and any $m$ \citep{genton_ccov}. Because our assumption is that $\covf(\bz) = \Cov(\calT)$, we need to show that $\Cov(\calT)$ is symmetric positive semidefinite. If $\bl \neq \bl'$, 
\begin{align*}
\Cov(\bl, \bl') &= \bSigma \odot \left( \bK(\bl, \bl') + \bD(\bl, \bl')  \right)\\
&= \bSigma \odot \begin{bmatrix} \bh_i(\bl) \bL_i \bL_j^\top \bh_j(\bl')^\top \end{bmatrix}_{i,j=1, \dots, q} \\
&= \bSigma \odot \begin{bmatrix} \bh_i(\bl') \bL_i \bL_j^\top \bh_j(\bl)^\top \end{bmatrix}_{i,j=1, \dots, q}^\top\\
&= \bSigma \odot \bK(\bl', \bl)^\top = \Cov(\bl', \bl)^\top.
\end{align*} 
The additional term $\bD(\bl,\bl')$ %(defined in Proposition \ref{prop:xcovmatf}) 
is zero if $\bl\neq \bl'$; otherwise,  
\[\bD(\bl, \bl) = \begin{bmatrix} \sqrt{r_1(\bl)} \\ \vdots \\ \sqrt{r_q(\bl)} \end{bmatrix} 
\begin{bmatrix} \sqrt{r_1(\bl)} & \cdots & \sqrt{r_q(\bl)} \end{bmatrix},\]
which is symmetric by construction.
As for positive semidefiniteness, define $\bH_i(\calT) = \rho_i(\calT, \calS) \rho_i(\calS)^{-1}$, i.e. the $m \times n$ matrix whose $j$th row is $\bh_i(\bt_j)$ and assume $\bSigma = \bA \bA^\top$ where $\bA$ is a $q\times k$ matrix of rank $k\leq q$. Then,
\begin{align*}
 \Cov(\calT) &= \{ \oplus \bH_i(\calT) \bL_i \} (\bSigma \otimes \bI_n) \{ \oplus \bL_i^\top \bH_i(\calT)^\top \} + (\bSigma \otimes \mathbf{1}_{m,m}) \odot \bD_T\\
 &= \{ \oplus \bH_i(\calT) \bL_i \} (\bA \otimes \bI_n) (\bA^\top \otimes \bI_n) \{ \oplus \bL_i^\top \bH_i(\calT)^\top \} + (\bSigma \otimes \mathbf{1}_{m,m}) \odot \bD_T \\ 
 &= \bG \bG^\top + (\bSigma \otimes \mathbf{1}_{m,m}) \odot \bD_T,
\end{align*}
where $\bG = \{ \oplus \bH_i(\calT) \bL_i \} (\bA \otimes \bI_n)$ and $\bD_T$ is a $mq \times mq$ block matrix whose $(i,j)$th block of dimension $m\times m$ is a diagonal matrix whose diagonal entries are $\sqrt{r_i(\bl_h) r_j(\bl_h)}$ for $h=1,\dots,m$. The first term is non-negative definite because $\bx^\top \bG \bG^\top \bx = \tilde{\bx}^\top \tilde{\bx} \ge 0$. As for the second term, $(\bSigma \otimes \mathbf{1}_{m,m})$ is non-negative definite because $\bSigma$ is assumed to be. Finally, suppose $\bP$ is a permutation matrix that rearranges rows and columns of $\bD_T$ so that $\bP\bD_T \bP^\top$ is a block diagonal matrix with $\bD(\bl_i, \bl_i)$ as its $i$th diagonal block, with $i=1,\dots,m$. $\bD(\bl_i, \bl_i)$ is non-negative definite rank 1. Therefore, $\bP\bD_T \bP^\top$ is non-negative definite of rank $m$, and so is $\bD_T$. Hadamard product and matrix sum retain non-negative definiteness.
%\end{proof}

\begin{customprop}{2.5}%\label{prop:separable}
In the separable specification, i.e., $\rho_i(\cdot, \cdot) = \rho(\cdot, \cdot)$ for all $i$, \modelname\ and \lmc\ coincide at $\calS$.
\end{customprop}
\textit{Proof. }%\begin{proof}
For \modelnames,
\begin{align*}
\Cov(\calS) &= \{\oplus \bL_i \} (\bSigma \otimes \bI_n) \{ \oplus \bL_i^\top \} \\
&= (\bI_q \otimes \bL) (\bSigma \otimes \bI_n) (\bI_q \otimes \bL^\top ) \\
&= \bSigma \otimes \bL\bL^\top = \bSigma \otimes \rho(\calS).
\end{align*}
For \lmc,
\begin{align*}
\Cov(\calS) &= (\bA \otimes \bI_n) \{ \oplus \rho_i(\calS) \} (\bA^\top \otimes \bI_n)\\
&= (\bA \otimes \bI_n) ( \bI_q \otimes \rho(\calS) ) (\bA^\top \otimes \bI_n)\\
&= \bA\bA^\top \otimes \rho(\calS) = \bSigma \otimes \rho(\calS).
\end{align*}
%\end{proof}

\begin{customprop}{2.6}%\label{prop:schur}
\noindent Suppose $\bl,\bl' \in \calS^c$, $\bl\neq \bl'$. Then, $\Cov(\bl, \bl') - \Cov(\bl, \calS) \Cov(\calS)^{-1} \Cov(\calS, \bl')  = \bzero$. %Let $\bX$ be a random $(n+2)\times q$ matrix whose $j$th column is $\bx_j = (x_j(\bl_1), \dots, x_j(\bl_n), x_j(\bl), x_j(\bl'))^\top$, and let $\bx = \text{vec}(\bX)$ with $E(\bx) = 0$ and $\covf(\bx) = \Cov(\calT)$ as in \eqref{prop:xcovmatf}. Then
\end{customprop}
\textit{Proof. }%\begin{proof}
We have $\Cov(\bl, \calS) =\{ \oplus \bh_i(\bl) \bL_i \} (\bSigma \otimes \bI_n) \{ \oplus \bL_i^\top \} $ and $\Cov(\calS)^{-1} = \{ \oplus \bL_i^{-\top} \} (\bSigma^{-1} \otimes \bI_n) \{ \oplus \bL_i^{-1} \}$. Together, these imply:
\begin{align*}
\Cov(\bl, \calS)\Cov(\calS)^{-1}\Cov(\calS, \bl') &= \{ \oplus \bh_i(\bl) \bL_i \} (\bSigma \otimes \bI_n) \{ \oplus \bL_i^\top  \bh_i(\bl')^\top \} = \Cov(\bl, \bl'). %\qedhere
\end{align*}
%\end{proof}

\begin{customprop}{3.1}[GP-\modelname\ log-density]%\label{prop:density}
    \begin{equation} \log p(\bw \mid \btheta, \bQ) = \text{const} + \frac{n}{2}\log\det(\bQ) + \sum_{ij}\log \bL_j^{-1}\rowcol{i}{i} -\frac{1}{2} \text{Tr}\left( \bV \bQ \bV^\top \right), \end{equation}
    where $\bL_j^{-1}\rowcol{i}{i}$ is the $i$th diagonal element of $\bL_j^{-1}$, and $\bV$ is the $n\times q$ matrix whose $j$th row is $\bv_j = \bL_j^{-1}\bw_j$. 
\end{customprop}
\textit{Proof. }%\begin{proof}
We start by noting that 
\begin{align*}
    \Cov^{-1} = \{ \oplus \bL_j^{-\top} \}(\bSigma^{-1} \otimes \bI_n) \{ \oplus \bL_j^{-1} \} = \{ \oplus \bL_j^{-\top} \}(\bQ^{-1} \otimes \bI_n) \{ \oplus \bL_j^{-1} \},
\end{align*}
and because $\det\left(\bL^{-1}_j\right) = \prod_i \bL^{-1}_j\rowcol{i, i}$, we get $-\frac{1}{2} \log \det( \Cov )=\frac{n}{2}\log\det(\bQ) +\sum_{ij} \log\bL_j^{-1}\rowcol{i}{i} $. Then,
\begin{align*}
    \log & p(\bw \mid \btheta, \bbeta, \bSigma) = \text{const} -\frac{1}{2} \log \det( \Cov ) -\frac{1}{2} \bw^\top \bC^{-1} \bw \\
    &= \text{const} + \frac{n}{2}\log\det(\bQ) + \sum_{ij} \log \bL_j^{-1}\rowcol{i}{i}  -\frac{1}{2} \bw^\top \{ \oplus \bL_j^{-\top} \}(\bQ \otimes \bI_n) \{ \oplus \bL_j^{-1} \} \bw \\
    &= \text{const} + \frac{n}{2}\log\det(\bQ) + \sum_{ij} \log \bL_j^{-1}\rowcol{i}{i}  -\frac{1}{2} \vecop( \bV )^\top (\bQ \otimes \bI_n) \vecop( \bV )\\
    &= \text{const} + \frac{n}{2}\log\det(\bQ) + \sum_{ij} \log \bL_j^{-1}\rowcol{i}{i}  -\frac{1}{2} \text{Tr}\left( \bV \bQ \bV^\top \right).
    %&= \text{const} + \frac{n}{2}\log\det(\bQ) \sum_{ij} \bL_i^{-1}\rowcol{j}{j} -\frac{1}{2} \bw^\top \{ \oplus \bL_i^{-\top} \}(\bU^\top \otimes \bI_n) (\bU \otimes \bI_n) \{ \oplus \bL_i^{-1} \} \bw \\
    %&= \text{const} + \frac{n}{2}\log\det(\bQ) \sum_{ij} \bL_i^{-1}\rowcol{j}{j} -\frac{1}{2} \sum_{i=1}^q \left( \sum_{j=1}^q U_{ij} \bL_j^{-1} \bw_j \right)^\top \left(\sum_{j=1}^q U_{ij} \bL_j^{-1} \bw_j \right) \\
    %&= \text{const} + \frac{n}{2}\log\det(\bQ) \sum_{ij} \bL_i^{-1}\rowcol{j}{j} -\frac{1}{2} \sum_{i=1}^q \left( \sum_{j=1}^q U_{ij} \bv_j \right)^\top \left(\sum_{j=1}^q U_{ij} \bv_j \right),
\end{align*}
%\end{proof}

\begin{customprop}{3.2}[Conditional density]%\label{prop:conditionals}
$p(\bw_j \mid \bw_{j^c}, \btheta, \bQ) = N(\bw_j; \bm_j, \bM_j)$, where $
\bm_j = - \bL_j \sum_{r\in j^c} \frac{Q_{jr}}{Q_{jj}} \bv_r$ and $\bM_j^{-1} = Q_{jj} \rho_j(\calS)^{-1}$. Then, we evaluate $\log N(\bw_j; \bm_j, \bM_j)$ as:
\begin{equation}%\label{eq:ioxconditional} 
    \begin{aligned}
\log N(\bw_j; \bm_j, \bM_j)
    &\propto  -\frac{1}{2} \log |\bM_j| -\frac{1}{2}(\bw_j - \bm_j)^\top \bM_j^{-1} (\bw_j - \bm_j) \\ 
    & = \frac{n}{2}\log Q_{jj} + \sum_{i=1}^n \log \bL_j^{-1}[i,i] -\frac{1}{2Q_{jj}} \bQ_{j\cdot} \bV^\top \bV \bQ_{j\cdot}^\top ,
\end{aligned}    
\end{equation}
where $Q_{jr}$ is the $(j,r)$th element and $\bQ_{j\cdot}$ is the $j$th row of $\bQ$.
\end{customprop}
\textit{Proof. } %\begin{proof}
Because the joint distribution of $\bw$ is Gaussian, we have $p(\bw_j \mid \bw_{j^c}, \btheta, \bQ) = N(\bw_j; \bm_j, \bM_j)$, where $\bM_j^{-1}$ is the $(j,j)$th block of the precision matrix $\Cov^{-1} = \left\{\oplus \bL_j^{-\top} \right\} (\bQ \otimes \bI_n) \left\{\oplus \bL_j^{-1} \right\}$. Therefore, we find $\bM_j^{-1} = Q_{jj} \rho_j(\calS)^{-1}$ following the same steps as in Proposition 2.1. Again because the joint is Gaussian, the conditional mean is $\bm_j = \Cov_{j, j^c} \Cov_{j^c}^{-1} \bw_{j^c}$, which becomes
\begin{align*}
\bm_j &= \bL_j (\bSigma_{j, j^c} \bSigma_{j^c}^{-1} \otimes \bI_n) \{ \oplus_{j^c} \bL_r^{-1} \} \bw_{j^c} = -\bL_j \left(\frac{\bQ_{j, j^c}}{Q_{jj}} \otimes \bI_n \right) \{ \oplus_{j^c} \bL_r^{-1} \} \bw_{j^c}\\
&= - \bL_j \sum_{r\in j^c} \frac{Q_{jr}}{Q_{jj}} \bL_r^{-1} \bw_r,
\end{align*}
where $\bQ_{j, j^c}$ is the $1 \times q-1$ vector obtained from the $j$th row and $j^c$ columns of $\bQ$ and we denote with $``\oplus_{j^c}''$ the direct sum operator over $j^c$ indices. 
As for evaluating the joint density, 
\begin{align*} 
    N(\bw_j; \bm_j, \bM_j) &\propto |\bM_j|^{-\frac{1}{2}} \exp\{ -\frac{1}{2} (\bw_j - \bm_j)^\top \bM_j^{-1} (\bw_j - \bm_j) \} \\
    &= |\bM_j|^{-\frac{1}{2}} \exp\{ -\frac{Q_{jj}}{2} (\bw_j - \bm_j)^\top \bL_j^{-\top} \bL_j^{-1} (\bw_j - \bm_j) \} \\
  (\star)\qquad  &= \sqrt{ |Q_{jj} \rho_j(\calS)^{-1}|} \exp\left\{ -\frac{1}{2Q_{jj}} \left( \sum_{r=1}^q Q_{jr} \bv_r \right)^\top \left( \sum_{r=1}^q Q_{jr} \bv_r \right) \right\}\\
    &= \sqrt{ |Q_{jj} \rho_j(\calS)^{-1}|} \exp\left\{ -\frac{1}{2Q_{jj}} \bQ_{j\cdot} \bV^\top \bV \bQ_{j\cdot}^\top \right\}.
    \end{align*}
where for $(\star)$ we use the following:
\begin{align*} 
    \bL_j^{-1} &(\bw_j - \bm_j) = \bL_j^{-1} \bw_j - \bL_j^{-1} \bm_j = \bL_j^{-1} \bw_j - \bL_j^{-1} \bm_j \\
    &= \bL_j^{-1} \bw_j + \left(\frac{\bQ_{j, j^c}}{Q_{jj}} \otimes \bI_n \right) \{ \oplus_{j^c} \bL_r^{-1} \} \bw_{j^c} = \bL_j^{-1} \bw_j + \sum_{r\in j^c} \frac{Q_{jr}}{Q_{jj}} \bL_r^{-1} \bw_r\\
     & = \sum_{r=1}^q \frac{Q_{jr}}{Q_{jj}} \bL_r^{-1} \bw_r =  \sum_{r=1}^q \frac{Q_{jr}}{Q_{jj}} \bv_r.
\end{align*}
Finally, by definition $\bL_j\bL_j = \rho_j(\calS)$ and we obtain the final result by taking logs.
%\end{proof}

\begin{customprop}{3.3}%\label{prop:predictions}
\noindent $\bH(\bt) = \{ \oplus \bh_i(\bt)\}$ and $\bR(\bt)= \bD(\bt, \bt) \bSigma \bD(\bt, \bt)$. For $\bt \notin \calS$, $p(w_i(\bt) \mid \bw, \btheta, \bSigma)$ only depends on $\rho_i(\cdot, \cdot)$ at $\calS$, $\sigma_{ii}$ and $\bw_i$.  
\end{customprop}
\textit{Proof. } %\begin{proof}
\begin{align*}
\bH(\bt) &= \Cov(\bt, \calS) \Cov(\calS)^{-1} \\
&= \{\oplus \bh_i(\bt) \bL_i \}(\bSigma\otimes \bI_n) \{ \oplus \bL_i^\top \} \{ \oplus \bL_i^{-\top} \} (\bSigma^{-1} \otimes \bI_n) \{ \oplus \bL_i^{-1} \} \\
&= \{\oplus \bh_i(\bt) \bL_i \} \{ \oplus \bL_i^{-1} \}  = \{ \oplus \bh_i(\bt) \},
\end{align*}
\begin{align*}
\bR(\bt) &= \Cov(\bt) - \bH(\bt) \Cov(\calS, \bt) \\
&= \{\oplus \bh_i(\bt) \bL_i \}(\bSigma\otimes \bI_n) \{\oplus \bL_i^\top \bh_i(\bt)^\top \} + \bD(\bt, \bt) \bSigma \bD(\bt, \bt) -\\ & \qquad\qquad - \{ \oplus \bh_i(\bt) \} \{ \oplus \bL_i \} (\bSigma\otimes \bI_n) \{\oplus \bL_i^\top \bh_i(\bt)^\top \} \\
&= \{\oplus \bh_i(\bt) \bL_i \}(\bSigma\otimes \bI_n) \{\oplus \bL_i^\top \bh_i(\bt)^\top \} + \bD(\bt, \bt) \bSigma \bD(\bt, \bt) -\\ & \qquad\qquad - \{ \oplus \bh_i(\bt)\bL_i \} (\bSigma\otimes \bI_n) \{\oplus \bL_i^\top \bh_i(\bt)^\top \} \\
&= \bD(\bt, \bt) \bSigma \bD(\bt, \bt).% = \bD(\bt, \bt) \bA \bA^\top \bD(\bt, \bt).
\end{align*}
The conclusion follows because the $i$th row of $\bH(\bt)$ (which is $\bh_i(\bt)$) and the $(i,i)$ element of $\bR(\bt)$ are constructed using $\rho_i(\cdot)$, $\calS$, $\sigma_{ii}$, and the predictive mean is thus $\bh_i(\bt)\bw_i$.
%\end{proof}

%\begin{proposition}\label{prop:predictions_S}
\begin{customprop}{3.4} \noindent
Suppose $q=2$ and $\calS = \{\bl_1, \bl_2\}$. Then $p(w_i(\bl_r) \mid \bw(\bl_s))$ depends on both $w_i(\bl_s)$ and $w_j(\bl_s)$, for each choice of $i,j,r,s \in \{1,2\}$, $i\neq j$ and $r\neq s$. 
\end{customprop}
\textit{Proof. }
%\begin{proof}
%---------------------------------------------------------------
%  Joint distribution (outcome-first vectorisation)
%---------------------------------------------------------------
\[
\bW=\begin{bmatrix}
  w_{1}(\bl_{1}) & w_{2}(\bl_{1}) \\
  w_{1}(\bl_{2}) & w_{2}(\bl_{2})
\end{bmatrix},
\qquad
\bw=\vecop(\bW)=
\left(
  w_{1}(\bl_{1}),
  w_{1}(\bl_{2}),
  w_{2}(\bl_{1}),
  w_{2}(\bl_{2})
\right)^{\!\top}.
\]

\[
\bw \sim N\left(\bmu = 
  \begin{bmatrix}\bmu_1 \\ \bmu_2\end{bmatrix}, \bG =
  \begin{bmatrix}
      \sigma_{11} \bC_{1}          & \sigma_{12} \bL_{1} \bL_{2}^{\top}\\
      \sigma_{12}\bL_{2}\bL_{1}^{\top} & \sigma_{22}\bC_{2}
  \end{bmatrix}
\right),
\]
with \(\bC_{j}= \rho_j(\{ \bl_1, \bl_2\}, \{\bl_1, \bl_2 \}; \btheta_j)= \bL_{j}\bL_{j}^{\top}\) for \(j=1,2\).
%---------------------------------------------------------------
%  Selection matrices (pick one location)
%---------------------------------------------------------------
Define 
\[
\bS_{1}=
\begin{bmatrix}
  1&0&0&0\\
  0&0&1&0
\end{bmatrix},
\qquad
\bS_{2}=
\begin{bmatrix}
  0&1&0&0\\
  0&0&0&1
\end{bmatrix}.
\]
Then, \(\bw(\bl_{1}) = \bS_{1} \bw\), \(\bw(\bl_{2})=\bS_{2}\bw\), $\bmu_{\bl_{i}} = \bS_{i}\bmu$, and $\bG_{\bl_{i},\bl_{j}} = \bS_{i} \bG \bS_{j}^{\top}$ for $i,j\in\{1,2\}$.
%---------------------------------------------------------------
%  Conditional distribution  Y(ℓ₁) | Y(ℓ₂)
%---------------------------------------------------------------
Then $\bw(\bl_{1}) \mid \bw(\bl_{2}) = \bv
\sim  N\left( \bm_{1|2}, \bV_{1|2} \right)$, where
\begin{align}
\bm_{1|2} &=
  \bmu_{\bl_{1}}
  +\bG_{\bl_{1},\bl_{2}}\,
   \bG_{\bl_{2},\bl_{2}}^{-1}\,
  \left(\bv-\bmu_{\bl_{2}}\right),\\[6pt]
\bV_{1|2} &=
  \bG_{\bl_{1},\bl_{1}}
  -\bG_{\bl_{1},\bl_{2}}\,
   \bG_{\bl_{2},\bl_{2}}^{-1}\,
   \bG_{\bl_{2},\bl_{1}}.
\end{align}
Because $\bG_{\bl_{1},\bl_{2}}\bG_{\bl_{2},\bl_{2}}^{-1}$ is generally a dense $2 \times 2$ matrix, the second element of $\bm_{1|2}$ will depend on the first element of $\bv$, and vice-versa. Therefore, marginal predictions at $\bw(\bl_1)$ depend on all data at $\bw(\bl_2)$ (as well as all covariance parameters).
%\end{proof}

%\begin{theoremEnd}[end, restate, text link=]{prop}
\begin{customprop}{3.5}
$p(\bw_m(\bt) \mid \bw_{o}(\bt), \bw) = N(\bw_m(\bt); \bh_{m\mid o}, \bR_{m \mid o}(\bt))$, where $\bh_{m\mid o} = \bH_m(\bt)\bw + \bH_{m\mid o}(\bt) (\bw_{o}(\bt) - \bH_{o}(\bt) \bw)$, $\bH_{o}(\bt)$ is the submatrix of $\bH(\bt)$ where we take the rows corresponding to $o$ (similarly $\bH_{m}(\bt)$ with rows $m$), and 
\begin{align*}
\bH_{m \mid o}(\bt) &= \bD_m(\bt, \bt) \bSigma_{m, o}  \bSigma_{o}^{-1} \bD^{-1}_{o}(\bt, \bt) = - \bD_m(\bt, \bt) \bQ^{-1}_m \bQ_{m,o} \bD^{-1}_{o}(\bt, \bt)\\
\bR_{m \mid o}(\bt) &= \bD_m(\bt, \bt) (\bSigma_{m} - \bSigma_{m, o} \bSigma_{o}^{-1} \bSigma_{o,m}) \bD_m(\bt, \bt) = \bD_m(\bt, \bt) \bQ_{m}^{-1} \bD_m(\bt, \bt),
\end{align*}
where $\bSigma_{m, o}$ subsets $\bSigma$ to its $m$ rows and $o$ columns, and similarly for $\bQ$. 
\end{customprop}
\textit{Proof. } %\begin{proof}
All the results are a consequence of the properties of multivariate Gaussians and the previous proposition.
%\end{proof}

\section{Section 1: \textit{Introduction}}

\subsection{Details on Figure 2}
The figure displays a 4-outcome multivariate GP with \modelname\ cross-covariance with the following settings:
\small \begin{align*} 
    \bSigma = \begin{bmatrix}1 & -0.9 & 0.7 & 0.8 \\
-0.9 & 1 & -0.5 & -0.7 \\
0.7 & -0.5 & 1 & 0.8 \\
0.8 & -0.7 & 0.8 & 1\end{bmatrix} & \quad & \begin{matrix*}[l]
    \rho_1(\bl, \bl') = \calM_{\nu=1}(\bl, \bl'; \phi_1 = 15)\\
    \rho_2(\bl, \bl') = \exp\{ - 15 \cdot \|\bl-\bl' \| \} \cos\{ 14.9 \cdot \|\bl-\bl'\|\} \\ 
    \rho_3(\bl, \bl') = \calM_{\nu_3=1.5}( \tilde{\bl}, \tilde{\bl}'; \phi_3=15 ) \quad\text{where}\quad \tilde{\bl} = \bl + \bu(\bl)\\
    \rho_4(\bl, \bl') = |1+v(\bl)| |1+v(\bl')| \calM_{\nu_4=1.5}(\bl, \bl'; \phi_4=15),
\end{matrix*}
\end{align*} \normalsize
where $\bu(\cdot) = (u_1(\cdot), u_2(\cdot))^\top$ is a bivariate GP with independent Mat\'ern margins with parameters $\sigmasq = 0.03, \phi = 1, \nu = 1$, and $v(\cdot)$ is a Mat\'ern GP with parameters $\sigmasq=0.5, \phi=30, \nu=1.5$. We generate data at a regular grid of 40,000 locations after applying a Vecchia approximation of each univariate GP using $m=40$ neighbors.  

\subsection{Key shortcomings of \lmcs}
The LMC is a tractable model of multivariate spatial dependence which leads to structured covariance matrices and gives rise to a multitude avenues for flexible and scalable models. Here, we provide more details about the limitations of \lmcs\ which we have mentioned in the Introduction section of the main article. \lmcs\ are inadequate in modeling outcomes with different smoothness because the smoothness of outcome $j$ is always equal to the smoothness of the roughest of the $k$ correlation functions \citep{genton_ccov}. The difficulty in interpretations and prior elicitation is exemplified by, e.g., a situation in which one builds a \lmc\ on $k$ exponential correlation functions: in this case, the spatial range of outcome $j$ is a non-linear function of all the $k$ spatial range parameters (\citealt{schmidtgelfand}, Section 3.4). For analogous reasons, scalable GP approximations using \lmcs\ are not easy to build: the lack of a direct link between the $k$ constituent spatial processes and each outcome means that it is not straightforward to implement, e.g., outcome-specific neighbor sets for building Vecchia approximations or NNGPs. 

\lmcs\ lack flexibility in modeling independent measurement errors because nugget effects in the constituent processes are linearly combined into correlated noise. One must then use \lmcs\ as priors for latent effects, leading to slower computations even when using scalable alternatives to GPs \citep{nngp_algos}. 
In a \lmc, nugget effects are linearly combined into correlated measurement error for the outcomes. In fact, the \lmc\ can be expressed as $\by(\bl) = \bA (\bv(\bl) + \bnu(\bl))$ where $\bv(\bl)$ is the $k$-variate spatial GP and $\bnu(\bl) \iidsim N(\bzero, \bD_{\nu})$. Then, $\by(\bl) = \bA \bv(\bl) + \bA \bnu(\bl)$ implies that the measurement error has covariance $\bA\bD_{\nu}\bA^\top$, resulting in considerable inflexibility due to its dependence on $\bA$. 
We mention two possible ways to address this issue: first, we can build a response model under certain identifiability constraints on the loadings matrix $\bA$ \citep{Ren2013} by letting the last of the $k$ factors be standard normal white noise. The data are directly modeled as a multivariate GP with \lmc\ cross-covariance: $\by(\cdot) \sim GP(\bx(\cdot)^\top \bbeta, \Cov_{\lmc}(\cdot, \cdot))$. Then, the square of the $(i,k)$ element of $\bA$ can be interpreted as the (independent) nugget effect for the $i$th outcome. This choice limits the number of spatial factors to $q-1$. If, instead, one wants to represent $q$ spatial outcomes as the linear combination of $q$ spatial factors, plus measurement error, then one must resort to an additional set of latent variables, leading to a multivariate regression model with latent effects. In this case, $\by(\cdot) = \bx(\cdot)^\top \bbeta + \bw(\cdot) + \beps(\cdot)$, where $\bw(\cdot) \sim GP(\bzero, \Cov_{\lmc}(\cdot, \cdot))$ and $\beps(\cdot)$ is independent Gaussian white noise with covariance $D$. The distinction between response and latent models is important when building scalable GPs; for instance, a NNGP approximation on the response model is more restrictive than the corresponding approximation on the spatial latent effects. On the other hand, computations with latent models tend to be more inefficient \citep{nngp_algos}. IOX introduces nugget effects more directly in response models and thus offers new ways for modeling multivariate spatially-correlated noisy data. Another scenario where a model incorporating nugget effects is preferable arises with multivariate spatial count data. In a log-Gaussian multivariate spatial GLM, accounting for overdispersion typically requires introducing an additional set of random effects, which complicates computation. By contrast, models based on IOX can accommodate overdispersion more directly: the covariance structure on the latent Gaussian field can include a nugget effect without the need for extra random effects.

A subtle feature of \lmcs\ is that they link the number of constituent spatial processes $k$ to the rank of $\bA\bA^\top = \Cov(\bl, \bl)$. The $q$ observed outcomes may each be characterized by specific spatial features (e.g., range,   smoothness, anisotropy, nonstationarity). However, there may also exist a low rank community structure describing cross-outcome dependence. A possible real-world example is a natural habitat where multiple species coexist and interact within the same spatial niches: although they may form interaction communities, each species may also spread spatially in unique ways. A related issue is that because \lmcs\ define both $C_{ii}(\cdot, \cdot)$ and $C_{ij}(\cdot, \cdot)$, $i\neq j$, based on the same set of $k$ functions, a reduction in $k$ reduces flexibility and expressiveness of both cross- and marginal covariances. 

The infill asymptotic properties of \lmcs\ remain poorly understood. \lmcs\ likely inherit inconsistencies of univariate Matérn models (\citealt{zhang2007environmetrics}, Section 6; see also \citealt{velandia17}). In the univariate case, however, the functional form of the microergodic parameter is known and can be estimated consistently \citep{stein90, zhang04}. This provides reassurance: even when individual covariance parameters are difficult or inefficient to estimate, the microergodic parameter remains identifiable. For \lmcs, no such functional form is currently available. As a result, in applied settings, one may encounter high variance in parameter estimates or inefficient posterior sampling, without a clear way to determine whether these issues stem from the sampling algorithm or from a fundamental lack of asymptotic identifiability of the model parameters.

IOX likely inherits infill behavior from the marginal correlation functions used to define it; an in-depth study of the asymptotic properties of IOX is beyond the scope of this article. The infill asymptotic scenario occurs in practical settings when the Mat\'ern spatial decay is too small relative to the dimension of the spatial domain. By setting $\phi_j$ large enough in simulations, we ensured its practical estimability. Specifically, we show that IOX performs well in estimating marginal parameters even in the misspecified setting where we generate data from a multivariate Mat\'ern. 

The general inability of a LMC to directly estimate outcome-specific parameters does not depend on whether or not those parameters are identifiable in an infill asymptotic setting. For example, consider the case of a multivariate model with squared exponential marginal correlations (i.e., Mat\'ern with $\nu\to\infty$). This model is asymptotically identifiable in an infill setting (we did not consider this model in simulations because it is typically considered too smooth for geostatistical applications), yet the LMC remains unable to estimate outcome-specific parameters.

\subsection{Comparison with multivariate Mat\'ern} \label{sec:multimatern}
The multivariate Mat\'ern model \citep{gneiting2010} defines each $C_{ij}(\cdot, \cdot)$ to be a Mat\'ern function parametrized via $\sigma_{ij}, \phi_{ij}, \nu_{ij}$, with the possible addition of nugget effects. The fact that $C_{ij}(\cdot, \cdot)$ are valid univariate covariance functions for each $i,j$ does not guarantee that the resulting cross-covariance matrix function will also be valid. The validity of the latter depends on the values of the entire set of parameters $\sigma_{ij}, \phi_{ij}, \nu_{ij}$ for $i,j=1, \dots, q$ in non-trivial ways. 
Finding the least restrictive constraints on the parameters is an active research area, see, e.g., \citep{apanasovich2012, Emery2022}. An additional problem is that if each $C_{ij}(\cdot, \cdot)$ has $m$ free parameters, then the total number of parameters to be estimated from the data is $mq(q+1)/2$ and the constraints on them become increasingly cumbersome to satisfy. 
The parsimonious Mat\'ern model of \cite{gneiting2010} resolves this issue by proposing to restrict the parameter space to guarantee the validity of the resulting cross-covariance matrix function. One defines the cross-covariances as
\begin{equation}\label{eq:parsim_matern}
\begin{aligned}
C_{ij}(\bl, \bl') &= \sigma_{ij} S(\nu_i, \nu_j, d) M_{\frac{1}{2}(\nu_i + \nu_j)}(\bl, \bl'; \phi),\\
\text{where}\quad S(\nu_i, \nu_j, d) &= \frac{\Gamma(\nu_i + d/2)^{\frac{1}{2}}}{\Gamma(\nu_i)^{\frac{1}{2}}}\frac{\Gamma(\nu_j + d/2)^{\frac{1}{2}}}{\Gamma(\nu_j)^{\frac{1}{2}}} \frac{\Gamma(\frac{1}{2}(\nu_i + \nu_j))}{\Gamma(\frac{1}{2}(\nu_i + \nu_j) + d/2)},
\end{aligned}
\end{equation}
and $M_{\nu}(\bl, \bl'; \phi)$ is the Mat\'ern correlation function with smoothness $\nu$ and range $1/\phi$ and $\bSigma = (\sigma_{ij})_{ij}$ is a covariance matrix. 
In other words, the parsimonious Mat\'ern forces the cross-covariances to have smoothness equal to the average smoothness of the variables they refer to, and all covariances (marginal and cross) to share a single range parameter. Furthermore, the scaling factor $0 < S(\nu_i, \nu_j, d) \leq 1$ operates to ensure validity of the cross-covariance matrix function. Like in \modelname, $\sigma_{ij}$ is the upper bound of $C_{ij}(\bl,\bl)$ which is achieved if $\nu_i = \nu_j$. Also similar to \modelname\ is the fact that the smoothness of $C_{ij}$ cannot be chosen freely. However, while \modelname\ allows different ranges, the multivariate Mat\'ern requires additional constraints to be imposed on other parameters to achieve the same result. The validity of \modelname\ is derived directly from the validity of the component correlation functions, whereas one needs to perform further checks in multivariate Mat\'ern models. This makes \modelname\ advantageous when implementing multivariate extensions of non-stationary correlations.

Finally, \modelname\ is advantageous in leading to a structured sample covariance matrix which we take advantage of in developing the models appearing in the main article. The multivariate Mat\'ern does not lead to similar structure.

\section{Section 2: \textit{Inside-out cross-covariance}}

\subsection{Constructive interpretation of IOX}\label{sec:prior_sampling}
We offer a constructive interpretation of \modelname. 
Let $\calT$ be sample locations. We consider two scenarios. The first is $\calS = \calT$, the second is $\calT \subset \calS^c$.
For the scenario $\calS = \calT$, from $\rho_j(\cdot, \cdot)$, compute $\bL_j$ for $j=1, \dots, q$. For all $\bl_i \in \calT$ let $\bv(\bl_i) \iidsim N(\bzero, \bSigma)$. Denote $\bV = [\bv(\bl_1)\ \cdots\ \bv(\bl_n)]^\top$---therefore, $\bV$ is a matrix-normal random matrix $\bV \sim MN(\bzero, \bI_n, \bSigma)$. Then, $\bw = \{\oplus \bL_j \} \text{vec}(\bV)$ is $\bw \sim N(\bzero, \{\oplus \bL_j \} (\bSigma \otimes \bI_n) \{ \oplus \bL_j^\top \})$ as desired.  
Figure 1 in the main article shows an example in which we generate $3$ spatially correlated outcomes with different smoothness via \modelnames. If only $\bL_i^{-1}$ is available for $i=1, \dots, q$, then from $\bv_j$ (the $j$th column of $\bV$), one obtains $\bw_j = \text{solve}(\bL_j^{-1}, \bv_j)$ without needing to compute $\bL_i$ directly.

\begin{figure}%[!ht]
    \centering
    \includegraphics[width=0.95\textwidth]{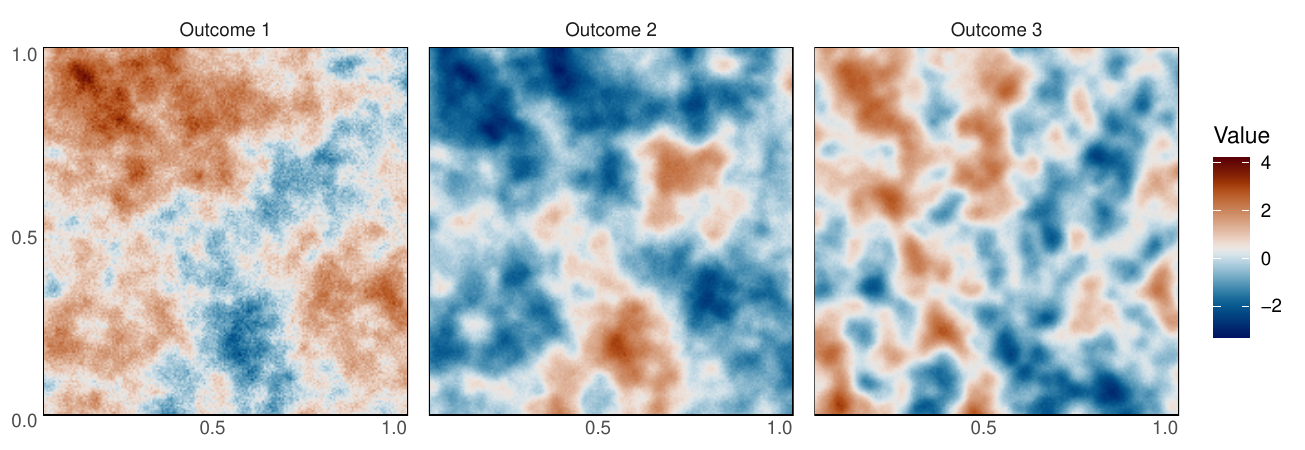}
    \caption{\footnotesize Three spatially correlated outcomes generated via GP-\modelnames\ at $n=$ 40,000 gridded locations from a reference set of size $n_S=2,500$. We let $\sigma_{ii}=1$ for $i\in\{1,2,3\}$ and $\sigma_{12}=-0.9, \sigma_{13}=0.7, \sigma_{23}=-0.5$ and choose $\rho_i(\cdot, \cdot)$ as Mat\'ern (Vecchia-approximated with $m=40$ neighbors) with range $1/5, 1/15, 1/30$ and smoothness $0.5, 1.2, 1.9$, respectively. \normalsize }
    \label{fig:prior_sample_U}
\end{figure}

For the second scenario, let $\calT = \{ \bt_1, \dots, \bt_N \}$ be sample locations and $\calS = \{ \bl_1, \dots, \bl_n \}$. We assume $\calT \subset \calS^c$. From $\rho_j(\cdot, \cdot)$, compute $\bL_j = \text{chol}(\rho_j(\calS))$ for $j=1, \dots, q$. For all $\bl_i \in \calS$ let $\bv(\bl_i) \iidsim N(\bzero, \bSigma)$. Denote $\bV = [\bv(\bl_1)\ \cdots\ \bv(\bl_n)]^\top$---therefore, $\bV$ is a matrix-normal random matrix $\bV \sim MN(\bzero, \bI_n, \bSigma)$. Then, $\bw_{\calS} = \{\oplus \bL_j \} \text{vec}(\bV)$ is $\bw \sim N(\bzero, \{\oplus \bL_j \} (\bSigma \otimes \bI_n) \{ \oplus \bL_j^\top \})$ as desired. Finally, for each $\bt\in \calT$ sample from $N(\bw(\bt); \bm_w(\bt), \bR(\bt))$, where $\bm_w(\bt) = \bH(\bt) \bw_{\calS}$, $\bH(\bt) = \{ \oplus \bh_j(\bt)\}$ and $\bR(\bt)= \bD(\bt, \bt) \bSigma \bD(\bt, \bt)$. 
We take $\calS$ as a 100$\times$100 regular grid in $\calD = [0,1]^2$ and sample a noise-free trivariate GP-\modelname\ at a regular grid of dimension 200$\times$200 using the same parameter settings as in Figure 1 in the main article. Figure \ref{fig:prior_sample_U} shows the resulting maps.

We outline an equivalent method for prior sampling. Call $\calG_S$ the DAG at $\calS$ and let the DAG for $\calS \cup \calT$, $\calG$, be such that each node from $\calT$ has $\calS$ as its set of parents. For Vecchia-GP models, the set of parents for $\bt \in \calT$ is restricted to the nearest neighbors within $\calS$. Based on the DAG, compute $\bL_j^{-1}$ from $j=1,\dots,q$. The elements of $\bL_j^{-1}$ can be directly computed in the following way. 
Let $\bh_l = \rho_i(\bl_l, \bl_{[l]}) \rho_i(\bl_{[l]})^{-1}$, $\br_l = \rho_i(\bl_l, \bl_l) - \bh_l \rho_i(\bl_{[l]}, \bl_l)$, where $[l]$ denotes the set of parents of node $l$ in the graph $\calG$. Then, the diagonal elements of $\bL_j^{-1}$ are $1/\sqrt{\br_{\bl}}$ for $\bl\in \calS\cup\calT$. The non-zero elements of row $a$ are $-\bh_a/\sqrt{\br_a}$. Moving on, for all $\bl \in \calS\cup \calT$, let  $\bv(\bl) \iidsim N(\bzero, \bSigma)$. Denote $\bV = [\bv(\bl_1)\ \cdots\ \bv(\bl_n), \bv(\bt_1), \dots, \bv(\bt_N)]^\top$---therefore, $\bV$ is a matrix-normal random matrix $\bV \sim MN(\bzero, \bI_{n+N}, \bSigma)$. Then, let $\bw = \{\oplus \bL_j \} \text{vec}(\bV)$ and build the $(n+N) \times q$ matrix $\bW$ from $\bw$ in column-major ordering. The last $N$ rows of $\bW$ have the desired marginal distribution.

\begin{figure}[H]
    \centering
\begin{tabular}{ccccc}
  \modelname: $\Cov(\calS)$ & = & $\{ \oplus \bL_j \}$ & $(\bSigma \otimes \bI_n)$ & $\{ \oplus \bL_j^\top \}$ \\
  \includegraphics[width=0.2\textwidth]{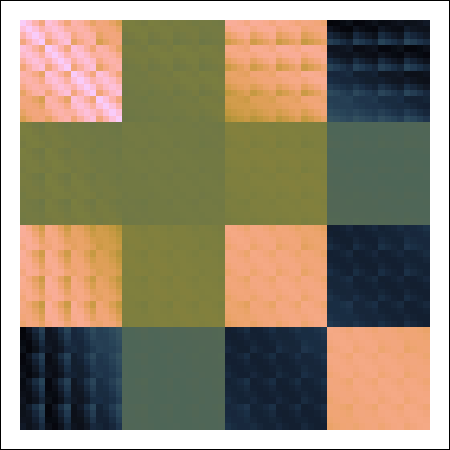} & & \includegraphics[width=0.2\textwidth]{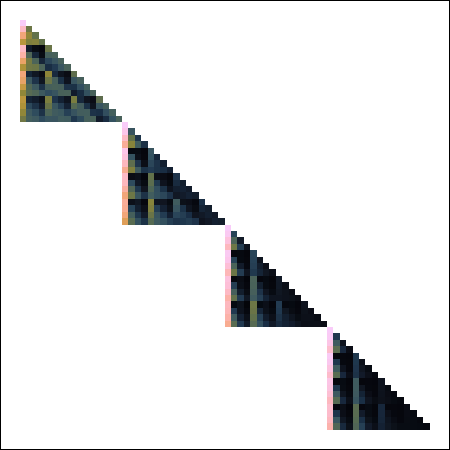} & \includegraphics[width=0.2\textwidth]{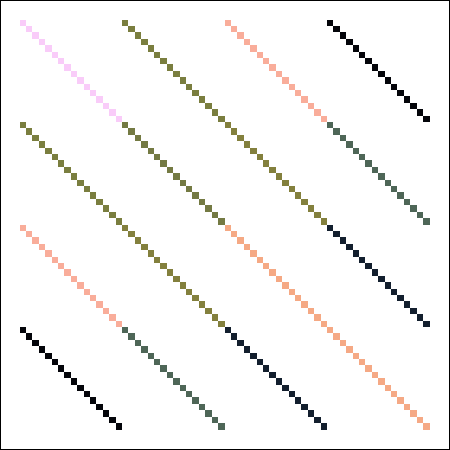} & \includegraphics[width=0.2\textwidth]{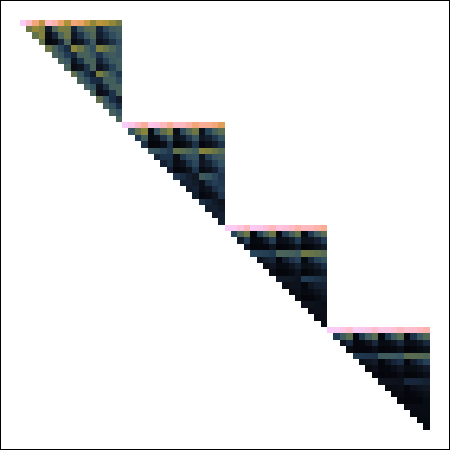} \\
  \lmc: $\Cov(\calS)$ & = & $(\bA \otimes \bI_n)$ & $\{ \oplus \rho_j(\calS) \}$ & $(\bA^\top \otimes \bI_n)$ \\
  \includegraphics[width=0.2\textwidth]{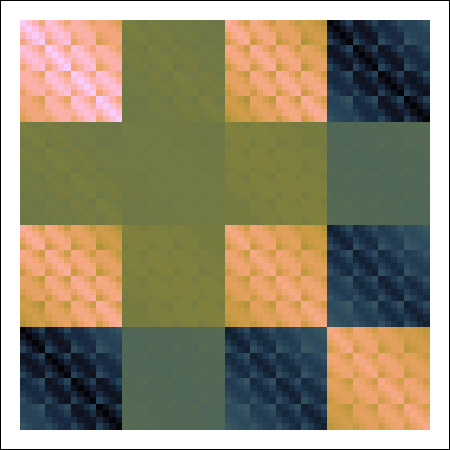} & & \includegraphics[width=0.2\textwidth]{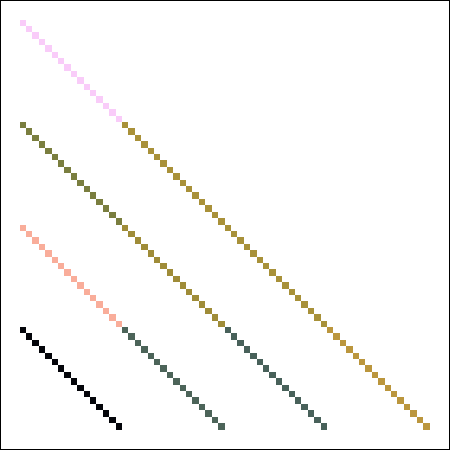} & \includegraphics[width=0.2\textwidth]{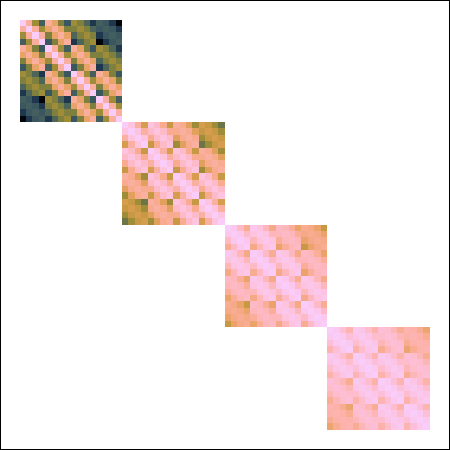} & \includegraphics[width=0.2\textwidth]{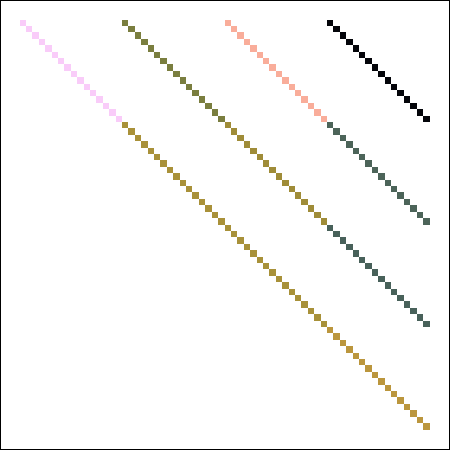}
\end{tabular}
\caption{Visualization of \modelname\ and \lmc\ sample covariance matrices and the sparsity patterns of their components.}  
    \label{fig:iox_lmc_matrices}  
\end{figure}

We can compare how IOX is defined constructively relative to a LMC. 
In \lmcs, one introduces spatial dependence first, and then multivariate dependence, whereas the order of these operations is inverted in \modelname. In Table \ref{tab:prior_sample_compare} we outline prior sampling for \modelname\ and \lmcs\ in a way that clarifies this point visually.  Figure \ref{fig:iox_lmc_matrices} visualizes the resulting sample covariances. 

\begin{table}[H]
\centering
% Using \arraystretch to prevent fractions/matrices from hitting the lines
\renewcommand{\arraystretch}{1.5} 
\begin{tabular}{|c|l|l|l|}
\hline
{\scriptsize Step} & \textbf{\modelname} & \textbf{Separable model} & \textbf{\lmc} \\ \hline
{\scriptsize 1} & \multicolumn{3}{c|}{$\calS$ sample locations}  \\ \hline
{\scriptsize 2} & \multicolumn{3}{c|}{$\bU \sim MN(\bzero, \bI_n, \bI_q)$} \\ \hline
{\scriptsize 3} & \multicolumn{3}{c|}{$\textcolor{Emerald}{\bA \bA^T = \bSigma}$} \\ \hline
{\scriptsize 4} & 
$\textcolor{WildStrawberry}{\bR_j} = \rho_j(\calS)$ for $j=1,\dots,q$ & 
$\textcolor{WildStrawberry}{\bR} = \rho(\calS)$ & 
$\textcolor{WildStrawberry}{\bR_j} = \rho_j(\calS)$ for $j=1,\dots,k$ \\ \hline
{\scriptsize 5} & 
$\textcolor{WildStrawberry}{\bL_j} = \text{chol}(\textcolor{WildStrawberry}{\bR_j})$ & 
$\textcolor{WildStrawberry}{\bL} = \text{chol}(\textcolor{WildStrawberry}{\bR})$ & 
$\textcolor{WildStrawberry}{\bL_j} = \text{chol}(\textcolor{WildStrawberry}{\bR_j})$ \\ \hline
{\scriptsize 6} & 
$\bv = \textcolor{Emerald}{(\bA \otimes \bI_n )} \text{vec}(\bU) $ & 
$\bW = \textcolor{WildStrawberry}{\bL} \bU \textcolor{Emerald}{\bA^\top}$ & 
$\bv = \textcolor{WildStrawberry}{ \{\oplus \bL_j \}} \text{vec}(\bU)$ \\ \hline
{\scriptsize 7} & 
$\bw = \textcolor{WildStrawberry}{ \{\oplus \bL_j \}} \bv$ & 
$\bw = \vecop(\bW)$ & 
$\bw = \textcolor{Emerald}{(\bA \otimes \bI_n )}  \bv$ \\ \hline
{\scriptsize 8} & 
$\bw \sim N(\bzero, \Cov)$ & 
$\bw \sim N(\bzero,  \Cov )$ & 
$\bw \sim N(\bzero,  \Cov ) $ \\
{\small $\Cov=$} & 
{ $\textcolor{WildStrawberry}{ \{\oplus \bL_j \}} \textcolor{Emerald}{(\bSigma \otimes \bI_n )} \textcolor{WildStrawberry}{ \{\oplus \bL_j^\top \}}$ } & 
{ $\textcolor{Emerald}{\bSigma} \otimes \textcolor{WildStrawberry}{\bR}$ }  & 
{ $\textcolor{Emerald}{(\bA \otimes \bI_n )}\textcolor{WildStrawberry}{ \{\oplus \bR_j \} } \textcolor{Emerald}{(\bA\otimes \bI_n )^\top}$ } \\ \hline
\end{tabular}
\caption{Comparison of prior sampling of $\bw(\cdot) \sim GP(\bzero, \Cov(\cdot, \cdot))$ at $\calS$ using IOX and coregionalization models. We color code the spatial components of the sample cross-covariance matrix in \textcolor{WildStrawberry}{\textbf{pink}}, the multivariate components in \textcolor{Emerald}{\textbf{emerald}}.}\label{tab:prior_sample_compare}
\end{table}

\subsection{Additional considerations on \texorpdfstring{$C_{ij}(\cdot, \cdot), i\neq j$}{Cij}}

Our \modelname\ cross-covariance matrix function is built on the idea that, in practice, we aim to primarily infer about $C_{ii}(\cdot, \cdot)$ while \textit{accounting} for cross-variable spatial dependence, without the latter being the primary inferential target. In \modelname\, the cross-covariance $C_{ij}(\cdot, \cdot)$ is not as directly interpretable as $C_{ii}(\cdot, \cdot)$. Here, we provide details on (1) how to interpret $C_{ij}(\cdot, \cdot)$ and (2) how to model $C_{ij}(\cdot, \cdot)$ more flexibly when explicit parametric models are needed.

For a given $\calS$ and a set of $\rho_i(\cdot, \cdot)$ and associated parameters, let $\calT$ be a set of $m$ test locations, and $\bh = (h_x, h_y)$. Then, compute
\begin{equation}\label{eq:cij_average}
\begin{aligned}
\tilde{C}_{ij}(\bh) &= \frac{1}{mn_a} \sum_{a = 1}^{n_a} \sum_{\bl \in \calT} C_{ij}(\bl, \bl + \bh_{\alpha} ),\qquad \text{where}\quad \bh_{\alpha} = \begin{bmatrix} h_x \cos(2\pi a/n_a) \\ h_y \sin(2\pi a/n_a)\end{bmatrix}.
\end{aligned}
\end{equation}

\noindent In particular, we may be interested in computing $\tilde{C}_{ij}(\bzero)$, which reduces \eqref{eq:cij_average} to
\begin{equation}\label{eq:cij_average_zero}
\begin{aligned}
\tilde{C}_{ij}(\bzero) &= \frac{1}{mn_a} \sum_{a = 1}^{n_a} \sum_{\bl \in \calT} C_{ij}(\bl, \bl  ).
\end{aligned}
\end{equation}

\modelname\ includes a scaling factor which as we showed in the main article results in $C_{ij}(\bl,\bl') < \sigma_{ij}$ for any $i,j$ such that $\rho_i(\cdot, \cdot) \neq \rho_j(\cdot, \cdot)$. This is unsurprising; suppose $\by_1$ and $\by_2$ are two noise-free, jointly \modelname\ variables at $\calS$. The limit case $\sigma_{12}\to 1$ implies near-collinearity of the spatially uncorrelated matrix-normal matrix $\bV$ used to generate $\bY$ (see Section \ref{sec:prior_sampling}), not of $\bY$ itself.

Our method restricts the cross-covariances $C_{ij}(\cdot, \cdot)$ through a product of the marginal Cholesky factors. Consequently, $C_{ij}(\cdot, \cdot)$ is parametrized directly via $\sigma_{ij}$ but implicitly on $\rho_i(\cdot, \cdot)$ and $\rho_j(\cdot, \cdot)$ and their parameters. For example, if $\rho_j(\cdot, \cdot)$ is Mat\'ern for all $j$, $C_{ij}(\cdot, \cdot)$ in \modelname\ is not Mat\'ern and is also not explicitly parametrized via range and smoothness parameters---it depends indirectly on $1/\phi_i, 1/\phi_j, \nu_i,\nu_j$. In a bivariate Mat\'ern, on the other hand, $C_{ij}(\cdot, \cdot)$ is Mat\'ern with smoothness $\nu_{ij}$ and range $1/\phi_{ij}$. We note that these parameters must be chosen carefully to ensure validity of the resulting cross-covariance matrix function \citep{Emery2022}, and interpreting them is also not straightforward \citep{kleiber2017coherence}. Refer to the discussion at Section \ref{sec:multimatern} comparing \modelname\ to multivariate Mat\'erns.

In situations in which it is desirable to model $C_{ij}(\cdot, \cdot)$ flexibly for some $i,j$, perhaps as a Mat\'ern, we can straightforwardly generalize \modelname. Without loss of generality, let $\xi = \{1,2\}$ be the pair of outcomes we wish to model more flexibly and $\brho_{\xi}(\cdot, \cdot)$ the marginal cross-correlation matrix function for the $\xi$ variables. $\brho_{\xi}(\cdot, \cdot)$ must be a valid cross-correlation matrix function with $\rho_{11}(\cdot, \cdot)$, $\rho_{12}(\cdot, \cdot)$ and $\rho_{22}(\cdot, \cdot)$ margins. Let $\bL_{\xi}$ the $2n \times 2n$ lower Cholesky factor of $\brho_{\xi}(\calS)$, and let $\bh_{\xi}(\bl) = \brho_{\xi}(\bl, \calS) \brho_{\xi}(\calS)^{-1}$, $\br_{\xi}(\bl) = \brho_{\xi}(\bl, \bl) - \bh_{\xi}(\bl) \brho_{\xi}(\calS, \bl)$. 
Finally, let 
\[ \bD_{\xi}(\bl, \bl') = \mathbb{1}_{\{\bl=\bl'\} }
\begin{bmatrix}
     \texttt{chol}(\br_{\xi}(\bl)) & & & \\
     & \sqrt{r_3(\bl)} & & \\
     & & \ddots & \\
     & & & \sqrt{r_q(\bl)}
\end{bmatrix}.
\]
Then, we can define the \modelname\ cross-covariance matrix function as 
\begin{align*} &\Cov(\bl, \bl') = \left(\bh_{\xi}(\bl) \bL_{\xi} \oplus \{ \oplus_{j>2} \bh_j(\bl) \bL_j \} \right) (\bSigma \otimes \bI_n)\left( \bL_{\xi}^\top \bh_{\xi}(\bl)^\top  \oplus \{ \oplus_{j>2} \bL_j^\top \bh_j(\bl)^\top  \} \right) +\\
     &\qquad \qquad + \bD_{\xi}(\bl, \bl')\bSigma \bD^\top_{\xi}(\bl, \bl')\\ 
&= \begin{bmatrix} \bh_{\xi}(\bl) \bL_{\xi} & \\ 
& \text{ \scriptsize \( \begin{matrix} \bh_3(\bl) \bL_3 & & \\
 & \ddots & \\
 & & \bh_q(\bl) \bL_q
\end{matrix} \) } \end{bmatrix} (\bSigma \otimes \bI_n) \begin{bmatrix} \bL_{\xi}^\top \bh_{\xi}(\bl)^\top & \\ 
     & \text{ \scriptsize \( \begin{matrix} \bL_3^\top \bh_3(\bl)^\top & & \\
      & \ddots & \\
      & & \bL_q^\top \bh_q(\bl)^\top 
     \end{matrix} \) } \end{bmatrix} + \\
& \qquad \qquad + \bD_{\xi}(\bl, \bl')\bSigma \bD^\top_{\xi}(\bl, \bl'),
\end{align*} 
where we essentially treat $\xi$ as a grouped variable. The computationally convenient structure of \modelname\ is lost at $\xi$ but retained for all other variables. Following Proposition 2.1, the marginal cross-covariance of $\xi$ will be a function of $\brho_{\xi}(\cdot, \cdot)$ as desired. 
We can extend this further to allow explicit models of any groups of variables, at additional computational costs. 

\subsection{Choice of \texorpdfstring{$\calS$}{S}}
\begin{figure}%[!ht]
    \centering
    \includegraphics[width=0.95\textwidth]{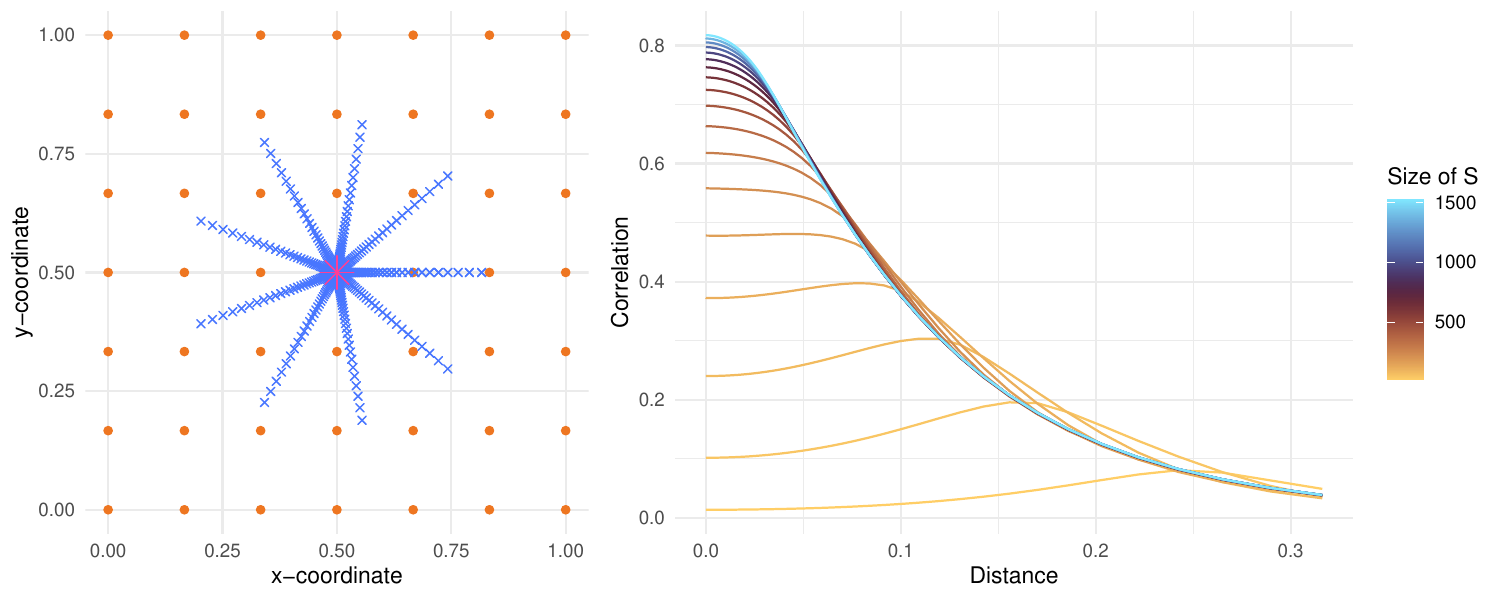}
    \caption{\footnotesize Left: orange points correspond to $\calS$ locations. Here, $\calS$ is computed as a regular grid on $7^2$ coordinates, minus the center point with coordinate $(1/2, 1/2)$. We compute $C_{ij}(\bl, \bl')$ where $\bl = (1/2, 1/2)$ and $\bl'$ is each blue location with ``\texttt{x}'' shape. We choose the marginal correlations as Mat\'ern with $\phi_i=10$, $\phi_j=20$, and smoothness $\nu_i=\nu_j=1$. We average $C_{ij}(\bl, \bl')$ across pairs of locations at the same distance to ultimately obtain the plot on the right, where each line corresponds to a different $\calS$. \normalsize }
    \label{fig:cij_sizeofS}
\end{figure}

Our recommendation in the main article is to take $\calS$ as the set of observed locations. This recommendation follows from the consideration that in doing so, IOX leads to straightforward interpretations of marginal covariance as in Proposition 2.1. The set $\calS$ is also where IOX allows cross-spatial-dependence, as in Proposition 3.4. However, one can in principle choose $\calS$ as any set of locations. In Figure \ref{fig:cij_sizeofS} we investigate the effect the size of $\calS$ on the resulting $C_{ij}(\cdot, \cdot)$, where $i \neq j$. We consider a scenario where we evaluate $C_{ij}(\cdot, \cdot)$ at pairs locations $\bl, \bl'$ neither of which are in $\calS$. This matches a situation in which, say, the data are observed irregularly in the domain, and we choose $\calS$ as a grid, akin to the typical knot setup of a ``predictive process'' \citep{gp_predictive_process}. In this setting, Figure \ref{fig:cij_sizeofS} shows that choosing $\calS$ too small leads to a shrinkage the cross-correlation function at all distances. 

Therefore, although a small $\calS$ facilitates computations, it also leads to a reduction in the allowed cross-spatial dependence. 

\subsection{Covariance-based definition of \modelname\ }
We have defined \modelname\ on the basis of $q$ univariate correlation functions in the main article, but we can equivalently define \modelname\ on a given list of $q$ covariance functions. 
Let $\bSigma = (\sigma_{ij})_{i,j=1}^q$ be a positive semidefinite matrix, $K_i(\cdot,\cdot) = s_{i}^2 \rho_i(\cdot, \cdot)$ for $i=1,\dots,q$, where $\rho_i(\cdot, \cdot)$ is a univariate correlation function and let $\calS$ be the reference set. Then, we define \modelname\ as
\[ C_{ij}(\bl, \bl') = \sigma_{ij} s_i s_j \left[ \bh_i(\bl) \bL_i \bL_j^\top \bh_j(\bl') + \varepsilon(\bl, \bl') \right], \]
where $s_j>0$ for all $j$, $\bh_i(\bl) = \rho_i(\bl, \calS) \rho_i(\calS)^{-1}$, $\bL_i$ is the lower Cholesky factor of $\rho_i(\calS)$, $\varepsilon(\bl, \bl') = \mathbb{1}_{\{ \bl=\bl'\} }\sqrt{r_i(\bl) r_j(\bl)}$, and $r_i(\bl)=\rho_i(\bl,\bl)-\bh_i(\bl)\rho_i(\calS,\bl)$.
The validity of the resulting \modelname\ model is verified straightforwardly following the main article. The only change here is that we use $\bSigma^* = \bD_s \bSigma \bD_s$, where $\bD_s = \text{diag}\{ s_i \}_{i=1}^q$, rather than $\bSigma$. Clearly, $\bSigma^*$ is positive semidefinite if $\bSigma$ is. All other components of the model remain the same. In the above model, $\bD_s$ and $\bSigma$ are not identifiable, but $\bSigma^*$ is. This suggests strategies for parameter expansion, which may improve the efficiency of posterior sampling algorithms; see, e.g., \cite{papasroberts07, grips}.

\section{Section 3: \textit{IOX Gaussian process models}}

\subsection{Latent factor models for the large \textit{q} case} \label{sec:sigmafactor}
\modelname\ can operate dimension reduction by assuming $\bSigma$ is low rank, i.e., $\bSigma = \bA \bA^\top$ where $\bA$ is a $q\times k$ matrix of full column rank. Changing $k$ has no impact on the marginal properties of \modelname\ because Proposition 2.1 does not depend on $k$. 
Let $\tilde{\by} = \{ \oplus \bL_j^{-1} \} (\by - \bX_q\bbeta)$ and $\tilde{\beps} = \{ \oplus \bL_j^{-1} \}\beps$. We can rewrite the latent model as
\begin{equation}\label{eq:factorq}
\begin{aligned}
\tilde{\by} &= \bv + \tilde{\beps},\qquad
\bv \sim N(\bzero, \bSigma \otimes \bI_n), \qquad
\tilde{\beps} \sim N(\bzero, \bB),
\end{aligned}
\end{equation}
where $\bB$ is a block-diagonal matrix whose $j$th block is $\bL_i^{-1}\bL_i^{-\top}$. Then, letting $\bu(\bl) \iidsim N(\bzero, \bI_{k})$ we have $\bv(\bl) = \bA\bu(\bl)$. In matrix form, $\tilde{\bY} = \bU \bA^\top + \tilde{\bE}$, where $\bU\sim MN(\bzero, \bI_n, \bI_{k})$ and $\tilde{\bE} \sim MN(\bzero, \bB, \bI_q)$. 
A low-rank $\bSigma$ leads to computational savings in sampling $\bw$. We obtain 
\begin{align*}
    \bM_{w\mid y}^{-1} &= \{ \oplus \bL_i^{-\top} \}(\bSigma^{-1} \otimes \bI_n) \{ \oplus \bL_i^{-1} \} + (\bDelta^{-1} \otimes \bI_n) \\
    %&=\{ \oplus \bL_i^{-\top} \}(\bA^{+\top} \bA^{+} \otimes \bI_n) \{ \oplus \bL_i^{-1} \} + (\bDelta^{-1} \otimes \bI_n)\\
    &=\{ \oplus \bL_i^{-\top} \}(\bA^{+\top} \otimes \bI_n)(\bA^{+} \otimes \bI_n) \{ \oplus \bL_i^{-1} \} + (\bDelta^{-1} \otimes \bI_n)\\
    &= \bK \bK^\top + (\bDelta^{-1} \otimes \bI_n)
\end{align*} 
where $\bK = \{ \oplus \bL_i^{-\top} \}(\bA^{+\top} \otimes \bI_n)$ is a $nq \times nk$ matrix, and $\bA^{+}$ is the $k\times q$ Moore-Penrose pseudoinverse of $\bA$. Although the prior on $\bw$ is singular if $k< q$, the posterior is well-defined, since $\bM_{w\mid y}$ is full rank. We can then take advantage of the Woodbury matrix identity in sampling $\bw$ a posteriori to make the cost for sampling $\bw$ depend on $k$ rather than $q$. Finally, to update $\bA$ we first deterministically compute $\bu = \bA^{+} \{\oplus \bL_i^{-1}\} \bw$, then, since $\tilde{\by}_i = \bU \bA\row{i}^\top + \tilde{\beps}_i$, we update each row of $\bA$ in parallel. Under a Gaussian prior on $\bA\row{i}^\top$ with prior covariance $\bM_A$, the full conditional distribution is Gaussian with precision matrix $\bU^\top \bL_i^\top \bL_i\bU + \bM_{A}^{-1}$. Here, computing $\bL_i\bU$ involves solving the linear system $\bL_i^{-1}\balpha = \bU$, which is cheap because $\bL_i^{-1}$ is triangular.

Due to the dimension of $\bM_{w\mid y}$, single block updates of $\bw$ may be computationally prohibitive in large $n$ settings, even if we assume a sparse DAG, $k$ is relatively small, and we use sparse Cholesky \citep{cholmod} or preconditioned conjugate gradient methods \citep{nishimura2022prior}. Instead of block sampling, we can take advantage of an assumed sparse DAG GP  and sample from the full conditional distributions at DAG nodes. 

\section{Section 4: \textit{Computations with GP-\modelname\ models}}

\subsection{Posterior sampling: response model}
We outline the remainder of the Gibbs sampler for the response model outlined in the main article. For $\bbeta$, we have $p(\bbeta \mid \by, \btheta, \bSigma) = N(\bm_{\beta\mid y}, \bM_{\beta\mid y})$, where
\begin{align*}
     \bM^{-1}_{\beta\mid y} &= \bM_{\beta}^{-1} + \bX_q^\top \Cov^{-1} \bX_q  \quad\text{and}\quad \bM^{-1}_{\beta\mid y} \bm_{\beta \mid y} = \bV_{\beta}^{-1}\bm_{\beta} + \bX_q^\top \Cov^{-1} \by.
\end{align*}
For updating $\bSigma$, first compute $\bU = \bY - \bX \bB$, where $\bB$ is the $p \times q$ matrix of regression coefficients. Then, compute the $nq$-dimensional vector $\bv = \{ \oplus \bL_i^{-1} \} \vecop(\bU)$, where dependence on $\btheta$ is through $\bL_i$, $i=1,\dots,q$. Then build the $n\times q$ matrix $\bV$ such that $\vecop(\bV) = \bv$. Notice that $\bV$ is the centered and spatially ``whitened'' version of $\bY$, and according to our model, $\bV \sim MN(\bzero, \bI_n, \bSigma)$. Then, with the prior $p(\bSigma) \propto \det(\bSigma)^{-\frac{\nu_{\Sigma} + q + 1}{2}} \exp \{-\frac{\text{tr}(\bPsi \bSigma^{-1})}{2} \}$, i.e., $\bSigma \sim \calW^{-1}(\nu_{\Sigma}, \bPsi^{-1})$, the conditional posterior is $p(\bSigma \mid \by, \btheta, \bbeta) = \calW^{-1}(\nu_{\Sigma\mid y}, \bPsi^{-1}_{\Sigma \mid y})$, where $\nu_{\Sigma \mid y} = \nu_{\Sigma} + n$ and $\bPsi_{\Sigma \mid y} = \bPsi + \bV^\top \bV$. 

\subsection{Posterior sampling: latent model}
We outline the remainder of the Gibbs sampler for the latent model outlined in the main article. For $\bbeta$, we have $p(\bbeta \mid \by, \bw, \bDelta) = N(\bm_{\beta\mid y}, \bM_{\beta\mid y})$, where
\begin{align*}
     \bM^{-1}_{\beta\mid y} &= \bM_{\beta}^{-1} + \bX_q^\top (\bDelta^{-1} \otimes \bI_n) \bX_q \quad\text{and}\quad \bM^{-1}_{\beta\mid y} \bm_{\beta \mid y} = \bM_{\beta}^{-1}\bm_{\beta} + \bX_q^\top (\bDelta^{-1}\otimes \bI_n)(\by - \bw).
\end{align*}
For $\bSigma$, the Bayesian DAG implies $p(\bSigma \mid \by, \bbeta, \bw, \bDelta, \btheta) = p(\bSigma \mid \bw, \btheta)$. We again let $\bSigma \sim \calW^{-1}(\nu_{\Sigma}, \bPsi^{-1})$. Compute the $nq$-dimensional vector $\bv = \{ \oplus \bL_i^{-1} \} \vecop(\bW)$, where dependence on $\btheta$ is through $\bL_i$, $i=1,\dots,q$. Then build the $n\times q$ matrix $\bV$ such that $\vecop(\bV) = \bv$. Then, $p(\bSigma \mid \bw) = \calW^{-1}(\nu_{\Sigma\mid w}, \bPsi^{-1}_{\Sigma \mid y})$, where $\nu_{\Sigma \mid w} = \nu_{\Sigma} + n$ and $\bPsi^{-1}_{\Sigma \mid w} = \bPsi + \bV^\top \bV$.
Finally, we update $\bDelta = \text{diag}\{ \delta_{ii} \}_{i=1, \dots, q}$. For each $j$, we sample $\delta_{jj} \mid \by_j, \bbeta_j, \bw_j$ from an Inverse Gamma distribution with parameters $a_{d\mid y} = a_{d} + n/2$ and $b_{d\mid y} = b_{d} + \frac{1}{2}\tilde{\by}_j^\top\tilde{\by}$, where $\tilde{\by}_j = \by_j - \bX \bbeta_j - \bw_j$.

\subsection{Sequential single-site sampler for sparse latent models}
Suppose we build \modelname\ with a Vecchia-approximated $\tilde{\rho}_j(\cdot, \cdot)$ for each $i=1, \dots, q$. As a consequence, $\tilde{\rho}_j(\calS)^{-1} = \tilde{\bL}_j^{-1} \tilde{\bL}_j^{-1}$ where $\tilde{\bL}_j^{-1}$ is sparse. Moving forward, we will denote $\bL_i = \tilde{\bL}_i$ to reduce the burden of notation and similarly for $\tilde{\rho}_j(\cdot, \cdot)$. A consequence of our sparsity assumption is that we want to work with $\bL_j^{-1}$ and $\rho_j(\calS)^{-1}$, and \textit{not} with $\bL_j$ and $\rho_i(\calS)$, because the former pair are sparse, whereas the latter are high dimensional and dense even when the DAG is sparse.

Without loss of generality, we assume a single sparse DAG for each $j=1,\dots,q$, resulting in $\bL_j^{-1}$ having the same sparsity pattern for all $j$. Assume a nearest-neighbor DAG for simplicity, so the first $m$ nodes in the DAG have $0, 1, \dots, m-1$ parents, respectively, and all remaining ones have $m$ parents. In this DAG, there is one node for each location of $\calS$. At $\calS$, the latent model is
\begin{align*}
\bw &\sim N(\bzero, \Cov) \qquad \Cov = \{ \oplus \bL_i \} (\bSigma \otimes \bI_n) \{ \oplus \bL_i^\top \} \\
\by &= \bw + \beps \quad \beps \sim N(\bzero, \bDelta \otimes \bI_n).
\end{align*}
We have discussed block-updating $\bw$ in the main article. Here, we target sequential single-site updates. 
Let $[i]$ denote the set of parent nodes of $i$. For simplicity, consider a location $\bl_i$ corresponding to node $i$ with $m$ parents. 
In this section we let $\by_i=\by(\bl_i) - (\bI_q \otimes \bx^\top(\bl_i)) \bbeta$ and $\bw_i = \bw(\bl_i)$, $\beps_i = \bw(\bl_i)$ to simplify notation. The single-site hierarchical model is
\begin{align*}
\bw_i &= \bH_i \bw_{[i]} + \bolds{\eta}_i\quad \bolds{\eta}_i \sim N(\bzero, \bR_i) \\
\by_i &= \bw_i + \beps_i \quad \beps_i \sim N(\bzero, \bDelta)
\end{align*}
where $\bl_{[i]} = \{ \bl_j\in\calS: j\to i \}$, $\bw_{[i]}$ is a $qm \times 1$ vector storing the values of the latent effects at $\bl_{[i]}$, and $\bH_i = \Cov(i, [i]) \Cov([i])^{-1}$ and $\bR_i = \Cov(i, i) - \bH_i \Cov(i, [i])^\top$, where we let $\Cov(i, [i]) = \Cov(\bl_i, \bl_{[i]})$. 
Here, we refer to $\Cov$ as the sample cross-covariance matrix as well as the \modelname\ cross-covariance matrix function since $\Cov(i, [i])$ is the cross-covariance computed at $\bl_i$ against $\bl_{[i]}$ and, equivalently, the $q \times mq$ matrix reading the rows and columns of $\Cov$ corresponding to $i$ and $[i]$, respectively. 

Single-site updates generally proceed by sequentially updating $\bw_i$ conditional on $\bw_{i^c}$, i.e., the value of $\bw$ at $\calS \setminus \{ i\}$. Let the $nq \times nq$ precision matrix be $\bP = \Cov^{-1}$ and let $i^c = \calS \setminus \{ i \}$. The properties of block matrix inverses imply that we can define
\begin{equation}\label{eq:seq_full_blocks}
\begin{aligned}
   \bB_i &:= \Cov(i, i) - \Cov(i, i^c) \Cov(i^c)^{-1} \Cov(i, i^c)^\top = \bP(i, i)^{-1} \\
   \bb_i &:= \Cov(i, i^c) \Cov(i^c)^{-1} = -\bP(i, i)^{-1} \bP(i, i^c),
\end{aligned}
\end{equation}
where $\bP(i, i)$ is a $q\times q$ and $\bP(i, i^c)$ a $q \times (n-1)q$ submatrix of $\bP$, which is sparse. 
Then, let $\bw_{i^c}$ be the vector of $\bw$ at $\calS\setminus \{ i \}$. The full conditional update for $\bw_i$ is $p(\bw_i \mid \bw_{i^c}, \by_i) = N(\bg_i, \bG_i)$, where, after seeing that $\bB_i^{-1} \bb_i = -\bP(i, i^c)$, we have
\begin{equation}\label{eq:fullcond_nomb}
\begin{aligned}
    \bG_i^{-1} = \bP(i, i) + \bDelta^{-1}, \quad \bG_i^{-1}\bg_i = - \bP(i, i^c) \bw_{i^c} + \bDelta^{-1}\by_i.
\end{aligned}
\end{equation}
This result follows from a conjugate update of the Gaussian ``prior'' $N(\bw_i; \bb_i \bw_{i^c}, \bB_i)$ on the Gaussian likelihood $N(\by_i; \bw_i, \bDelta)$. 

Without sparsity assumptions on the DAG, single-site updates are as costly if not more costly than block updates, due to the size of $i^c$ which leads to large costs in computing and sampling from each $p(\bw_i \mid \bw_{i^c})$. 
But because we assume a sparse DAG on $\bw$, we can write $p(\bw_i \mid \bw_{i^c}) = p(\bw_i \mid \bw_{b_i})$, where $b_i$ is the Markov blanket of $i$ in the DAG. The Markov blanket of $i$ is the set of nodes (locations) $j$ such that $j$ is either (1) a parent of $i$, that is $j\to i$ in the DAG, (2) a child of $i$: $i\to j$, or (3) a co-parent of $i$: there is $r$ such that $i\to r$ and $j \to r$. Based on the precision matrix $\rho_r(\calS)^{-1}$, we find $b_i$ as the list of non-zero indices of its $i$th column (or row), minus $i$ itself. Then, \eqref{eq:seq_full_blocks} simplifies with $\bb_i = -\bP(i, i)^{-1} \bP(i, b_i)$ and similarly in \eqref{eq:fullcond_nomb} where we replace $\bP(i, i^c)$ with $\bP(i, b_i)$.

Further, we can avoid calculating $\bP$ and instead directly use $\bL_j^{-1}$ for each $j=1,\dots, q$ for all computations. Denote as $\bLi_j\rowcol{:}{i}$ the $i$th column of $\bLi_j = \bL_j^{-1}$. Then, 
\begin{equation}\label{eq:seq_full_hadamard}
\begin{aligned}
   \bP(i, i)& = \{ \oplus \bLi_j\rowcol{:}{i}^{\top} \} (\bSigma^{-1} \otimes \bI_n) \{ \oplus \bLi_j\rowcol{:}{i} \} = \bSigma^{-1} \odot \begin{bmatrix} \bLi_r\rowcol{:}{i}^\top \bLi_s\rowcol{:}{i}  \end{bmatrix}_{r,s=1,\dots,q} \\
   \bP(i, b_i) &= \{ \oplus \bLi_j\rowcol{:}{i}^{\top} \} (\bSigma^{-1} \otimes \bI_n) \{ \oplus \bLi_j\rowcol{:}{b_i} \} = (\bSigma^{-1} \odot \mathbf{1}_{1, |b_i|}) 
   \begin{bmatrix} \bLi_r\rowcol{:}{i}^\top \bLi_s\rowcol{:}{b_i}  \end{bmatrix}_{r,s=1,\dots,q},
\end{aligned}
\end{equation}
where $|b_i|$ is the dimension of $b_i$ and we use a similar argument to Proposition 2.3 in the first row to express $\bP(i,i)$ as a Hadamard product. Computing \eqref{eq:seq_full_hadamard} for each $i=1, \dots, n$ as well as $\bG_i$ (and its Cholesky factor) in \eqref{eq:fullcond_nomb} can be performed in parallel. 

If $q$ is large and we factorize $\bSigma = \bA\bA^\top$ where $\bA$ has $k\ll q$ columns, then we have shown in the main article how we can achieve speedups in the block sampler via the Woodbury matrix identity. We can do something similar here.
First, denote $c_i = \{i\} \cup \{j: i \to j \}$, i.e., the set of children of $i$ along with $i$ itself; the dimension of $c_i$ is $|c_i|$. Then, we can write
\begin{equation*}
\begin{aligned}
   \bP(i, i)& = \{ \oplus \bLi_j\rowcol{c_i}{i}^{\top} \} (\bSigma^{-1} \otimes \bI_{|c_i|}) \{ \oplus \bLi_j\rowcol{c_i}{i} \} 
\end{aligned}
\end{equation*}
which is useful in reducing the inner matrix dimension. Then, $\bSigma^{-1}$ does not exist, but the unique pseudoinverse $\bSigma^+ = \bA^{+\top} \bA^+$ does, and is such that $\bA^+$ is of dimension $k \times q$. Then, in \eqref{eq:fullcond_nomb} we find 
\begin{equation} \label{eq:seq_full_woodbury}\begin{aligned}
     \bG_i^{-1} &= (\bP(i, i) + \bDelta^{-1})^{-1} =\left( \{ \oplus \bLi_j\rowcol{c_i}{i}^{\top} \} (\bA^{+\top}  \otimes \bI_{|c_i|})( \bA^+ \otimes \bI_{|c_i|}) \{ \oplus \bLi_j\rowcol{c_i}{i} \}  + \bDelta^{-1} \right)^{-1} \\
     &= (\bJ \bJ^\top + \bDelta^{-1})^{-1} = \bDelta - \bDelta \bJ(\bI_{|c_i|k} + \bJ^\top \bDelta \bJ)^{-1}\bJ^\top \bDelta,
\end{aligned} \end{equation}
where we let $\bJ =  \{ \oplus \bLi_j\rowcol{c_i}{i}^{\top} \} (\bA^{+\top}  \otimes \bI_{|c_i|} )$, which is of dimension $|c_i|q \times |c_i|k$ and we use the Woodbury matrix identity to find the last equality. The required inverse is of dimension $|c_i|k$, meaning that for \eqref{eq:seq_full_woodbury} to be advantageous compared to \eqref{eq:seq_full_hadamard}, we need $|c_i|k < q$. In nearest-neighbor DAGs, one controls the number of parents $m$ whereas the number of children varies across locations and may be large. For this reason, the radial neighbors DAG \citep{radgp} may be a convenient choice in this case; in a radial neighbors DAG, all neighbors of $\bl_i$ within radius $r$ are either parents or children of $i$ in the DAG. Consequently, $r$ indirectly bounds $|c_i| \leq | \{ \bl \in \calS : |\bl - \bl_i| \leq r\} |$. 

What we have just shown is that the structure induced by \modelname\ leads to computations of the conditional mean and covariance of $p(\bw_i \mid \bw_{b_i}, \by_i)$ by either reading elements of the precision matrix $\bP$, or via simple dot products between columns of the sparse matrices $\bLi_j = \bL_j^{-1}$ and without ever directly computing $\bL_i$ or $\rho_i(\calS)$. 

\subsection{Compute cost of proposed scalable algorithms}
We detail the computational complexity of the scalable algorithms outlined in the main article. Our assumption here is that $n$, the size of $\calS$, is very large and $n \gg q$, leading to computational cost being dominated by operations along the spatial sites.
We also assume a sparse DAG GP using $m$ nearest neighbors. We compute $\bL_i$ at $O(n m^3)$ flops for each $i=1,\dots, q$, for a total of $O(qnm^3)$.
%With clustering of the correlation functions into $k_1$ groups, this operation reduces to $O(k_1nm^3)$.
Determinant and inverse of $\bSigma$ are cheap to compute in requiring $O(q^3)$ flops, independent of $n$. 

In the block sampler for the latent model $\bM_{w\mid y}$ is a $nq \times nq$ sparse matrix. One can use iterative matrix solvers such as conjugate gradient (CG); at each iteration, CG performs a matrix-vector multiplication (cost $O(n_{\text{nz}})$), for a total number of iterations that is guaranteed to be less than $nq$ and can be reduced via preconditioning. The cost for solving $\bM_{w\mid y}$ is in the order of $O(n_{\text{nz}} c)$ where $n_{\text{nz}}$ is the number of nonzeros of $\Cov^{-1}$, and $c$ is the condition number of $\bM_{w\mid y}$. 

\section{Applications and software}

\subsection{Simulation setup: trivariate data}
In all 120 datasets, we let $\bSigma$ be a correlation matrix with off-diagonal entries $\sigma_{12} = -0.9, \sigma_{13} = 0.7, \sigma_{23} = -0.5$. The decay, smoothness, and nugget effect parameters of the three outcomes are set as $\phi_1=\phi_2=\phi_3=30$, $\nu_1=0.5, \nu_2=0.8, \nu_3=1.2$, and $\tau^2_1=\tau^2_2=\tau^2_3=10^{-3}$. No other parameters are necessary for \modelname\ datasets. For the multivariate Mat\'ern, we let the cross-variable decay be fixed at $\phi_{ij}=30$ and the smoothness as $\nu_{ij}=\frac{\nu_i + \nu_j}{2}$ for all $i,j$ and we fix the cross-variable nugget effects at $\tau^2_{ij}=10^{-3}$ for all $i,j$, resulting in a parsimonious Mat\'ern specification as in \cite{gneiting2010}. We evaluate the performance in estimating the marginal parameters as well as the cross-correlations at zero distance $\rho_{ij}$, which are computed in the multivariate Mat\'ern cases as $\rho_{ij} = \sigma_{ij} \frac{\sqrt{ \Gamma(\nu_i+1) }}{\sqrt{ \Gamma(\nu_i)}} \frac{\sqrt{ \Gamma(\nu_j+1) }}{\sqrt{\Gamma(\nu_j)}}  
\frac{ \Gamma(\nu_i/2 + \nu_j/2)} { \Gamma(\nu_i/2 + \nu_j/2 + 1) }$, whereas in the \modelname\ cases we average the values of $C_{ij}(\bzero)$ at each observed location to compute the resulting scaling factor (the Supplement details this procedure for interpreting $C_{ij}$). In \lmcs\, the cross-correlation at zero is $\bD_S^{-1} \bSigma \bD_S^{-1}$, where $\bSigma = \bA \bA^\top + \bI_q$, $\bD_S = \text{diag}\{ \sqrt{ \bSigma_{jj} } \}$, and $\bA$ is the loadings matrix. 

\subsection{Simulation setup: data with 24 outcomes}
Of the 40 datasets, we generate 20 by sampling from a GP-\modelname\ such that the $j$th variable has Mat\'ern marginal correlation with range $\phi_j=30$, nugget effect parameter $\tau^2_j=10^{-3}$; we sample $\nu_j$ independently from a discrete uniform prior on $\{0.5, 0.8, 1.1, 1.4, 1.7, 2.0\}$ and $\bSigma \sim \calW^{-1}_{q+1}\left(\frac{1}{2} \bI_q \right)$. The figure in the main article shows one of the \modelname\ synthetic datasets. The other 20 datasets are generated as GPs with \lmc\ cross-covariance with rank $k=8$ such that each of the $8$ constituent processes is Mat\'ern with smoothness $\nu_j=1$, no nugget effect, and spatial range $\phi_j$ sampled uniformly for each $j=1,\dots,8$ from a discrete distribution on a sequence of length 10 starting from $5$ and ending at $20$ with equal spacing. We generate the $24\times 8$ loading matrix $\bA$ by independently sampling each of its elements as $a_{ij} \sim N(0,1)$. Finally, we add independent Gaussian measurement error with unit variance to each of the outcomes. 

\subsection{Additional details on simulated data applications}
We report boxplots that expand on the results of Tables 1 and 2 in the main article. 

\begin{figure}[H]
    \centering
    \includegraphics[width=0.95\textwidth]{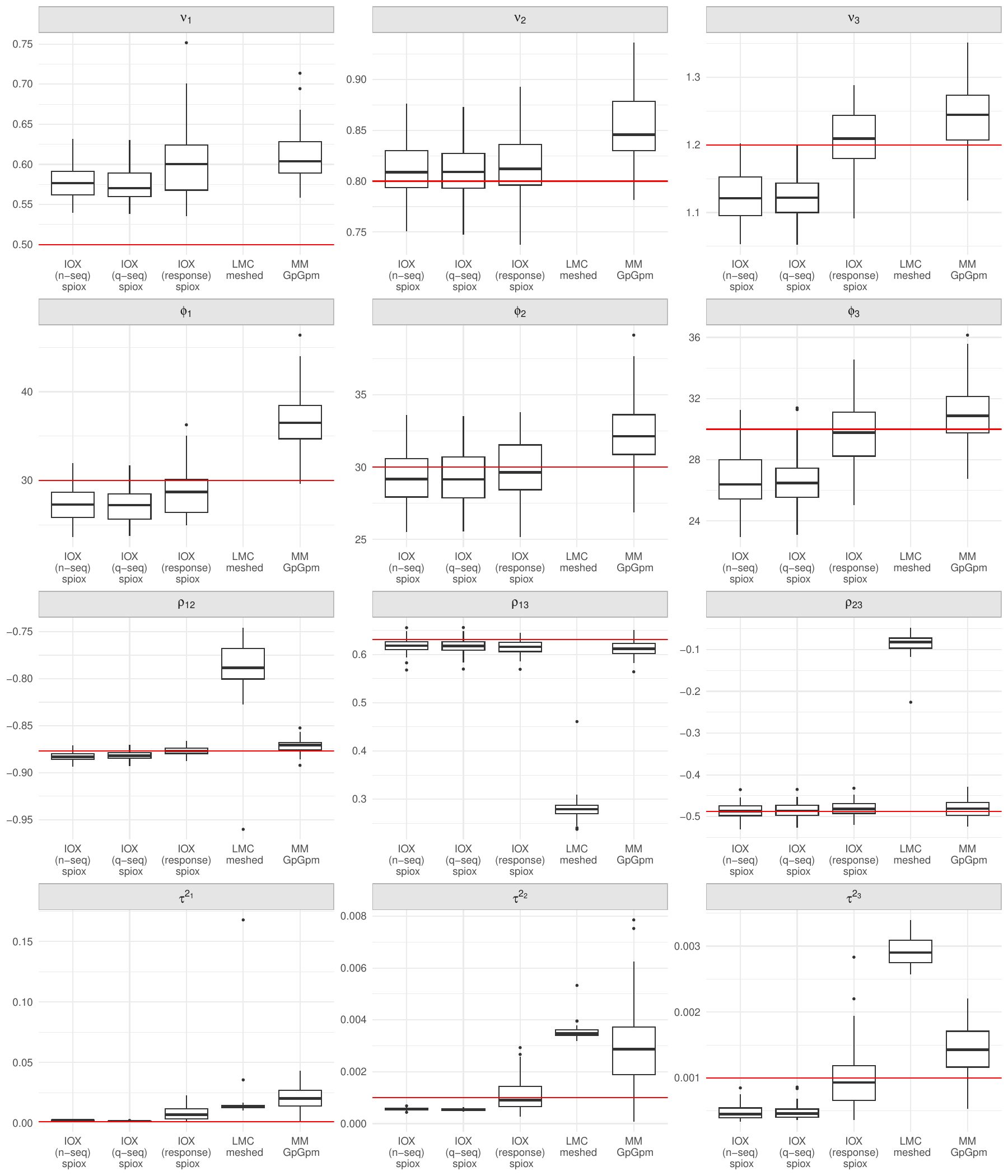}
    \caption{\footnotesize Box plots summarizing each tested model's point estimates of the corresponding covariance parameter across all 60 IOX-generated datasets. True values are in red. Table 1 in the main article reports averages based on these boxplots. \normalsize }
    \label{fig:trivariate_iox_boxplots}
\end{figure}

\begin{figure}[H]
    \centering
    \includegraphics[width=0.95\textwidth]{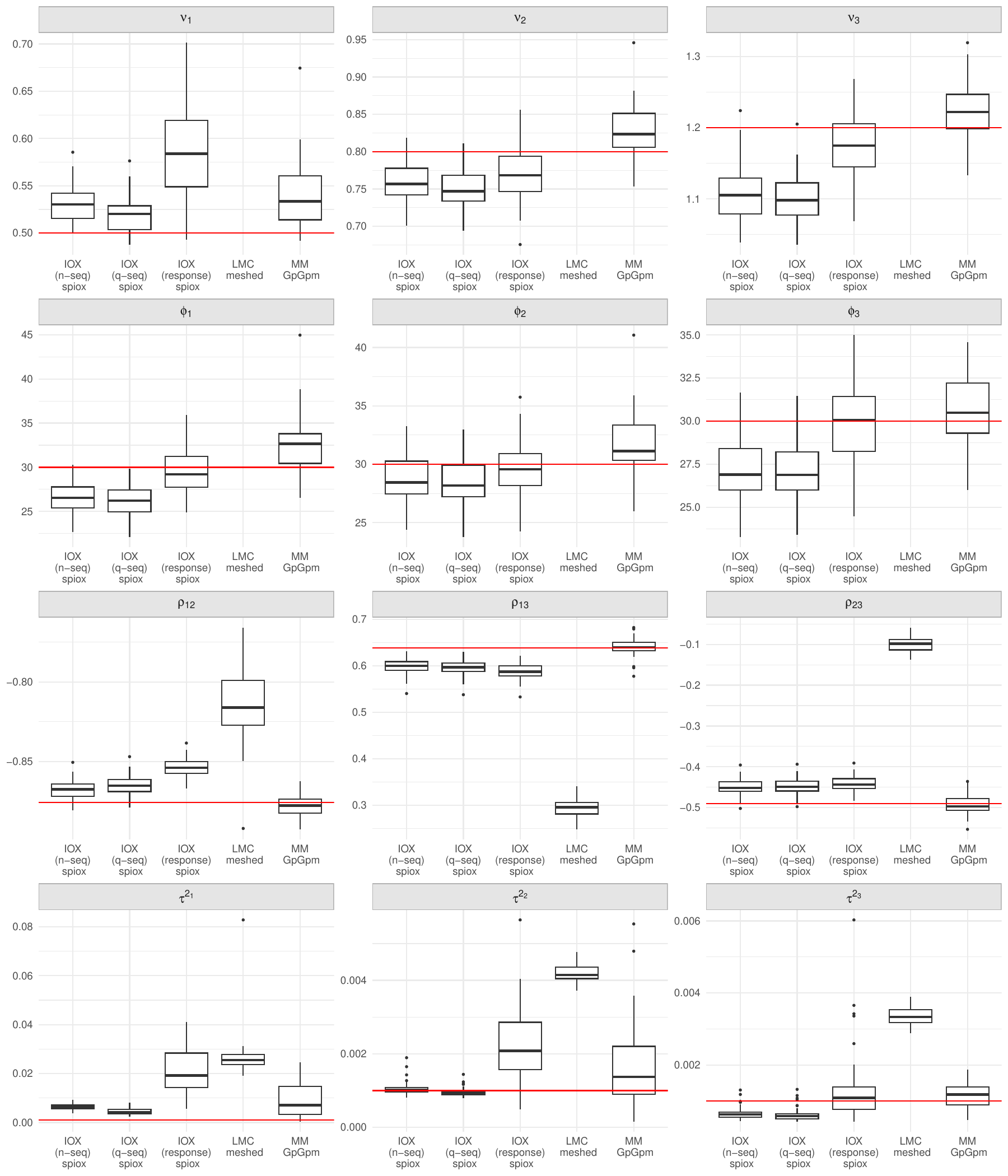}
    \caption{\footnotesize  Box plots summarizing each tested model's point estimates of the corresponding covariance parameter across all 60 multivariate Mat\'ern-generated datasets. True values are in red. Table 1 in the main article reports averages based on these boxplots. \normalsize }
    \label{fig:trivariate_matern_boxplots}
\end{figure}

\begin{figure}[H]
    \centering
    \includegraphics[width=0.95\textwidth]{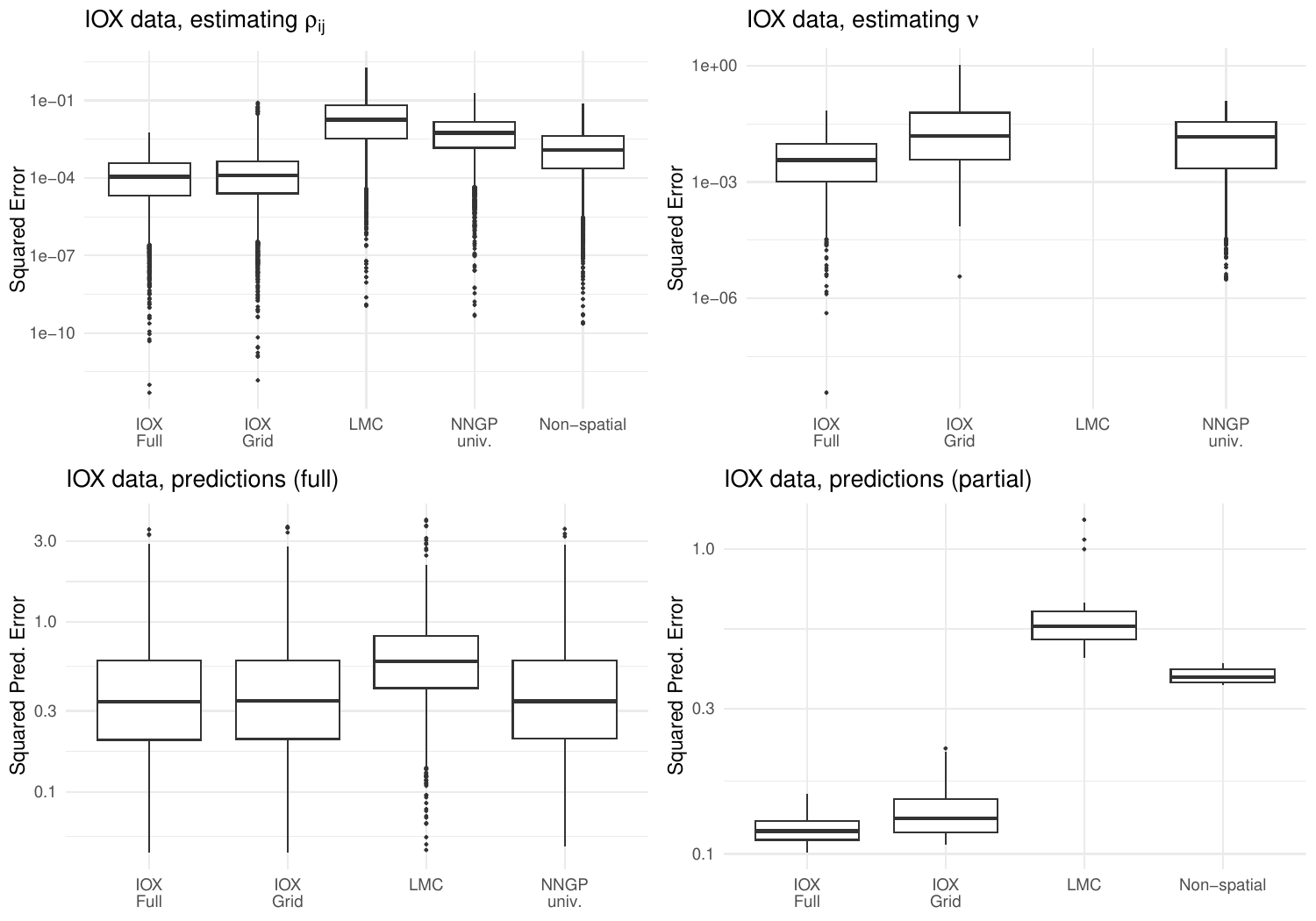}
    \caption{\footnotesize Box plots summarizing each tested model's squared error in estimating the corresponding covariance parameters across all IOX-generated datasets. True values are in red. Table 2 in the main article reports averages based on these boxplots. \normalsize }
    \label{fig:multivariate_iox_boxplots}
\end{figure}

\begin{figure}[H]
    \centering
    \includegraphics[width=0.95\textwidth]{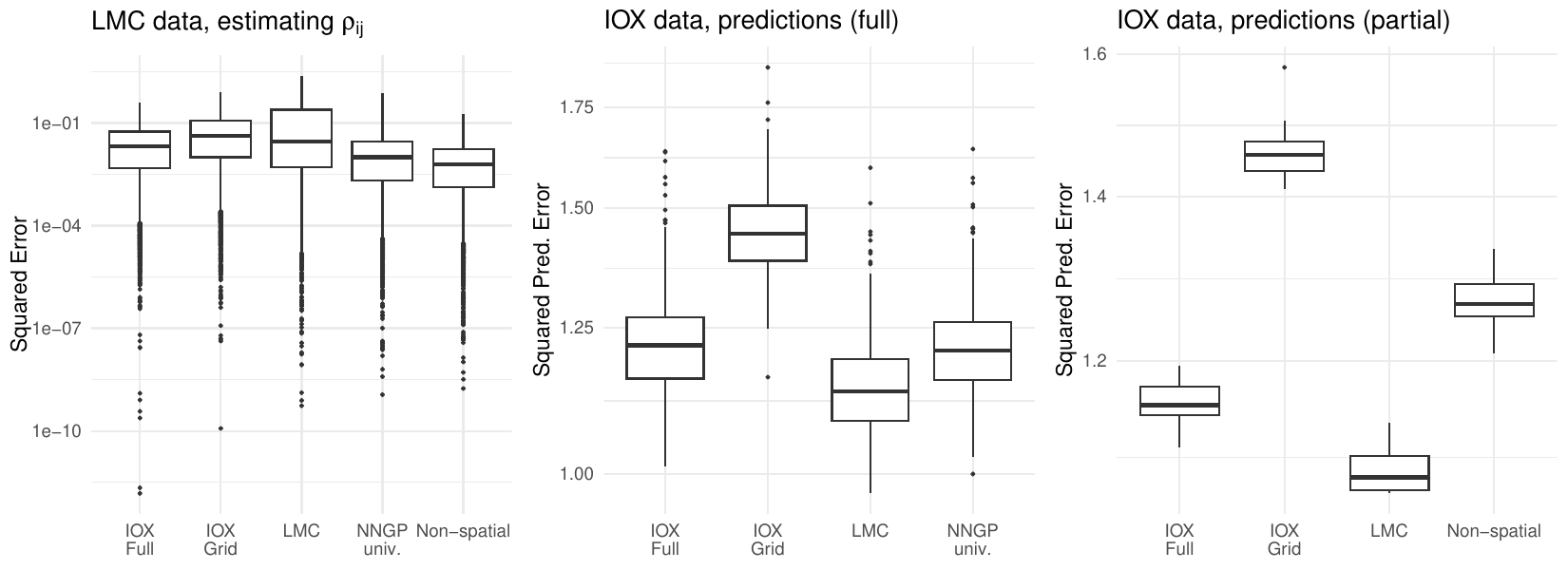}
    \caption{\footnotesize Box plots summarizing each tested model's squared error in estimating the corresponding covariance parameters across all LMC-generated datasets. True values are in red. Table 2 in the main article reports averages based on these boxplots. \normalsize }
    \label{fig:multivariate_lmc_boxplots}
\end{figure}

\subsection{\texttt{spiox} R package and code to reproduce analyses}
The supplementary material includes code to generate figures and run analyses. The R package (\repourl) is general purpose and fits the response model as well as the latent model via the three algorithms we outlined (block, $q$ sequential, $n$ sequential). The R package is implemented in C++ using the Armadillo library \citep{armadillo} via RcppArmadillo \citep{rcpparmadillo}. Parallel operations are performed via OpenMP \citep{dagum1998openmp}. For fast linear algebra, we recommend to link R with efficient BLAS/LAPACK libraries \citep{lapack99} such as OpenBLAS \citep{openblas}. All our analyses are run on a workstation equipped with a 16-core AMD Ryzen 9 7950X3D CPU, 192GB memory, using Intel Math Kernel Library version 2024.2 \citep{intel2019mkl} for fast linear algebra and SuperLU \citep{superlu} as a sparse solver.

%% file: biblio.bib
@Article{	  polyagamma,
  author	= {Nicholas G. Polson and James G. Scott and Jesse Windle},
  journal	= {Journal of the American Statistical Association},
  note		= {\doib{10.1080/01621459.2013.829001}},
  pages		= {1339-1349},
  title		= {Bayesian inference for logistic models using
		  {P}olya-{G}amma latent variables},
  volume	= {108},
  year		= {2013}
}

@Article{	  albertchib93,
  author	= {James H. Albert and Siddhartha Chib},
  journal	= {Journal of the American Statistical Association},
  note		= {\doib{10.2307/2290350}},
  number	= {422},
  pages		= {669-679},
  title		= {Bayesian Analysis of Binary and Polychotomous Response
		  Data},
  volume	= {88},
  year		= {1993}
}

@Article{	  katzfuss_jasa17,
  author	= {Matthias Katzfuss},
  issue		= {517},
  journal	= {Journal of the American Statistical Association},
  note		= {\doib{10.1080/01621459.2015.1123632}},
  pages		= {201-214},
  title		= {A Multi-Resolution Approximation for Massive Spatial
		  Datasets},
  volume	= {112},
  year		= {2017}
}

@Article{	  nngp,
  author	= {Abhirup Datta and Sudipto Banerjee and Andrew O. Finley
		  and Alan E. Gelfand},
  issue		= {514},
  journal	= {Journal of the American Statistical Association},
  note		= {\doib{10.1080/01621459.2015.1044091}},
  pages		= {800-812},
  title		= {Hierarchical Nearest-Neighbor {Gaussian} Process Models
		  for Large Geostatistical Datasets},
  volume	= {111},
  year		= {2016}
}

@Article{	  gp_predictive_process,
  author	= {Banerjee, Sudipto and Gelfand, Alan E. and Finley, Andrew
		  O. and Sang, Huiyan},
  issue		= {4},
  journal	= {Journal of the Royal Statistical Society, Series B},
  note		= {\doib{10.1111/j.1467-9868.2008.00663.x}},
  pages		= {825--848},
  title		= {Gaussian predictive process models for large spatial data
		  sets},
  volume	= {70},
  year		= {2008}
}

@Article{	  gp_pp_biasadj,
  author	= {Sudipto Banerjee and Andrew O. Finley and Patrik Waldmann
		  and Tore Ericsson},
  journal	= {Journal of American Statistical Association},
  note		= {\doib{10.1198/jasa.2009.ap09068}},
  number	= {490},
  pages		= {506-521},
  title		= {Hierarchical Spatial Process Models for Multiple Traits in
		  Large Genetic Trials},
  volume	= {105},
  year		= {2010}
}

@Misc{		  conjnngp,
  author	= {Shirota, Shinichiro and Finley, Andrew O. and Cook, Bruce
		  D. and Banerjee, Sudipto},
  note		= {\arXiv{1907.10109}},
  title		= {Conjugate Nearest Neighbor {Gaussian} Process Models for
		  Efficient Statistical Interpolation of Large Spatial Data},
  year		= {2019}
}

@Article{	  katzfuss_vecchia,
  author	= {Katzfuss, Matthias and Guinness, Joseph},
  journal	= {Statistical Science},
  title		= {A general framework for {Vecchia} approximations of
		  {Gaussian} processes},
  year		= {2021},
  volume	= {36},
  number	= {1},
  pages		= {124-141},
  note		= {\doib{10.1214/19-STS755}}
}

@Article{	  vecchia88,
  author	= {Vecchia, A. V.},
  issue		= {2},
  journal	= {Journal of the Royal Statistical Society, Series B},
  note		= {\doib{10.1111/j.2517-6161.1988.tb01729.x}},
  pages		= {297-312},
  title		= {Estimation and Model Identification for Continuous Spatial
		  Processes},
  volume	= {50},
  year		= {1988}
}

@Article{	  steinetal2004,
  author	= {Stein, Michael L. and Chi, Zhiyi and Welty, Leah J.},
  issue		= {2},
  journal	= {Journal of the Royal Statistical Society, Series B},
  note		= {\doib{10.1046/j.1369-7412.2003.05512.x}},
  pages		= {275-296},
  title		= {Approximating likelihoods for large spatial data sets},
  volume	= {66},
  year		= {2004}
}

@Article{	  genton_ccov,
  author	= {Genton, Marc G. and Kleiber, William},
  issue		= {2},
  journal	= {Statistical Science},
  note		= {\doib{10.1214/14-STS487}},
  pages		= {147-163},
  title		= {Cross-Covariance Functions for Multivariate
		  Geostatistics},
  volume	= {30},
  year		= {2015}
}

@Article{	  heaton2019,
  author	= "Heaton, Matthew J. and Datta, Abhirup and Finley, Andrew
		  O. and Furrer, Reinhard and Guinness, Joseph and
		  Guhaniyogi, Rajarshi and Gerber, Florian and Gramacy,
		  Robert B. and Hammerling, Dorit and Katzfuss, Matthias and
		  Lindgren, Finn and Nychka, Douglas W. and Sun, Furong and
		  Zammit-Mangion, Andrew",
  title		= "A Case Study Competition Among Methods for Analyzing Large
		  Spatial Data",
  journal	= "Journal of Agricultural, Biological and Environmental
		  Statistics",
  year		= "2019",
  month		= "Sep",
  day		= "01",
  volume	= "24",
  number	= "3",
  pages		= "398--425",
  note		= {\doib{10.1007/s13253-018-00348-w}}
}

@Article{	  modifiedpp,
  title		= {Improving the performance of predictive process modeling
		  for large datasets},
  author	= {Finley, Andrew O. and Sang, Huiyan and Banerjee, Sudipto
		  and Gelfand, Alan E.},
  journal	= {Computational Statistics and Data Analysis},
  year		= {2009},
  volume	= {53},
  month		= {June},
  pages		= {2873-2884},
  note		= {\doib{10.1016/j.csda.2008.09.008}}
}

@Article{	  prates,
  author	= {Quiroz, Zaida C. and Prates, Marcos O. and Dey, Dipak K.
		  and Rue, H\mathring{a}vard},
  journal	= {Statistics and Computing},
  volume	= {33},
  number	= {2},
  pages		= {54},
  note		= {\doib{10.1007/s11222-023-10227-1}},
  title		= {Fast {Bayesian} inference of Block Nearest Neighbor
		  {Gaussian} process for large data},
  year		= {2023}
}

@Article{	  apanasovich_genton2010,
  author	= {Apanasovich, Tatiyana V. and Genton, Marc G.},
  title		= {Cross-covariance functions for multivariate random fields
		  based on latent dimensions},
  journal	= {Biometrika},
  volume	= {97},
  issue		= {1},
  year		= {2010},
  pages		= {15-30},
  note		= {\doib{10.1093/biomet/asp078}}
}

@Article{	  apanasovich2012,
  title		= {A valid {M}at{\'e}rn class of cross-covariance functions
		  for multivariate random fields with any number of
		  components},
  author	= {Apanasovich, Tatiyana V. and Genton, Marc G. and Sun,
		  Ying},
  journal	= {Journal of the American Statistical Association},
  volume	= {107},
  number	= {497},
  pages		= {180--193},
  year		= {2012},
  publisher	= {Taylor \& Francis},
  note		= {\doib{10.1080/01621459.2011.643197}}
}

@Article{	  spnngp_rpack,
  author	= {Finley, Andrew O. and Datta, Abhirup and Banerjee,
		  Sudipto},
  title		= {spNNGP R Package for Nearest Neighbor {Gaussian} Process
		  Models },
  year		= {2022},
  volume	= {103},
  number	= {5},
  note		= {\doib{10.18637/jss.v103.i05}}
}

@Article{	  taylor2019spatial,
  title		= {Spatial factor models for high-dimensional and large
		  spatial data: An application in forest variable mapping},
  author	= {Taylor-Rodriguez, Daniel and Finley, Andrew O and Datta,
		  Abhirup and Babcock, Chad and Andersen, Hans Erik and Cook,
		  Bruce D and Morton, Douglas C and Banerjee, Sudipto},
  journal	= {Statistica Sinica},
  volume	= {29},
  number	= {3},
  pages		= {1155--1180},
  year		= {2019},
  publisher	= {Institute of Statistical Science},
  note		= {\doib{10.5705/ss.202018.0005}}
}

@Article{	  meshedgp,
  author	= {Peruzzi, Michele and Banerjee, Sudipto and Finley, Andrew
		  O.},
  title		= {Highly Scalable {Bayesian} Geostatistical Modeling via
		  Meshed {Gaussian} Processes on Partitioned Domains},
  journal	= {Journal of the American Statistical Association},
  year		= {2022},
  volume	= {117},
  number	= {538},
  pages		= {969-982},
  publisher	= {Taylor & Francis},
  note		= {\doib{10.1080/01621459.2020.1833889}}
}

@Article{	  katzfussgong2019,
  author	= {Katzfuss, Matthias and Gong, Wenlong},
  journal	= {Statistica Sinica},
  title		= {A class of multi-resolution approximations for large
		  spatial datasets},
  volume	= {30},
  pages		= {2203-2226},
  note		= {\doib{10.5705/ss.202018.0285}},
  year		= 2019
}

@Article{	  spbayes15,
  title		= {{spBayes} for Large Univariate and Multivariate
		  Point-Referenced Spatio-Temporal Data Models},
  author	= {Andrew O. Finley and Sudipto Banerjee and Alan E.Gelfand},
  journal	= {Journal of Statistical Software},
  year		= {2015},
  volume	= {63},
  number	= {13},
  pages		= {1--28},
  note		= {\doib{10.18637/jss.v063.i13}}
}

@Article{	  spamtrees,
  author	= {Peruzzi, Michele and Dunson, David B.},
  title		= {Spatial Multivariate Trees for Big Data {Bayesian}
		  Regression},
  journal	= {Journal of Machine Learning Research},
  year		= {2022},
  volume	= {23},
  number	= {17},
  pages		= {1-40},
  note		= {\url{http://jmlr.org/papers/v23/20-1361.html}}
}

@Article{	  matheron82,
  author	= {Matheron, G.},
  journal	= {Technical report N.732, Centre de G\'eostatistique},
  year		= {1982},
  title		= {Pour une analyse krigeante des donn\'ees r\'egionalis\'ees}
}

@Article{	  finley2008,
  author	= {Finley, Andrew O. and Banerjee, Sudipto and Ek, Alan R.
		  and McRoberts, Ronald E.},
  journal	= {Journal of Agricultural, Biological, and Environmental
		  Statistics},
  title		= {Bayesian multivariate process modeling for prediction of
		  forest attributes},
  year		= {2008},
  volume	= {13},
  pages		= {60},
  note		= {\doib{10.1198/108571108X273160}}
}

@Book{		  wackernagel03,
  author	= {Wackernagel, Hans},
  publisher	= {Springer, Berlin},
  title		= {Multivariate Geostatistics: An Introduction with
		  Applications},
  note		= {\doib{10.1007/978-3-662-05294-5}},
  year		= {2003}
}

@Article{	  gneiting2010,
  journal	= {Journal of the American Statistical Association},
  title		= {Mat\'ern Cross-Covariance Functions for Multivariate
		  Random Fields},
  author	= {Gneiting, Tilmann and Kleiber, William and Schlather,
		  Martin},
  year		= {2010},
  volume	= {105},
  number	= {491},
  pages		= {1167-1177},
  note		= {\doib{10.1198/jasa.2010.tm09420}}
}

@Article{	  zhangbanerjee20,
  author	= {Zhang, Lu and Banerjee, Sudipto},
  title		= {Spatial Factor Modeling: A {Bayesian} Matrix-Normal
		  Approach for Misaligned Data},
  journal	= {Biometrics},
  year		= {2022},
  volume	= {78},
  number	= {2},
  pages		= {560-573},
  note		= {\doib{10.1111/biom.13452}}
}

@Article{	  girolamicalderhead11,
  author	= {Girolami, Mark and Calderhead, Ben},
  title		= {Riemann manifold {Langevin} and {Hamiltonian Monte Carlo}
		  methods},
  year		= {2011},
  journal	= {Journal of the Royal Statistical Society: Series B},
  pages		= {123-214},
  volume	= {73},
  number	= {2},
  note		= {\doib{10.1111/j.1467-9868.2010.00765.x}}
}

@Article{	  schmidtgelfand,
  author	= {Schmidt, Alexandra M. and Gelfand, Alan E.},
  title		= {A {Bayesian} coregionalization approach for multivariate
		  pollutant data},
  journal	= {Journal of Geophysical Research},
  year		= {2003},
  volume	= {108},
  pages		= {D24},
  note		= {\doib{10.1029/2002JD002905}}
}

@Article{	  deyetal20,
  author	= {Dey, Debangan and Datta, Abhirup and Banerjee, Sudipto},
  title		= {Graphical {Gaussian} Process Models for Highly
		  Multivariate Spatial Data},
  journal	= {Biometrika},
  year		= {2022},
  volume	= {109},
  number	= {4},
  pages		= {993-1014},
  note		= {\doib{10.1093/biomet/asab061}}
}

@Article{	  melange,
  author	= {Michele Peruzzi and David B. Dunson},
  title		= {Spatial meshing for general {Bayesian} multivariate
		  models},
  journal	= {Journal of Machine Learning Research},
  year		= {2024},
  volume	= {25},
  number	= {87},
  pages		= {1--49},
  note		= {\url{http://jmlr.org/papers/v25/22-0083.html}}
}

@Article{	  sampsonguttorp1992,
  author	= {Sampson, Paul D. and Guttorp, Peter},
  title		= {Nonparametric Estimation of Nonstationary Spatial
		  Covariance Structure},
  journal	= {Journal of the American Statistical Association},
  volume	= {87},
  number	= {417},
  pages		= {108-119},
  year		= {1992},
  publisher	= {Taylor & Francis},
  note		= {\doib{10.1080/01621459.1992.10475181}}
}

@Article{	  gelnonstat,
  title		= {Nonstationary multivariate process modeling through
		  spatially varying coregionalization},
  author	= {Gelfand, Alan and Schmidt, Alexandra and Banerjee, Sudipto
		  and Sirmans, C. F.},
  year		= {2004},
  journal	= {Test},
  volume	= {13},
  number	= {2},
  pages		= {263-312},
  note		= {\doib{10.1007/BF02595775}}
}

@Article{	  radgp,
  title		= {Radial Neighbours for Provably Accurate Scalable
		  Approximations of {Gaussian} Processes},
  author	= {Zhu, Yichen and Peruzzi, Michele and Li, Cheng and Dunson,
		  David B.},
  year		= {2024},
  volume	= {111},
  number	= {4},
  pages		= {1151-1167},
  journal	= {Biometrika},
  note		= {\doib{10.1093/biomet/asae029}}
}

@Article{	  zbf21,
  author	= {Zhang, Lu and Banerjee, Sudipto and Finley, Andrew O.},
  title		= {High-dimensional multivariate geostatistics: A {Bayesian}
		  matrix-normal approach},
  journal	= {Environmetrics},
  volume	= {32},
  number	= {4},
  pages		= {e2675},
  note		= {\doib{10.1002/env.2675}},
  year		= {2021}
}

@Misc{		  alie24,
  title		= {Computational Considerations for the Linear Model of
		  Coregionalization},
  author	= {Alie, Renaud and Stephens, David A. and Schmidt, Alexandra
		  M.},
  note		= {\arXiv{2402.08877}},
  year		= {2024}
}

@Article{	  lmc_neural,
  title		= {Scalable multi-task {Gaussian} processes with neural
		  embedding of coregionalization},
  journal	= {Knowledge-Based Systems},
  volume	= {247},
  pages		= {108775},
  year		= {2022},
  note		= {\doib{10.1016/j.knosys.2022.108775}},
  author	= {Haitao Liu and Jiaqi Ding and Xinyu Xie and Xiaomo Jiang
		  and Yusong Zhao and Xiaofang Wang}
}

@Article{	  hmsc_package,
  title		= {Joint species distribution modelling with the {R}-package
		  Hmsc},
  author	= {Tikhonov, Gleb and Opedal, Oystein H. and Abrego, Nerea
		  and Lehikoinen, Aleksi and de Jonge, Melinda M. J. and
		  Oksanen, Jari and Ovaskainen, Otso},
  journal	= {Methods in Ecology and Evolution},
  year		= {2020},
  volume	= {11},
  pages		= {442-447},
  number	= {3},
  note		= {\doib{10.1111/2041-210X.13345}}
}

@Article{	  townes2023engelhardt,
  title		= {Nonnegative spatial factorization applied to spatial
		  genomics},
  author	= {Townes, Forest W. and Engelhardt, Barbara E.},
  journal	= {Nature Methods},
  volume	= {20},
  pages		= {229--238},
  year		= {2023},
  publisher	= {Springer Nature},
  note		= {\doib{10.1038/s41592-022-01687-w}}
}

@InProceedings{	  moreno18neurips,
  author	= {Moreno-Mu\~{n}oz, Pablo and Art\'{e}s, Antonio and
		  \'{A}lvarez, Mauricio},
  booktitle	= {Advances in Neural Information Processing Systems},
  editor	= {S. Bengio and H. Wallach and H. Larochelle and K. Grauman
		  and N. Cesa-Bianchi and R. Garnett},
  publisher	= {Curran Associates, Inc.},
  title		= {Heterogeneous Multi-output {Gaussian} Process Prediction},
  volume	= {31},
  year		= {2018}
}

@Article{	  velandia17,
  author	= {Daira Velandia and Fran{\c{c}}ois Bachoc and Moreno
		  Bevilacqua and Xavier Gendre and Jean-Michel Loubes},
  title		= {{Maximum likelihood estimation for a bivariate Gaussian
		  process under fixed domain asymptotics}},
  volume	= {11},
  journal	= {Electronic Journal of Statistics},
  number	= {2},
  pages		= {2978 -- 3007},
  year		= {2017},
  doi		= {10.1214/17-EJS1298},
  note		= {\doib{10.1214/17-EJS1298}}
}

@Article{	  zhang2007environmetrics,
  title		= {Maximum-likelihood estimation for multivariate spatial
		  linear coregionalization models},
  author	= {Hao Zhang},
  journal	= {Environmetrics},
  volume	= {18},
  number	= {2},
  pages		= {125--139},
  year		= {2007},
  note		= {\doib{10.1002/env.807}}
}

@Article{	  emery2022,
  author	= {Xavier Emery and Emilio Porcu and Philip White},
  title		= {New Validity Conditions for the Multivariate {M}at\'ern
		  Coregionalization Model, with an Application to Exploration
		  Geochemistry},
  journal	= {Mathematical Geosciences},
  year		= {2022},
  volume	= {54},
  number	= {6},
  pages		= {1043--1068},
  node		= {\doib{10.1007/s11004-022-10000-6}}
}

@Article{	  alvarez2011jmlr,
  author	= {Mauricio A. {{\'A}}lvarez and Neil D. Lawrence},
  title		= {Computationally Efficient Convolved Multiple Output
		  {Gaussian} Processes},
  journal	= {Journal of Machine Learning Research},
  year		= {2011},
  volume	= {12},
  number	= {41},
  pages		= {1459--1500}
}

@Article{	  fricker2013multivariate,
  title		= {Multivariate {Gaussian} Process Emulators With
		  Nonseparable Covariance Structures},
  author	= {Fricker, T. E. and Oakley, J. E. and Urban, N. M.},
  journal	= {Technometrics},
  volume	= {55},
  number	= {1},
  pages		= {47--56},
  year		= {2013},
  publisher	= {Taylor \& Francis},
  node		= {\doib{10.1080/00401706.2012.715835}}
}

@Article{	  schafer2024sparse,
  title		= {Sparse {Cholesky} Factorization by {Kullback}--{Leibler}
		  Minimization},
  author	= {Schäfer, Florian and Katzfuss, Matthias and Owhadi,
		  Houman},
  journal	= {SIAM Journal on Scientific Computing},
  note		= {\doib{10.1137/20M1336254}},
  year		= {2024}
}

@InProceedings{	  teh2005semiparametric,
  author	= {Teh, Yee Whye and Seeger, Matthias and Jordan, Michael
		  I.},
  title		= {Semiparametric Latent Factor Models},
  editor	= {Cowell, Robert G. and Ghahramani, Zoubin},
  booktitle	= {Proceedings of the Tenth International Workshop on
		  Artificial Intelligence and Statistics},
  pages		= {333--340},
  publisher	= {Society for Artificial Intelligence and Statistics},
  year		= {2005},
  note		= {\small\url{http://proceedings.mlr.press/r5/teh05a/teh05a.pdf}\normalsize}
}

@Article{	  majumdar2007convol,
  author	= {Majumdar, Anandamayee and Gelfand, Alan E.},
  title		= {Multivariate Spatial Modeling for Geostatistical Data
		  Using Convolved Covariance Functions},
  journal	= {Mathematical Geology},
  volume	= {39},
  number	= {2},
  pages		= {225--245},
  year		= {2007},
  note		= {\doib{10.1007/s11004-006-9072-6}}
}

@Article{	  gaspari1999construction,
  author	= {Gaspari, Gregory and Cohn, Stephen E.},
  title		= {Construction of Correlation Functions in Two and Three
		  Dimensions},
  journal	= {Quarterly Journal of the Royal Meteorological Society},
  volume	= {125},
  number	= {554},
  pages		= {723--757},
  month		= {January},
  year		= {1999},
  note		= {\doib{10.1002/qj.49712555417}}
}

@Misc{		  yarger24matern,
  author	= {Drew Yarger and Stilian Stoev and Tailen Hsing},
  title		= {Multivariate {M}at\'ern Models -- A Spectral Approach},
  note		= {\arXiv{2309.02584}},
  year		= {2024}
}

@Article{	  fahmy2022vecchia,
  title		= {Vecchia Approximations and Optimization for Multivariate
		  {M}at\'ern Models},
  author	= {Youssef Fahmy and Joseph Guinness},
  journal	= {Journal of Data Science},
  volume	= {20},
  number	= {4},
  pages		= {475--492},
  year		= {2022},
  month		= {October},
  note		= {\doib{10.6339/22-JDS1074}}
}

@Article{	  berrocal2010bivariate,
  title		= {A bivariate space--time downscaler under space and time
		  misalignment},
  author	= {Berrocal, Veronica and Gelfand, Alan E and Holland, David
		  M},
  journal	= {The Annals of Applied Statistics},
  volume	= {4},
  number	= {4},
  pages		= {1942--1975},
  year		= {2010},
  notes		= {\doib{10.1214/10-AOAS351}}
}

@Article{	  gelfand2003spatial,
  title		= {Spatial Modeling With Spatially Varying Coefficient
		  Processes},
  author	= {Gelfand, Alan E. and Kim, Hyon J. and Sirmans, C. F. and
		  Banerjee, Sudipto},
  journal	= {Journal of the American Statistical Association},
  volume	= {98},
  number	= {462},
  pages		= {387--396},
  year		= {2003},
  note		= {\doib{10.1198/016214503000170}}
}

@Article{	  de_iaco_choosing_2019,
  title		= {Choosing suitable linear coregionalization models for
		  spatio-temporal data},
  author	= {De Iaco, Santina and Palma, Michele and Posa, Domenico},
  journal	= {Stochastic Environmental Research and Risk Assessment},
  volume	= {33},
  pages		= {1419--1434},
  year		= {2019},
  publisher	= {Springer},
  note		= {\doib{10.1007/s00477-019-01701-2}}
}

@Article{	  reich_bayesian_2010,
  title		= {Bayesian Variable Selection for Multivariate Spatially
		  Varying Coefficient Regression},
  author	= {Reich, Brian J. and Fuentes, Montserrat and Herring, Amy
		  H. and Evenson, Kelly R.},
  journal	= {Biometrics},
  volume	= {66},
  number	= {3},
  pages		= {772--782},
  year		= {2010},
  month		= {September},
  publisher	= {Wiley},
  note		= {\doib{10.1111/j.1541-0420.2009.01333.x}}
}

@Article{	  finazzifasso,
  title		= {{D-STEM}: A Software for the Analysis and Mapping of
		  Environmental Space-Time Variables},
  volume	= {62},
  note		= {\doib{10.18637/jss.v062.i06}},
  number	= {6},
  journal	= {Journal of Statistical Software},
  author	= {Finazzi, Francesco and Fass\`o, Alessandro},
  year		= {2014},
  pages		= {1–29}
}

@Article{	  gstat,
  title		= {Multivariable geostatistics in {S}: the {gstat} package},
  author	= {Pebesma, Edzer J.},
  journal	= {Computers \& Geosciences},
  volume	= {30},
  number	= {7},
  pages		= {683--691},
  year		= {2004},
  publisher	= {Elsevier},
  doi		= {10.1016/j.cageo.2004.03.012}
}

@Book{		  krainski_advanced_2019,
  title		= {Advanced Spatial Modeling with Stochastic Partial
		  Differential Equations Using {R} and {INLA}},
  author	= {Krainski, Elias T. and G\'omez-Rubio, Virgilio and Bakka,
		  Haakon and Lenzi, Amanda and Castro-Camilo, Daniela and
		  Simpson, Daniel and Lindgren, Finn and Rue,
		  H\mathring{a}vard},
  publisher	= {Chapman \& Hall/CRC Press},
  year		= {2019}
}

@Article{	  schurch2020,
  author	= {Sch\"urch, Christian M. and Bhate, Salil S. and Barlow,
		  Graham L. and Phillips, Darci J. and Noti, Luca and Zlobec,
		  Inti and Chu, Pauline and Black, Sarah and Demeter, Janos
		  and McIlwain, David R. and Kinoshita, Shigemi and Samusik,
		  Nikolay and Goltsev, Yury and Nolan, Garry P.},
  title		= {Coordinated Cellular Neighborhoods Orchestrate Antitumoral
		  Immunity at the Colorectal Cancer Invasive Front},
  journal	= {Cell},
  year		= {2020},
  volume	= {182},
  number	= {5},
  pages		= {1341-1359.e19},
  note		= {\doib{10.1016/j.cell.2020.07.005}}
}

@Article{	  bradley_multivariate_2015,
  title		= {Multivariate spatio-temporal models for high-dimensional
		  areal data with application to {Longitudinal}
		  {Employer}-{Household} {Dynamics}},
  volume	= {9},
  issn		= {1932-6157, 1941-7330},
  note		= {\doib{10.1214/15-AOAS862}},
  number	= {4},
  urldate	= {2025-07-29},
  journal	= {The Annals of Applied Statistics},
  author	= {Bradley, Jonathan R. and Holan, Scott H. and Wikle,
		  Christopher K.},
  month		= dec,
  year		= {2015},
  pages		= {1761--1791}
}

@Article{	  bradley_computationally_2018,
  title		= {Computationally {Efficient} {Multivariate}
		  {Spatio}-{Temporal} {Models} for {High}-{Dimensional}
		  {Count}-{Valued} {Data} (with {Discussion})},
  volume	= {13},
  issn		= {1936-0975, 1931-6690},
  note		= {\doib{10.1214/17-BA1069}},
  number	= {1},
  urldate	= {2025-07-29},
  journal	= {Bayesian Analysis},
  author	= {Bradley, Jonathan R. and Holan, Scott H. and Wikle,
		  Christopher K.},
  month		= mar,
  year		= {2018},
  pages		= {253--310}
}

@Article{	  krock_modeling_2023,
  title		= {Modeling {Massive} {Highly} {Multivariate} {Nonstationary}
		  {Spatial} {Data} with the {Basis} {Graphical} {Lasso}},
  volume	= {32},
  issn		= {1061-8600},
  number	= {4},
  urldate	= {2025-07-21},
  journal	= {Journal of Computational and Graphical Statistics},
  author	= {Krock, Mitchell L. and Kleiber, William and Hammerling,
		  Dorit and Becker, Stephen},
  month		= oct,
  year		= {2023},
  pages		= {1472--1487},
  note		= {\doib{10.1080/10618600.2023.2174126}}
}

@Article{	  hanahan_hallmarks_2011,
  title		= {Hallmarks of {Cancer}: {The} {Next} {Generation}},
  volume	= {144},
  issn		= {0092-8674, 1097-4172},
  shorttitle	= {Hallmarks of {Cancer}},
  note		= {\doib{10.1016/j.cell.2011.02.013}},
  language	= {English},
  number	= {5},
  urldate	= {2025-07-12},
  journal	= {Cell},
  author	= {Hanahan, Douglas and Weinberg, Robert A.},
  month		= mar,
  year		= {2011},
  pmid		= {21376230},
  pages		= {646--674}
}

@Article{	  de_visser_evolving_2023,
  title		= {The evolving tumor microenvironment: {From} cancer
		  initiation to metastatic outgrowth},
  volume	= {41},
  issn		= {1878-3686},
  shorttitle	= {The evolving tumor microenvironment},
  note		= {\doib{10.1016/j.ccell.2023.02.016}},
  language	= {eng},
  number	= {3},
  journal	= {Cancer Cell},
  author	= {de Visser, Karin E. and Joyce, Johanna A.},
  month		= mar,
  year		= {2023},
  pmid		= {36917948},
  pages		= {374--403}
}

@Article{	  yuan_spatial_2016,
  title		= {Spatial {Heterogeneity} in the {Tumor}
		  {Microenvironment}},
  volume	= {6},
  issn		= {2157-1422},
  note		= {\doib{10.1101/cshperspect.a026583}},
  language	= {eng},
  number	= {8},
  journal	= {Cold Spring Harbor Perspectives in Medicine},
  author	= {Yuan, Yinyin},
  month		= aug,
  year		= {2016},
  pmid		= {27481837},
  pmcid		= {PMC4968167},
  pages		= {a026583}
}

@Article{	  giraldo_clinical_2019,
  title		= {The clinical role of the {TME} in solid cancer},
  volume	= {120},
  copyright	= {2018 Cancer Research UK},
  issn		= {1532-1827},
  note		= {\doib{10.1038/s41416-018-0327-z}},
  language	= {en},
  number	= {1},
  urldate	= {2025-02-11},
  journal	= {British Journal of Cancer},
  author	= {Giraldo, Nicolas A. and Sanchez-Salas, Rafael and Peske,
		  J. David and Vano, Yann and Becht, Etienne and Petitprez,
		  Florent and Validire, Pierre and Ingels, Alexandre and
		  Cathelineau, Xavier and Fridman, Wolf Herman and
		  Sautès-Fridman, Catherine},
  month		= jan,
  year		= {2019},
  pages		= {45--53}
}

@Article{	  tsujikawa_prognostic_2020,
  title		= {Prognostic significance of spatial immune profiles in
		  human solid cancers},
  volume	= {111},
  issn		= {1347-9032},
  note		= {\doib{10.1111/cas.14591}},
  number	= {10},
  urldate	= {2025-02-11},
  journal	= {Cancer Science},
  author	= {Tsujikawa, Takahiro and Mitsuda, Junichi and Ogi, Hiroshi
		  and Miyagawa-Hayashino, Aya and Konishi, Eiichi and Itoh,
		  Kyoko and Hirano, Shigeru},
  month		= oct,
  year		= {2020},
  pmid		= {32726495},
  pmcid		= {PMC7540978},
  pages		= {3426--3434}
}

@Article{	  nngp_algos,
  author	= {Finley, Andrew O. and Datta, Abhirup and Cook, Bruce D.
		  and Morton, Douglas C. and Andersen, Hans E. and Banerjee,
		  Sudipto},
  issue		= {2},
  journal	= {Journal of Computational and Graphical Statistics},
  note		= {\doib{10.1080/10618600.2018.1537924}},
  pages		= {401-414},
  title		= {Efficient Algorithms for {Bayesian} Nearest Neighbor
		  {Gaussian} Processes},
  volume	= {28},
  year		= {2019}
}

@Article{	  armadillo,
  title		= {Armadillo: a template-based {C++} library for linear
		  algebra},
  author	= {Sanderson, Conrad and Curtin, Ryan},
  journal	= {Journal of Open Source Software},
  volume	= {1},
  pages		= {26},
  year		= {2016}
}

@Article{	  rcpparmadillo,
  title		= {{RcppArmadillo}: Accelerating {R} with high-performance
		  {C++} linear algebra},
  author	= {Dirk Eddelbuettel and Conrad Sanderson},
  journal	= {Computational Statistics and Data Analysis},
  year		= {2014},
  volume	= {71},
  month		= {March},
  pages		= {1054--1063},
  note		= {\doib{10.1016/j.csda.2013.02.005}}
}

@Article{	  dagum1998openmp,
  author	= {Dagum, Leonardo and Menon, Ramesh},
  journal	= {Computational Science \& Engineering, IEEE},
  number	= {1},
  pages		= {46--55},
  publisher	= {IEEE},
  title		= {{OpenMP}: an industry standard API for shared-memory
		  programming},
  volume	= {5},
  year		= {1998}
}

@Article{	  cholmod,
  author	= {Chen, Yanqing and Davis, Timothy A. and Hager, William W.
		  and Rajamanickam, Sivasankaran},
  title		= {Algorithm 887: {CHOLMOD}, Supernodal Sparse {Cholesky}
		  Factorization and Update/Downdate},
  year		= {2008},
  volume	= {35},
  number	= {3},
  note		= {\doib{10.1145/1391989.1391995}},
  journal	= {ACM Trans. Math. Softw.}
}

@Article{	  ren2013,
  title		= {Hierarchical Factor Models for Large Spatially Misaligned
		  Data: A Low-Rank Predictive Process Approach},
  author	= {Ren, Qian and Banerjee, Sudipto},
  note		= {\doib{10.1111/j.1541-0420.2012.01832.x}},
  journal	= {Biometrics},
  number	= {1},
  pages		= {19--30},
  pmid		= {23379832},
  volume	= {69},
  year		= {2013}
}

@Book{		  lapack99,
  author	= {Anderson, E. and Bai, Z. and Bischof, C. and Blackford, S.
		  and Demmel, J. and Dongarra, J. and Du Croz, J. and
		  Greenbaum, A. and Hammarling, S. and McKenney, A. and
		  Sorensen, D.},
  title		= {{LAPACK} Users' Guide},
  edition	= {Third},
  publisher	= {Society for Industrial and Applied Mathematics},
  year		= {1999},
  address	= {Philadelphia, PA}
}

@Manual{	  openblas,
  title		= "{An Optimized BLAS Library Based on GotoBLAS2.}",
  author	= "Zhang, Xianyi",
  url		= "https://github.com/xianyi/OpenBLAS/",
  year		= "2020"
}

@Article{	  zhang04,
  title		= {Inconsistent Estimation and Asymptotically Equal
		  Interpolations in Model-Based Geostatistics},
  author	= {Zhang, Hao},
  year		= {2004},
  volume	= {99},
  number	= {465},
  pages		= {250-261},
  journal	= {Journal of the American Statistical Association},
  note		= {\doib{10.1198/016214504000000241}}
}

@Article{	  stein90,
  title		= {Uniform Asymptotic Optimality of Linear Predictions of a
		  Random Field Using an Incorrect Second-Order Structure},
  author	= {Stein, Michael},
  year		= {1990},
  volume	= {18},
  number	= {2},
  pages		= {850-872},
  journal	= {The Annals of Statistics},
  note		= {\doib{10.1214/aos/1176347629}}
}

@Article{	  papasroberts07,
  author	= {Papaspiliopoulos, Omiros and Roberts Gareth O. and Sköld,
		  Martin},
  journal	= {Statistical Science},
  title		= {A General Framework for the Parametrization of
		  Hierarchical Models},
  year		= {2007},
  volume	= {22},
  number	= {1},
  pages		= {59-73},
  note		= {\doib{10.1214/088342307000000014}}
}

@Article{	  grips,
  title		= {Gridding and parameter expansion for scalable latent
		  {Gaussian} models of spatial multivariate data},
  issn		= {1936-0975, 1931-6690},
  notes		= {\doib{10.1214/25-BA1515}},
  urldate	= {2025-05-08},
  journal	= {Bayesian Analysis},
  publisher	= {International Society for Bayesian Analysis},
  author	= {Peruzzi, Michele and Banerjee, Sudipto and Dunson, David
		  B. and Finley, Andrew O.},
  month		= jan,
  year		= {2025},
  pages		= {1--27}
}

@Article{	  nishimura2022prior,
  title		= {Prior-Preconditioned Conjugate Gradient Method for
		  Accelerated {Gibbs} Sampling in ``Large n, Large p''
		  {Bayesian} Sparse Regression},
  author	= {Nishimura, Akihiko and Suchard, Marc A.},
  journal	= {Journal of the American Statistical Association},
  volume	= {118},
  number	= {544},
  pages		= {2468--2481},
  year		= {2023},
  note		= {\doib{10.1080/01621459.2022.2057859}}
}

@Article{	  kleiber2017coherence,
  title		= {Coherence for Multivariate Random Fields},
  author	= {Kleiber, William},
  journal	= {Statistica Sinica},
  volume	= {27},
  number	= {4},
  pages		= {1675--1697},
  year		= {2017},
  note		= {\doib{10.5705/ss.202015.0309}}
}

@Manual{	  intel2019mkl,
  title		= {Intel Math Kernel Library},
  author	= {{Intel Corporation}},
  year		= {2024},
  note		= {Version 2024-2, Software available at
		  \url{https://software.intel.com/content/www/us/en/develop/tools/math-kernel-library.html}}
}

@Article{	  superlu,
  author	= {Xiaoye S. Li},
  title		= {An Overview of {SuperLU}: Algorithms, Implementation, and
		  User Interface},
  journal	= {ACM Transactions on Mathematical Software},
  volume	= {31},
  number	= {3},
  month		= {September},
  year		= {2005},
  pages		= {302-325}
}
